\documentclass[aps,prl,twocolumn,showpacs,floatfix,superscriptaddress,nofootinbib]{revtex4-2}

\usepackage{graphicx}
\usepackage{amsmath,amssymb,bm}
\usepackage{xcolor}

\newcommand{\seff}{\sigma_{\mathrm{eff}}}
\newcommand{\Qel}{Q_{\mathrm{el}}}
\usepackage[colorlinks=false,hidelinks]{hyperref}

\newcommand{\sA}{\sigma_{y}}
\newcommand{\sApl}{\sigma_{y}^{\,(\mathrm{pl})}}   
\newcommand{\hmo}{h_{-1}}
\newcommand{\hV}{h_{-1}\,V_{\varepsilon}}
\newcommand{\dm}{\delta m}
\newcommand{\ee}[1]{\times10^{#1}}
\newcommand{\Veps}{V_{\varepsilon}}
\newcommand{\tauopt}{\tau_{\mathrm{opt}}}

\makeatletter
\AtBeginDocument{%
  \long\def\@makefntext#1{%
    \def\baselinestretch{1}%
    \parindent\z@
    \noindent\@makefnmark\,#1\par}%
  \begingroup\footnotesize
    \global\setlength{\footnotesep}{\dimexpr 1.5\baselineskip-\dp\strutbox\relax}%
  \endgroup
}

\makeatother
\graphicspath{{../}{./}}

\makeatletter
\renewcommand{\date}[1]{\gdef\@date{#1}}
\makeatother

\makeatletter
\def\frontmatter@RRAPformat#1{%
 \removelastskip
 \begingroup
  \addvspace{0.2\baselineskip}%
  \frontmatter@RRAP@format
  #1\par
 \endgroup
}%
\makeatother

\begin{document}
\raggedbottom

\title{$1/f$ frequency noise in mechanical resonators\\
scales inversely with volume, not with quality factor}

\author{M.~L.~Roukes}
\email{roukes@caltech.edu}
\affiliation{Kavli Nanoscience Institute and Departments of Physics, Applied Physics, and Bioengineering,\\
California Institute of Technology, Pasadena, California 91125, USA}

\date{Submitted: 15 September 2026 / v.~R2}

\begin{abstract}
\setlength{\parindent}{0pt}%
Mechanical frequency-shift sensors have never reached the resolution their thermomechanical or quantum limits predict: a $1/f$ frequency-noise floor keeps them 1--4 orders short of it. Across the published record, the product of the floor's flicker coefficient $\hmo$ and the strain-weighted resonator volume, $\Veps$, is size-independent over 13 decades. It is also independent of measured $Q$; I find that $Q$--volume covariance underlies reported $Q$-dependence. It is \textit{defect density} that imposes the frequency-noise floor: prestress can dilute it, but only in fixed-end devices. Mass resolution follows as $|\dm|\propto\sqrt{V}$.
\end{abstract}

\maketitle

Resonant sensors convert a perturbation --- an adsorbed mass, an axial force or strain, a temperature change, an absorbed photon --- into shifts of their resonance frequencies. The resolution of such frequency-shift measurements is bounded by the resonator's own frequency fluctuations: these are characterized by the Allan deviation $\sA(\tau)$ at the measurement averaging time, $\tau$. Accordingly, the best performance a frequency-shift sensor can provide is obtained by operating at the minimum, $\sA(\tauopt)$. This experimental optimum is attainable if the value of $\tauopt$ permits sufficient bandwidth to extract the measurand's information. 

However, the minimum observed in experiments, $\sA(\tauopt)$, has \emph{never} reached the thermomechanical or quantum
limits;\footnote{Operation near the thermomechanical limit has been reported, but only: (\textit{i}) at reduced drive, which raises that limit itself~\cite{Sadeghi2020}; (\textit{ii}) in an extrinsically gas-damped beam, where drift closes the averaging window before the $1/f$ floor the same beam shows in vacuum is reached~\cite{Roy2018}; or (\textit{iii}) with a control loop employing a second mode, which cancels common-mode signals, \emph{the measurand included}, along with the noise~\cite{Gavartin2013} (SM~S20).} a $1/f$ frequency-noise floor determines the attainable resolution (Supplemental Material~\cite{SM}, hereafter SM, S20). This minimum generally manifests as a \textit{plateau} that is roughly constant over a range of $\tau$.\footnote{Thermomechanical noise, imposed by the fluctuation--dissipation theorem~\cite{ClelandRoukes2002,SchmidBook2016}, sets a limit that lies 1--4 decades \textit{below} the $1/f$ frequency-noise plateau. The branches of $\sA(\tau)$ and the plateau are depicted schematically in Fig.~\ref{fig:allan} of End Matter.}   

This $1/f$ \textit{plateau} in $\sA$ has been reported for more than two decades. Across every material, device size, and transduction scheme yet examined, it has remained 1--4 orders above the fundamental limits~\cite{Bachtold2022} (SM~S5). No explanation has been offered that holds across platforms, and, as argued below, it has never been properly \emph{scaled} in any previous analysis. As this plateau is \textit{parametric} in origin, its underlying fluctuations are transduced identically to the signal --- so neither drive, readout, nor $Q$ can remediate them (see End Matter).

\begin{figure*}[!t]
\includegraphics[width=\textwidth]{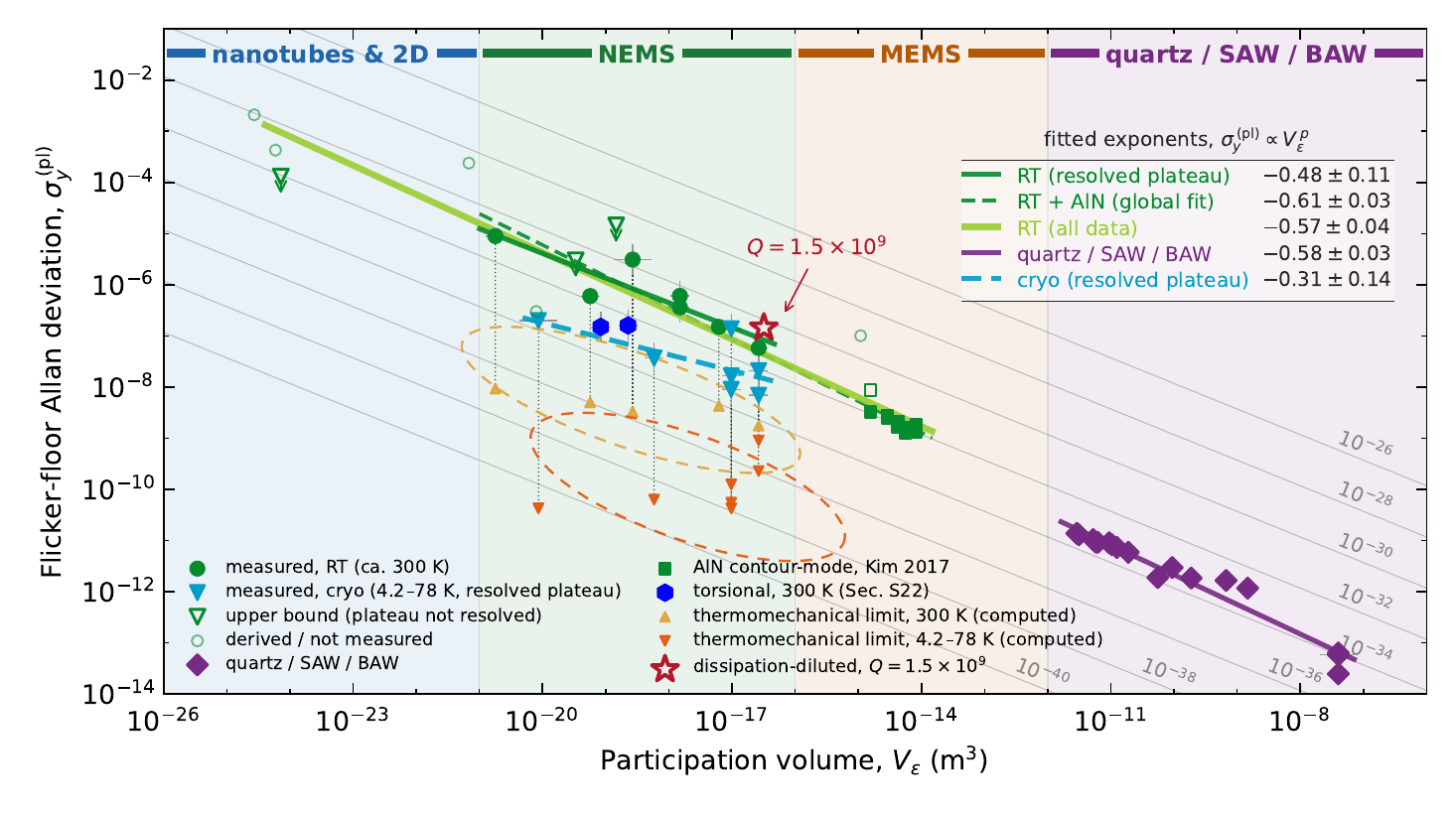}
\caption{\label{fig:survey}(Color online)
The Allan deviation on the $1/f$ plateau, $\sApl$, for every resonator in the published record (admitted by the uniformly applied rules of SM~S1), against participation volume, $\Veps$. As seen, no device reaches its own thermomechanical limit: the dashed ellipses enclosing computed limits lie below the measured room-temperature floors (the ellipses are schematic guides, not fits). Fine dotted lines join resolved floors (filled markers) to their own thermomechanical limits, computed at each device's temperature and averaging time (SM~S5, Table~S4); these gaps span multiple decades. In each family, $\sApl$ falls with increasing $\Veps$, yielding a slope near $-1/2$ (key, upper right), and no systematic $Q$ ordering is resolved (SM~S17). The open star, a dissipation-diluted resonator with $Q=1.5\ee{9}$ (an upper bound only, not fitted; SM~S4) sits, perhaps surprisingly, \textit{above} the room-temperature trend, despite its high $Q$. Solid green and purple and dashed cyan lines fit resolved plateaus only; the dashed dark-green line adds the AlN contour-mode devices to the room-temperature fit; the solid yellow-green line fits the full room-temperature population. Filled hexagons represent the torsional family (SM~S22). Open markers --- upper bounds, devices without a measured $\sA$, and the star --- enter no fit. The light grey diagonals are reference contours of constant $\hV$. }
\end{figure*}

\emph{Frequency fluctuation noise.} --- Frequency stability is conventionally decomposed using the power-law basis introduced in~\cite{Barnes1971} and codified in IEEE Std.~1139~\cite{IEEE1139}: 
\begin{equation}
    S_y(f)=\sum_\alpha h_\alpha f^\alpha \ .
\end{equation} 
Here $S_y(f)$ is the spectral density of the fractional frequency fluctuations,
$y=(f-f_0)/f_0$ the fractional frequency deviation, $f$ the frequency offset from  carrier, $f_0$ the resonance frequency, and $\alpha$ an integer. This basis is a convention rather than a completeness theorem --- for example, surface diffusion of adsorbates yields $S_y\propto f^{-1/2}$ and, hence, $\alpha=-1/2$ \cite{Yang2011}. The separate branches of $\sA$, which appear over different ranges of $\tau$, are associated with different mechanisms of frequency instability (End Matter, Fig.~\ref{fig:allan}).

\textit{The anomalous noise arises from the} $\alpha=-1$ \textit{branch.} What is called $1/f$ frequency noise can be characterized as $S_y(f)=\hmo/f^{\beta}$; in experiments $\beta \sim1$. The associated Allan variance\footnote{The transformation is \ \(\sigma _{y}^{2}(\tau )=2\int _{0}^{\infty }S_{y}(f)\frac{\sin ^{4}(\pi f\tau )}{(\pi f\tau )^{2}}df \)\ \ \cite{Allan1966}.} is 
\begin{equation}
\sA^2 \;=\; 2\ln2\;\hmo \ ,
\label{eq:flicker}
\end{equation}
a constant. Accordingly, this $\tau$-independent value, $\sApl\equiv\sqrt{2\ln2\,\hmo}$, \textit{is the absolute height} of the $1/f$ plateau in $\sA(\tau)$ (Fig.~\ref{fig:allan}, End Matter).

\emph{Why $1/f$?} --- Although a two-state fluctuator with correlation time $\tau_c$ contributes only a single Lorentzian to the power spectral density~\cite{Machlup1954}, an \textit{ensemble} of fluctuators with correlation times distributed log-uniformly, $P(\tau_c)\propto1/\tau_c$, will sum to $1/f$~\cite{Bernamont1937,vanderZiel1950}. Thermal activation with uniformly distributed barrier heights~\cite{duPre1950,DuttaHorn1981} and tunneling through barriers of uniformly distributed width~\cite{McWhorter1957} both generate such weighting.\footnote{Asymmetry in the potential wells changes the weighting of each fluctuator's amplitude, but leaves the spectral shape unchanged. The amplitude is largest for near-degenerate well minima, when the offset is small compared with $k_BT$.} This is the origin of the $\tau$-independent plateau in $\sA$ (SM S12). While the spectrum alone does not help identify the fluctuators~\cite{Weissman1988}, the invariant $\hV$ defined below serves to constrain them.

\emph{New survey.} --- The disparity in $\sA$ between data and predictions was previously examined in \cite{Sansa2016}, on which I am an author. There, Fig.~1 showed a trend in $\sA (\tau)$ versus resonator mass for 25 devices, which was left unexplained. The new compilation here includes 41 additional devices --- 66 in total, with data for each carefully extracted from 38 original studies. Reconstruction from primary sources is necessary, as reported  averaging times and mass conventions are mixed and not recoverable from \cite{Sansa2016} (SM S2). Also, $\sApl$ carries both geometric and material factors; the product $\hV$, analyzed here against the \textit{strain participation volume} $\Veps\equiv(\int\varepsilon^{2}dV)^{2}/\!\int\varepsilon^{4}dV$ --- the volume the modal strain $\varepsilon(\mathbf r)$ actually samples, evaluated per mode class (see End Matter) --- separates them.

\textit{The product $\hV$ is volume invariant.} --- Within each \textit{family} of the record assembled here, $\hV$ is consistent with a single value, independent of device size; together the families span 13 decades of $\Veps$. This is the first application of a volume invariant to nanoelectromechanical systems (NEMS). An invariant for surface-acoustic-wave (SAW) resonators, $\hmo^{\rm (vel)}V_{\xi}$, was advanced in~\cite{Parker1993}; it is the product of a coherence volume $V_{\xi} = \xi^{3}$ ($\xi$ is the coherence length) and the flicker coefficient for one coherence volume, $\hmo^{\rm (vel)}$. A related definition was proposed for bulk-acoustic-wave (BAW) devices \cite{Sthal2013}. Each was advanced for a single platform and only over a relatively narrow range of device size.

Figure~\ref{fig:survey} displays the results of this new and general analysis. In addition to the devices of \cite{Sansa2016}, it incorporates quartz, SAW, and BAW devices~\cite{Parker1993,Parker1994,Driscoll1993,Salzenstein2010,Rubiola2007}; cryogenic NEMS measurements~\cite{Maillet2018,Naik2009,Fong2012,FengThesis2007}; piezoelectric microelectromechanical-systems (MEMS)~\cite{Kim2017}; torsional NEMS~\cite{Zhang2013,Duraffourg2018}; strain-engineered tuning forks~\cite{Wang2020}; and dissipation-diluted NEMS \cite{Ghadimi2018,Beccari2022,Cupertino2024,Bereyhi2022}. To avoid conflating different device families, provenance here is explicit: Fig.~\ref{fig:survey} separates data with resolved plateaus from those where they are hidden, and identifies studies lacking any frequency stability measurements. (The latter solely provide upper bounds.) Crucially, the ordinate in Fig.~\ref{fig:survey} represents values of $\sA(\tau)$ where it is $\tau$-independent: namely, \textit{on the} $1/f$ \textit{plateau}, $\sApl$.

\textit{Mechanism engendering the $1/f$ plateau.} --- In Ref.~\cite{ClelandRoukes2002}, Cleland modeled frequency noise as arising from a thermally activated reorientation of elastic dipoles, switching at a rate $\Gamma_d=\nu_0\,e^{-\Delta g^{*}/k_BT}$, with $\nu_0$ an attempt frequency and $\Delta g^{*}$ a free-energy barrier. The switching of a defect can perturb the modal resonance frequency through two coupling routes (see End Matter).\\ (\textit{i})~\emph{Modulus coupling}~\cite{ClelandRoukes2002}. In this case, the local elastic modulus differs between the two states by $\delta E\equiv E_{+}-E_{-}$ over a defect volume $v_d$, the region affected by the transition. A defect at $\mathbf r$ thus shifts the fractional frequency by $\delta_1(\mathbf r)=v_C\,w(\mathbf r)/V$, with effective coupling volume $v_C\equiv(v_d/2)\,\delta E/E$. Here, $E$ is the unperturbed modulus and $w(\mathbf r)$ the fraction of the modal stiffness at the site: for a mode set by body forces it is the normalized strain-energy density $\varepsilon^{2}(\mathbf r)/\langle\varepsilon^{2}\rangle$, whereas it is uniform for one set by tension (SM~S16, S23). Averaged over uniformly distributed defects, $\langle\delta_1^{2}\rangle=\langle v_C^{2}\rangle/(V\Veps)$.\\
(\textit{ii})~\emph{Eigenstrain coupling}~\cite{KanellopulosSchmid2026}. In this case, the stress-free strain (``\textit{eigenstrain}''~\cite{Eshelby1957,NowickBerry,Mura1987}) of the defect differs between its states by $\delta\varepsilon^{*}$ --- it is an elastic dipole of strength $\gamma\equiv E\,v_d\,\delta\varepsilon^{*}$. Here, $v_d$ is as defined above. In a resonator whose span is fixed by its supports (a doubly clamped beam or string, or a clamped membrane), the fixed boundary forces the rest of the resonator to absorb the switching effects. Force balance makes the resulting stress uniform, $\delta\sigma=-\gamma/V$, irrespective of the defect's  location. The fractional frequency then shifts by $\delta_1=-\,\gamma/(2\seff V)$, where $\seff=\sigma_0+\sigma_c$ is the prestress plus the bending-equivalent stress $\sigma_c$ ($3.4\,E(h/L)^{2}$ for an unstressed beam; SM~S23)~\cite{KanellopulosSchmid2026}. In a free body there is no such constraint: the defect simply expands, and what remains of this coupling is a dimensional change smaller by $E/2\seff$ (EM). 

\textit{Evaluating $\sApl=\sqrt{2\ln2\,\hmo}$.} --- Summing over $N=nV$ uncorrelated fluctuators in quadrature yields
\begin{equation}
\hmo=\underbrace{\frac{n\,\langle v_C^{2}\rangle}{\Veps}}_{\text{modulus}}
\;+\;
\underbrace{\frac{E\,k_BT}{2\pi\,\seff^{2}\,V\,\Qel}}_{\text{eigenstrain}} .
\label{eq:twochannel}
\end{equation}
Here, the eigenstrain term is written in the fluctuation--dissipation form of Ref.~\cite{KanellopulosSchmid2026}, with $\Qel^{-1}\equiv\phi_{\rm el}$ the intrinsic (undiluted) elongation loss angle of the eigenstrain fluctuators~\cite{Saulson1990} --- the internal friction of Nowick and Berry~\cite{NowickBerry} --- at the flicker frequency. The two-term form of Eq.~\eqref{eq:twochannel}, its weight function, and its free-body limit are derived in SM~S23. In the modulus term, $n$ is the fluctuator density per unit $\ln\tau_c$ of correlation time, and the occupancy factor $p(1-p)$, $\tfrac14$ for symmetric wells, is absorbed in $\langle v_C^{2}\rangle$ (SM~S12). The first term is stress-independent and \textit{is present in every mode class}; the second exists \textit{solely} in fixed-end devices and dilutes as $\seff^{-2}$. Both scale as $1/V$ for a bulk population, so the invariant $\hV$ is the sum of the two.

A single defect (correlation time $\tau_c$, of order $1/\Gamma_d$) gives a Lorentzian $S_y$; an ensemble of defects with a \textit{distribution} of barriers superposes the independent Lorentzians into $S_y\propto1/f$, the $\tau$-independent plateau in $\sA(\tau)$ [Eqs.~(67) and (69)--(70)
of~\cite{ClelandRoukes2002}]. The size dependence is determined by whether $N$ scales with volume or surface area:
\begin{equation}
\hmo \;\propto\; N\,\delta_1^2 \;\propto\; \frac{N}{V^2}
\;\rightarrow\;
\begin{cases}
N\propto V:&\!\!\hV = \text{const},\\[2pt]
N\propto S:&\!\!\hV \propto S/V ;
\end{cases}
\label{eq:scaling}
\end{equation}
here, $S$ is the resonator surface area.

\begin{figure}[!tbp]
\includegraphics[width=\columnwidth]{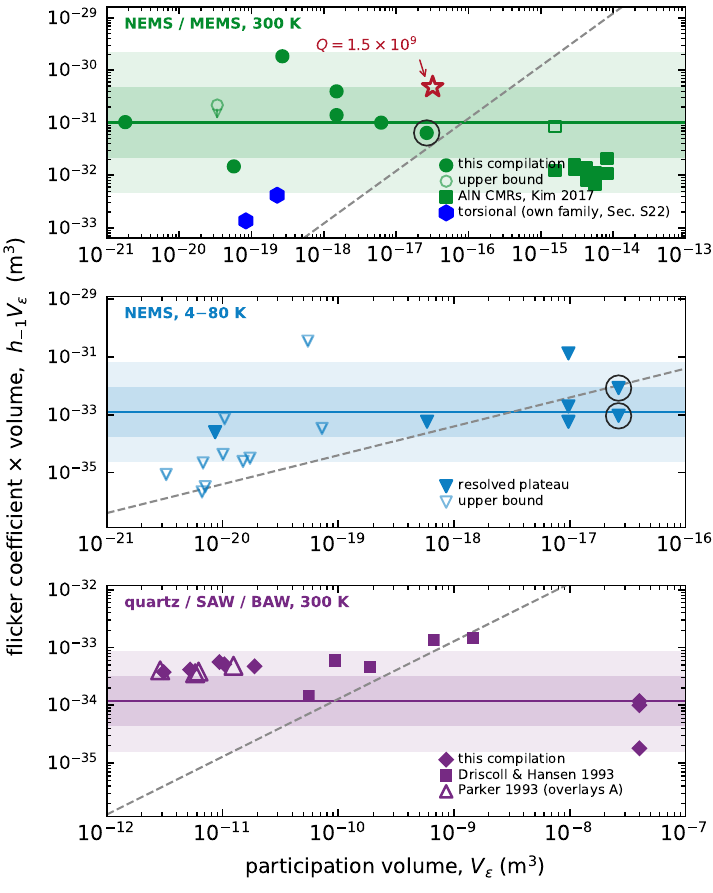}
\caption{\label{fig:invariant}(Color online)
The data of Fig.~\ref{fig:survey} replotted as the invariant $\hV\equiv(\sApl)^{2}V_\varepsilon/(2\ln2)$, with one family per panel. Within each family $\hV$ is consistent with a single value: the horizontal line is the family reference (the median over independent studies, one value per study) and the shaded bands are $\pm1\sigma$ and $\pm2\sigma$ of the scatter about it ($\pm0.7$, $\pm0.9$, $\pm0.4$ decades, top to bottom); the gray dashed slope-$+1$ line, which $\hV$ would follow were $\hmo$ volume-independent, crosses each family. The solid colored reference lines, from top to bottom, are at: $\hV=1.0\ee{-31}$, $1.3\ee{-33}$, $1.2\ee{-34}$~m$^{3}$ (SM~S17). Hexagons (top): the torsional family, excluded from that panel's median (SM~S22). Open markers with arrows are upper bounds and enter no median. The panels are not isometric. Devices outside their frames are explicitly enumerated in SM. }
\end{figure}

In Fig.~\ref{fig:invariant}, which plots $\hV$ against $\Veps$,
bulk-distributed fluctuators appear with slope \textit{zero}; volume-independent processes---readout noise, drift, \textit{etc.}---yield slope $+1$. Contributions from a surface population, by contrast, depend on how the surface fluctuators self-average. If each surface defect's influence is averaged over \textit{volume}, $\delta_1\propto1/V$, then $\hmo\propto S/V^{2}\propto V^{-4/3}$, and slope $-1/3$ will be manifested. If the surface contribution instead self-averages over  \textit{surface} sites, $\delta_1\propto1/S$, then $\hmo\propto1/S$ and slope $+1/3$ is displayed. I extract $S(V)$ from the original sources (SM~S14). For Fig.~\ref{fig:survey} these cases predict $\sApl\propto V_{\varepsilon}^p$ with slope $p=-2/3$ and $p=-1/3$, respectively --- which brackets $p=-1/2$ for bulk fluctuators.

Figure~\ref{fig:invariant} separates the data, one panel per device family (platform), so that trends are evaluated \textit{within} each family. Fitting a common exponent with per-family prefactors to all $42$ resolved-plateau devices (EM) yields
\begin{equation}
\sApl \propto \Veps^{\,p},\qquad p=-0.581\pm0.063 ,
\label{eq:exponent}
\end{equation}
across thirteen decades of participation volume. The volume-independent
alternative, $p=0$, lies $9.2\sigma$ outside the fit, whose bootstrap 95\%
interval is $[-0.66,-0.40]$. (Herein, $\sigma$ always denotes the study-cluster bootstrap standard error of $p$, not a Gaussian tail probability; SM~S6.) Of the two surface conventions, site self-averaging, $p=-1/3$, lies $3.9\sigma$ out; the volume-weighted case, $p=-2/3$, is $1.4\sigma$ out --- thus it is not excluded. Bulk self-averaging, $p=-1/2$, lies $1.3\sigma$ from the fit: the data of this compendium do not separate it from volume-weighting.

\textit{An electrical-domain precedent.} --- Hooge's law for $1/f$ resistance noise, $S_R(f)/R^{2}=\alpha_H/(N_c f)$ with $N_c$ the carrier count~\cite{Hooge1969}, identifies $\alpha_H$ as an invariant for each material, one that falls as quality improves. The mechanical invariant has the same structure. With defect count $N=nV$, number density $n$, and effective coupling volume $v_C$ as defined above, Eq.~\eqref{eq:twochannel} gives
\begin{equation}
\hV \;=\; n\,\langle v_C^{2}\rangle \;+\;\eta\,\frac{E\,k_BT}{2\pi\,\seff^{2}\,\Qel} \ ,
\label{eq:nv1}
\end{equation}
the defect density times the average mean-square coupling volume, plus, for fixed-end devices only, the eigenstrain term ($\eta\equiv\Veps/V$ is the participation ratio).\footnote{Here,
$\langle v_C^{2}\rangle=\tfrac14\langle v_d^{2}(\delta E/E)^{2}\rangle$ is averaged over the defect population's distribution in volume $v_d$ and modulus change $\delta E$. Only this second moment enters, because independent fluctuators add in quadrature. The same squaring makes strain enter $\Veps$ as $\varepsilon^{4}$.} Both terms are constitutive, independent of device size; \textit{they are set by material quality and processing}; the second term also depends on the prestress.

\emph{The single-defect fractional frequency shift, $\delta_1$.} --- Our recent work at mK temperatures on lithium niobate (LN) NEMS resonators resolves \textit{individual}, intrinsic two-level systems (TLS) that are strongly coupled to a mechanical mode. Stress and electrical tuning of an individual TLS through resonance with the mode yields its \textit{strain coupling} $|\gamma|=0.84$~eV~\cite{Yuksel2026} --- the deformation potential by which strain shifts the defect's level splitting. A single coupled defect produces random telegraph jumps in the resonance frequency with fractional frequency shifts $\delta_1\approx3\ee{-6}$ for an associated mode volume of $1~\mu$m$^3$ \cite{Maksymowych2025}. The zero-point rms strain of a mode is $\varepsilon_{\rm zp}=\sqrt{\hbar\omega_0/(2EV)}\propto V^{-1/2}$, so the coupling $g=|\gamma|\varepsilon_{\rm zp}/\hbar$ scales as $V^{-1/2}$ and the dispersive shift $g^2/\Delta$ as $1/V$. The proportionality $\delta_1\propto1/V$ of Eq.~\eqref{eq:scaling} is anchored upon this \textit{measured} coupling together with the mode-volume scaling of $\varepsilon_{\rm zp}$. This shift is dispersive, not the classical jump of the same defect far from resonance; for the eigenstrain channel the latter would be $\gamma/(2\seff V)\approx2\ee{-10}$ at $\seff=0.4$~GPa in $1~\mu$m$^{3}$. Accordingly, the room-temperature $1/f$ floor of silicon nitride requires $10^{8}$--$10^{9}$ fluctuators per $\mu$m$^{3}$ within the band $\ln R$ of correlation times (End Matter). The corresponding density, $10^{20}$--$10^{21}$~cm$^{-3}$, matches the two-level-system density in glasses, and it is this population whose relaxation gives $\Qel\sim10^{4}$ (SM~S12, S23).

$\hV$ \textit{is scale-free, but is not universal.} --- Within a device or material platform, $\hV$ is consistent with a single value across volume; its \emph{value} reflects material quality or
temperature. The quartz/SAW/BAW platform sits three decades
($\sim\!850\times$, $2.9\pm0.5$~dec) below the room-temperature NEMS
reference, and the record BVA resonator is $7\times$ further
below~\cite{Salzenstein2010}. A separate, cryogenically measured population
sits about $80\times$ ($1.9\pm0.6$~dec) below the room-temperature band ---
as is expected for thermally activated fluctuators.

\textit{Cooling lowers $\hV$ only while the fluctuators are thermally
activated.} In~\cite{Fong2012}, measurements on a single stoichiometric SiN
string resonator (fixed $V$) show the $1/f$ amplitude follows
$T^{0.94\pm0.10}$ from 296 to 5~K, and $\hV$ decreases from the
room-temperature to the cryogenic offset, while $Q$ rises by a factor of
$5.5$. Over that span the floor falls $46\times$ following $T^{0.94}$,
whereas $Q^{-1}$ would give $6\times$ and $Q^{-3.5}$ $\sim400\times$. At
higher temperatures, elastic dipoles can surmount barriers of
$0.01$--$0.8$~eV --- presumably this is an $N\!\gg\!1$ ensemble of the
fluctuators that are resolved individually in \cite{Yuksel2026}.  Below the activated regime ($T\lesssim1$~K) this temperature scaling must fail, and it does: the frequency noise of LN devices \emph{rises} on further cooling, because these same elastic dipoles tunnel rather than hop~\cite{Maksymowych2025}. This crossover presumably underlies the cryogenic exponent in Fig.~\ref{fig:survey}, $-0.31\pm0.14$, the shallowest of the three families. (SM S11 offers an alternative interpretation.)

\textit{Scaling alone cannot identify the fluctuators.} The slope of $\sApl$
versus $V_{\varepsilon}$ can distinguish one spatial distribution of fluctuators from another, and the intercept fixes only the invariant $n\,\langle v_C^{2}\rangle$ --- but nothing more. The assertion here, that the fluctuators are material defects, is predicated on three facts: (i) the temperature dependence stated above, (ii) the ordering of the plateau heights in relation to the (material and fabrication) quality and the temperature of the devices, and (iii) the results from LN NEMS at mK temperatures, where individual two-level systems are
unambiguously resolved~\cite{Yuksel2026,Maksymowych2025}. 

\emph{The $1/f$ floor is $Q$-independent.} --- This has historically been
contested ground. SAW resonators are reported to follow $1/(\text{size})$
while being independent of $Q$~\cite{Parker1993}; BAW resonators are
reported to follow $Q^{-4}$~\cite{Gagnepain1981,Walls1992}, and
piezoelectric contour-mode literature reports a $Q^{-2.7}$ to $Q^{-3.8}$
dependence~\cite{Kim2017,Kim2018}. This contrasting behavior is attributed
to separate mechanisms in the different device types. I assert these arise
from the same mechanism, reflected in two coordinate systems: within a device family, $Q$ and volume \textit{co-vary} with frequency --- the material's Akhiezer $f\!\cdot\!Q$ ceiling ties $Q$ to frequency, and the acoustic wavelength ties frequency to
volume~\cite{Tabrizian2009}. A volume law then appears as a steep $Q$-law, but the reported exponents fall within the same bulk-to-surface band resolved directly against $\Veps$ in Fig.~\ref{fig:invariant}. The two coordinates are only \textit{nearly} degenerate: the covariance is exact within a family but broken across platforms, where the floors separately follow the power law in $\Veps$ (EM).

\textit{Dissipation dilution.} --- Through two decades of effort, dissipation dilution (DD) has raised mechanical $Q$ by four orders, \textit{e.g.}, up to $6.6\ee{9}$ for a centimeter-scale device~\cite{Cupertino2024} and $1.3\ee{10}$ for a strained crystalline resonator~\cite{Beccari2022}. At the critical amplitude, to which each device is referenced here (see End Matter), raising $Q$ leaves the thermomechanical limit to frequency stability
\textit{unchanged} ($Q$ cancels against $z_c^{2}$) while lowering the detector-limited (white) branch of $\sA(\tau)$ by just $1/\sqrt{Q}$. The flicker plateau is likewise independent of the \emph{measured} $Q$: neither term of Eq.~(\ref{eq:twochannel}) contains the \textit{dilution factor} $D_Q=\seff/\sigma_c$, the ratio of total to bending energy --- through which dilution multiplies $Q$. The modulus term does not dilute because the modulus scales all factors in the dispersion relation. For a resonator held at fixed length, $\sigma_0=E\varepsilon_0$ and the bending stiffness $\propto EI$ both scale with $E$. Accordingly, $\omega\propto\sqrt{E}$, and this holds however the modal energy is divided between tension and bending; thus, $\delta f/f_0=\tfrac12\,\delta E/E$ irrespective of dilution. The eigenstrain term is set by the prestress and the intrinsic loss, and not by $D_Q$: raising $Q$ at fixed $\sigma_0$ by soft clamping, phononic shielding or perimeter modes leaves it unchanged. Two measurements show this directly. Membranes of equal size whose $Q$ differ tenfold through clamping loss have the same  floor to within a factor of two~\cite{KanellopulosSchmid2026}, and the only stability
measurement on a $Q=1.5\ee{9}$ perimeter mode sits within the room-temperature band, actually \textit{above} its median~\cite{Bereyhi2022} (SM~S4). What does move the eigenstrain term is the stress itself: raising $\sigma_0$ from $357$ to $440$~MPa within one film is seen to lower the floor by a factor of $1.78\pm0.20$, as compared with the predicted $(440/357)^{2}=1.52$~\cite{KanellopulosSchmid2026}. This survey cannot resolve that dependence --- the NEMS string resonators in this compendium span $0.2$ decades of $\sigma_0$, and its unstressed beams have $\sigma_c=0.3$--$5$~GPa, comparable to the strings' $\sigma_0$ --- which is why the two groups share the invariant. Accordingly, here it is $\hmo$, and not $Q$, that sets the resolution of frequency-shift sensing.

\emph{Why the measured $Q$ and the frequency noise decouple.} --- A thermally activated fluctuator with barrier $\varepsilon_a$ switches at rate
$\tau_c^{-1}=\tau_0^{-1}\exp(-\varepsilon_a/k_BT)$. A measurement with
characteristic timescale, $\tau$, is affected solely by defects whose inverse transition rate matches $\tau$. Within the full distribution of fluctuators, $D(\varepsilon_a)$, this is the sub-population with barrier energies near
\begin{equation}
\varepsilon_a = k_BT\ln(\tau/\tau_0) .
\label{eq:window}
\end{equation}
Here, the assumed attempt time is $\tau_0\sim10^{-12}$\,s (SM~S11). In typical experiments $\tau\sim10$~ms--$100$~s, giving $\ln(\tau/\tau_0)\approx 23\text{--}32$;
the defects that dominate dissipation, by contrast, switch near the resonance period, $\tau_c\sim1/\omega_0$, yielding $\ln(\tau_c/\tau_0)\approx7$. At $78$~K the
frequency-stability slice sits at $155$--$217$~meV and the dissipation slice near $50$~meV: \textit{these are disjoint sub-populations of fluctuators}, separated by $\sim\!135$~meV ($\sim\!20\,k_BT$). The fluctuation--dissipation form of the eigenstrain term ties it to the elongation loss angle \emph{at the flicker frequency}, $\phi_{\rm el}(f)$, which the measured $Q$ at $\omega_0$ does not determine. Only when $D(\varepsilon_a)$ is flat between the two slices is $\phi_{\rm el}(f)\simeq\phi_{\rm el}(\omega_0)$ and the floor predictable from the intrinsic loss. For silicon nitride at room temperature it is, to within $\times2$--$3$~\cite{KanellopulosSchmid2026}; evaluated on this survey's own strings with $\Qel$ approximated by the intrinsic $Q$ at $\omega_0$ ($Q/D_Q$), $\hmo V\seff^{2}\Qel$ is $0.3$--$1.6\times Ek_BT/2\pi$ at $4$--$300$~K, with one device at $5$~K at $3.8\times$ (SM~S23).

\emph{Consequence for sensing.} --- With
$\sApl=\sqrt{2\ln 2\,\hmo}$ and $m_{\mathrm{eff}}=\alpha\rho V$ ($\alpha$ is the effective modal mass fraction), the mass resolution of a resonant detector factorizes as
\begin{equation}
  \label{eq:dm}
  |\dm| = 2\,m_{\mathrm{eff}}\,\sApl
  = \frac{2\alpha\rho}{\sqrt{\eta}}\,\sqrt{2\ln 2}\,\sqrt{(\hV)\,V}.
\end{equation}
This holds for a point mass adsorbed at an antinode, and is proportional to one material factor and one geometric
factor. \textit{Note that }$Q$ \emph{is absent from both factors}.
With fixed material, processing, prestress, temperature, and mode shape, $\hV$ is fixed and $|\dm|\propto\sqrt{V}$: shrinking a resonator improves mass resolution as the square root of its volume  and, within these constraints, by no other route. It is striking that four orders of improvement in $Q$ by DD still leave Eq.~\eqref{eq:dm} essentially unchanged.

\emph{A practical suggestion for future reporting.} --- As demonstrated here, the $1/f$ frequency-noise floor cannot be inferred from the measured $Q$, nor $Q$ from the floor. \textit{Both must be measured}. A recommendation follows: $\sA(\tau)$ should be reported with $\tau$ stated, the response time $\tau_r$ denoted, and the participation volume $V_{\varepsilon}$, the effective stress $\seff$ (for fixed-end devices), and modal $Q$-values provided. These determine $\hV$, and for frequency-shift sensing, this is the true target for optimization.

\begin{acknowledgments}
   \textbf{Acknowledgments:} This work has benefited immensely from many enlightening discussions with Prof.~Mark I.~Dykman. Also critical to this analysis were comments and input from Andrew Cleland, Alex Eichler, Nils Engelsen, Bhavesh Kharbanda, Kostas Kanellopulos, Joon Kim, Gianluca Piazza, Mark Sansa, and, especially, Silvan Schmid. Support from an NIH Director's Transformative Research Award (RF1MH136394) and the Moore Foundation's \textit{Quantum Imaging, Sensing, and Metrology} program (GBMF 12214) is gratefully acknowledged.
\end{acknowledgments}

\appendix
\section{End Matter}

\begin{figure}[!tb]
\includegraphics[width=0.96\columnwidth]{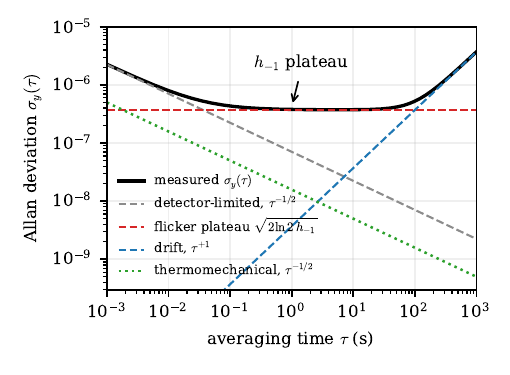}
\caption{\label{fig:allan}Orientation: the schematic Allan deviation $\sA(\tau)$ of a mechanical resonator, with a detector-limited $\tau^{-1/2}$ branch, the $\tau$-independent flicker plateau $\sApl=\sqrt{2\ln2\,\hmo}$ analyzed in this Letter, and a drift branch rising as $\tau^{+1}$. The thermomechanical $\tau^{-1/2}$ limit lies 1--4 decades below the plateau.}
\end{figure}

\emph{Compilation and curation.} --- Two rules govern admission into the cohort compared here: a device is included only if its $\sA$ is traceable to a measurement in the primary source, and only if its linewidth is set by internal dissipation --- that is, only if $f_0/Q$ is fixed by loss in the resonator and its supports rather than in the surrounding gas or liquid (the internal-dissipation rule (2) of SM~S1). Cantilevers operated in air and a capacitive transducer designed to radiate into it are excluded on this ground; suspended-microchannel devices are retained, since the fluid is inside an embedded channel and the resonator itself is in vacuum. No macroscopic resonator enters: for the high-$Q$ macroscopic classes --- gravitational torsion pendulums, helium torsional oscillators, resonant bars, vibratory gyroscopes --- no resolved flicker floors have been published; their records are solely white-noise- and then drift-limited, and the bounds they imply sit $5$--$17$ decades \textit{above} the invariant (SM S4). These rules, the devices each rule excludes, and the full per-device records are given in SM~S1 and S4.

\emph{The multiplicative-noise branch and critical amplitude.} --- Amplitude noise is converted into frequency noise by nonlinearity; in mechanical devices the Duffing coefficient generally sets the conversion, and this amplitude-to-phase-noise (APN) conversion becomes appreciable once the critical amplitude, $z_c$, is exceeded~\cite{Bachtold2022}. Devices reported above $z_c$ are therefore excluded from the fits: the branch is identified by its operating-point dependence, so reducing the drive is the diagnostic that can separate APN from the parametric floor.

\emph{Participation volume.} --- $\hV$ denotes the invariant of Eq.~\eqref{eq:nv1}; Fig.~\ref{fig:invariant} plots this quantity directly; tables in SM record the raw $V$ and $\sA$, related to it by the per-class factor $\eta=V_{\varepsilon}/V$ (SM S16). $V$ is the total device volume, excluding substrate, supports, and any inert attached mass. It is \textit{not} the modal volume $V_{\rm eff}^{(n)}\equiv m_{\rm eff}/\rho=\alpha V$, which is weighted by displacement rather than by strain. $\eta$ depends on mode shape alone; it is constant within a geometry class and can only move a fit's intercept rather than vary its slope. Across classes it could tilt the fit, since geometry correlates with size  --- flexural beams and strings at the small-volume end, extensional plates and bulk-acoustic devices at the large end. SM S16 evaluates it for every class present: $0.300$ for a doubly clamped beam, $0.237$ for a cantilever, $1$ for a tension-dominated string (where built-in stress makes the weighting uniform), $2/3$ for both the extensional and the thickness-shear plate, and $0.62$/$0.57$ for the two torsion-rod cross-sections, by Saint-Venant shear weighting (SM S22); a surface wave has no bounded material volume, so $\Veps$ for SAW devices is taken directly from the Rayleigh field. \emph{The weighting is also not load-bearing:} refitting the entire record against bare $V$ gives $p=-0.584\pm0.058$ against $-0.581\pm0.063$ for $\Veps$ --- a shift of a twentieth of a cluster error --- with per-family scatter unchanged to $0.03$~dec and the exclusions marginally \emph{stronger} against $V$ (volume independence at $10.0\sigma$). This is expected: $\eta$ spans $0.237$--$1$ ($1.07$ for the SAW devices, whose $\Veps$ is referred to the tabulated nominal volume $A\lambda/2$ rather than to a bounded $V$; SM~S16), at most $0.65$~dec between classes, against $13.35$ decades of volume. $\Veps$ is retained because the derivation produces it (and it contains the operative physics) --- the invariant's value $n\,\langle v_C^{2}\rangle$ is defined by the strain weighting --- not because the record resolves it; the $\le0.65$-dec class offsets $\eta$ predicts are a target for future within-class versus cross-class comparisons (SM S16).

\emph{Thermomechanical limits.} --- Each thermomechanical point of Fig.~\ref{fig:survey} is computed at its own temperature, at each device's own measurement time $\tau$ (assumed to be $\sim$1~s if none was stated), and referenced to the Duffing critical amplitude, at which $Q$ cancels from $Qz_c^{2}$. The four-rule hierarchy used to obtain $Qz_c^{2}$ per device is provided in SM S5. Because that branch falls as $\tau^{-1/2}$ while the plateau is flat, the invariant required to reach it depends on $\tau$: at $\tau=1$~s it is $\hV\approx6\ee{-35}$ to $10^{-37}$~m$^{3}$ for the strings and beams of this record, three to six decades below the room-temperature reference (SM~S5).

\emph{$Q$-independence: covariance and the survey.} --- Within a device family the Akhiezer ceiling $f\!\cdot\!Q=\text{const}$ and $V\propto\lambda^{3}$ give $V\propto Q^{3}$, so a floor $\hmo\propto V^{-q}$ appears as $\hmo\propto Q^{-3q}$: a bulk law ($q=1$) reads as $Q^{-3}$ and a surface law ($q=4/3$) as $Q^{-4}$ --- the same bulk-to-surface band found directly against $\Veps$ (Fig.~\ref{fig:invariant}), and comparable to the reported $Q^{-4}$ (BAW) and $Q^{-2.7}$--$Q^{-3.8}$ (contour-mode) exponents (the shallowest lying just above the bulk value). Because the covariance is only approximate, the two laws separate across platforms: the cross-platform analysis spans three decades of $Q$ among resolved plateaus --- nearly six decades if the dissipation-diluted \textit{bound} is admitted --- and a fitted $Q$ covariate excludes the steep historical laws ($Q^{-3.5}$ lies $9.7$ cluster errors away) while resolving no $Q$ dependence of its own (SM~S17). The single-device test of~\cite{Sage2013} complements these, sweeping $Q$ by $6.4\times$ at fixed displacement while leaving the $1/f$ frequency-noise plateau unmoved (SM~S7). A second single-device test varies the operating point rather than $Q$: the soft-clamped resonator of Ref.~\cite{Kharbanda2026} shows its intrinsic $1/f$ floor independent of both the optical readout power and the mechanical drive.

\emph{Why a plateau may be hidden, or absent.} --- A $1/f$ floor is visible only over the range of $\tau$ where the $\alpha=-1$ term predominates, bounded by the additive branch below, by drift above, and by the response time $\tau_r=Q/\pi f_0$. SM~S3 sets out these ``closures'' in full; the one result used in the body is that the additive and drift closures both \emph{raise} the apparent floor, so a hidden plateau gives an upper bound and never a lower one (SM S9).

\emph{Per-family exponents and leverage.} --- The three families agree pairwise within $1.9\sigma$: $p=-0.48\pm0.11$ at room temperature ($-0.61\pm0.03$ with the AlN contour-mode devices included), $-0.58\pm0.03$ for quartz/SAW/BAW, and $-0.31\pm0.14$ for cryogenic data; all $42$ fitted values trace to named primary sources (SM~S18). The two-device torsional family is placed by its offset --- its $\hV$ sits $1.6\pm0.4$ decades below the flexural room-temperature reference --- which I attribute to the higher-order strain multipole of a torsional mode reducing $v_C$ (SM~S22). Although the invariant spans $13.35$ decades of $\Veps$, the exponent is fixed by within-family variation, an effective lever arm of only $3.4$ decades, with $\sigma(p)=0.063$ from a study-cluster bootstrap (a jackknife gives $0.045$, a hierarchical fit $p=-0.57$; SM~S6, S17).

\emph{The two couplings, in detail.} --- In the modulus channel the weight $w(\mathbf r)$ of Eq.~(\ref{eq:twochannel}) is the fraction of the modal stiffness the defect controls: $w=1-(1-\hat w)/D_Q$ with $\hat w=\varepsilon_b^{2}/\langle\varepsilon_b^{2}\rangle$ the normalized bending strain energy, so that $\eta=\Veps/V=[1+(\eta_b^{-1}-1)/D_Q^{2}]^{-1}$ interpolates between the bending value $\eta_b$ at $D_Q=1$ and $1$ for a tension-dominated string ($0.87$--$0.99$ for the strings of this record; SM~S23). In the eigenstrain channel the induced stress is uniform, so the shift is position-independent and the relevant volume is $V$, not $\Veps$; the term equals the microscopic sum $n\langle\gamma^{2}p(1-p)\rangle/(4\seff^{2}V\ln R)$ over the band of correlation times, $\ln R$, to which the relaxation loss of the same fluctuators is tied~\cite{KanellopulosSchmid2026}. The bending stress $\sigma_c$ keeps the term finite at zero prestress, where it scales as $(L/h)^{4}/(EV\Qel)$; it diverges only at buckling, $\sigma_0\to-\sigma_c$, where $\omega\to0$. In a free body the eigenstrain relaxes; what remains is the dimensional change $\gamma/EV$, smaller than the clamped response by $E/2\seff\sim10^{2}$--$10^{3}$ and, thus, negligible (SM~S23).

\emph{Two constraint classes.} --- Equation~(\ref{eq:twochannel}) predicts a
structure within the invariant that the record can test: devices whose span
is fixed carry both terms, free bodies only the first. In
the room-temperature family the fixed-end members --- SiN strings, Si and AlN
beams, five studies --- have median $\hV=1.0\ee{-31}$~m$^{3}$; the free bodies --- AlN contour-mode plates and a Si cantilever, two studies ---
$1.4\ee{-32}$~m$^{3}$. The offset, $0.9$ decades, is in the predicted direction, exceeds the $\le0.65$~dec that $\eta$ can produce, survives against bare $V$ ($1.1$~dec), and every room-temperature fixed-end cluster lies above every free body of the record, torsional and quartz included (SM~S23). Its size is what the two terms give: the eigenstrain term evaluated with the intrinsic loss of SiN reproduces the strings, and the free-body level supplies the modulus term. Measured on a doubly clamped beam and a cantilever of one film, the same offset would give the ratio of the two couplings directly; on the present record it rests on two free-body studies and is a target, not a result.

\emph{Predictions and tests.} --- Equation~(\ref{eq:twochannel}) makes six
statements, each with an outcome that would falsify it.
(1)~Within a family, $\hV$ is independent of size: a resolved slope of
$\hV$ against $\Veps$ at fixed material and class falsifies the bulk law
(a slope of $-1/3$ identifies a surface population). \textit{Observed:} Fig.~\ref{fig:invariant}. 
(2)~The floor does not follow the measured $Q$: at fixed $\sigma_0$ and film, soft clamping, phononic shielding or perimeter modes raise $Q$ by $10^{2}$--$10^{4}$ and leave $\hmo$ unchanged; a floor falling with $D_Q$ falsifies both terms. \textit{Observed:}  Refs.~\cite{Sage2013,Bereyhi2022,KanellopulosSchmid2026}.
(3)~Fixed-end floors dilute with stress: at fixed film and volume,
$\hmo\propto\seff^{-2}$; free-body floors do not move with stress. \textit{Observed} for strings over $357$--$440$~MPa~\cite{KanellopulosSchmid2026}; but it is \textit{untested} for free bodies.
(4)~A doubly clamped beam and a cantilever of one film differ by the eigenstrain term, $\hmo^{\rm beam}-\hmo^{\rm cant}=Ek_BT/(2\pi\seff^{2}V\Qel)$, and for unstressed beams this term scales as $(L/h)^{4}/V$ at fixed film: an aspect-ratio series is the sharpest geometric test. For stoichiometric SiN at $1$~GPa the measured loss bounds the cantilever floor at $\hV\le6\ee{-33}$~m$^{3}$ (SM~S23). \textit{Untested.}
(5)~The floor is linear in $T$ where the barrier distribution is flat, and the fluctuation--dissipation relation to $\Qel$ holds there; a floor tracking the resonance $Q$ rather than $T$ falsifies the two-slice picture. \textit{Observed:} $T^{0.94}$ from $296$ to $5$~K~\cite{Fong2012}, with the $\Qel$ relation holding to $\times2$--$3$ above $78$~K. A corollary with no free parameter: each fluctuator species produces its Debye loss peak at $T_Q\approx3\,T_h$, three times the temperature at which it contributes most to the floor (SM~S11, S24).
(6)~Lowering $\hV$: at fixed geometry, $n\langle v_C^{2}\rangle$ and $\Qel^{-1}$ move together with defect density (annealing, surface treatment, crystallinity); a $1/f$ floor unchanged by a treatment that raises $\Qel$ tenfold at fixed $\sigma_0$ would falsify the eigenstrain term's loss relation. \textit{Untested.} A suggested validation program follows: one film, four devices --- a plain string, a soft-clamped string, a doubly clamped beam and a cantilever --- measured at $300$ and $78$~K before and after
anneal, would resolve (2)--(6) on a single chip.

\emph{Why a single-channel scheme cannot remove the $1/f$ floor.} --- A parametric fluctuation enters the equation of motion just as the signal does, and \textit{the three standard routes to a better signal-to-noise ratio all fail on this fact.} For a frequency servo it is a theorem: the mass resolution $|\dm|\propto\sqrt{D_\phi}/\mathcal{R}$ is invariant under the loop gain, which drives the phase diffusion $D_\phi$ and the
responsivity $\mathcal{R}$ to zero together~\cite{Dankowicz2026}. Phase-noise cancellation schemes~\cite{Kenig2012} preserve responsivity but
exempt this channel by construction. And common-mode rejection has been tried
and reported: Ref.~\cite{Sansa2016} corrected one mode against a second used as a
thermometer, degrading the white branch, improving the drift, but leaving the
plateau untouched. SM~S8 discusses all three in full.
Correlated multimode readout evades the single-channel hypothesis, not
the invariant (SM~S20, S24). Its scope is set by the inter-mode correlation of the floor, which Eq.~(\ref{eq:twochannel}) predicts from the overlap of the site weights: it is complete for tension-set modes, where both couplings are uniform ($r=0.999998$ at $D_Q=80$, up to $55$~dB of cancellation), but only partial for bending-set modes ($r=0.74$ for the first two modes of a clamped beam, $3.5$~dB at most). This is why use of a second mode can stabilize a nanostring to the thermomechanical level~\cite{Gavartin2013} while leaving the plateau of a beam untouched~\cite{Sansa2016}.


\begin{thebibliography}{99}

\bibitem{Bachtold2022} A. Bachtold, J. Moser, and M.I. Dykman, ``Mesoscopic physics of nanomechanical resonators'', \textit{Rev. Mod. Phys.} \textbf{94}, 045005 (2022).

\bibitem{ClelandRoukes2002} A.N. Cleland and M.L. Roukes, ``Noise processes in nanomechanical resonators,'' \emph{J. Appl. Phys.} \textbf{92}, 2758--2769 (2002).

\bibitem{SchmidBook2016} S. Schmid, L.G. Villanueva and M.L. Roukes, \emph{Fundamentals of Nanomechanical Resonators} (Springer, 2nd Ed., 2023), doi:10.1007/978-3-031-29628-4.

\bibitem{Barnes1971}
J.A. Barnes, A.R. Chi, L.S. Cutler, D.J. Healey, D.B. Leeson,
T.E. McGunigal, J.A. Mullen, Jr., W.L. Smith, R.L. Sydnor, R.F.C. Vessot
and G.M.R. Winkler, ``Characterization of frequency stability,''
\emph{IEEE Trans. Instrum. Meas.} \textbf{IM-20}, 105--120 (1971).

\bibitem{IEEE1139} \emph{IEEE Standard Definitions of Physical Quantities for Fundamental Frequency and Time Metrology---Random Instabilities}, IEEE Std 1139-2022 (IEEE, New York, 2022).

\bibitem{Yang2011}
Y.T. Yang, C. Callegari, X.L. Feng and M.L. Roukes, ``Surface adsorbate
fluctuations and noise in nanoelectromechanical systems,'' \emph{Nano Lett.}
\textbf{11}, 1753 (2011).

\bibitem{Allan1966}
D.W. Allan, ``Statistics of Atomic Frequency Standards,''
Proceedings of the IEEE, Vol. 54, No. 2, pp.~221--230 (1966).

\bibitem{Machlup1954}
S. Machlup, ``Noise in semiconductors: spectrum of a two-parameter random
signal,'' \emph{J. Appl. Phys.} \textbf{25}, 341 (1954).

\bibitem{Bernamont1937}
J. Bernamont, ``Fluctuations de potentiel aux bornes d\textquotesingle un
conducteur m\'etallique de faible volume parcouru par un courant,''
\emph{Ann. Phys. (Paris)} \textbf{7}, 71 (1937).

\bibitem{vanderZiel1950}
A. van der Ziel, ``On the noise spectra of semi-conductor noise and of
flicker effect,'' \emph{Physica} \textbf{16}, 359 (1950).

\bibitem{duPre1950}
F.K. du Pr\'e, ``A suggestion regarding the spectral density of flicker
noise,'' \emph{Phys. Rev.} \textbf{78}, 615 (1950).

\bibitem{DuttaHorn1981}
P. Dutta and P.M. Horn, ``Low-frequency fluctuations in solids: $1/f$
noise,'' \emph{Rev. Mod. Phys.} \textbf{53}, 497 (1981).

\bibitem{McWhorter1957} A.L. McWhorter, ``$1/f$ noise and germanium surface properties,'' in \emph{Semiconductor Surface Physics}, edited by R.H. Kingston (Univ. of Pennsylvania Press, Philadelphia, 1957), pp.~207--228.

\bibitem{Weissman1988}
M.B. Weissman, ``$1/f$ noise and other slow, nonexponential kinetics in
condensed matter,'' \emph{Rev. Mod. Phys.} \textbf{60}, 537 (1988).

\bibitem{Sansa2016} M. Sansa, E. Sage, E.C. Bullard, M. G\'ely, T. Alava, E. Colinet, A.K. Naik, L.G. Villanueva, L. Duraffourg, M.L. Roukes, G. Jourdan and S. Hentz, ``Frequency fluctuations in silicon nanoresonators,'' \emph{Nature Nanotech.} \textbf{11}, 552 (2016), and Supplementary Information.

\bibitem{Parker1993} T.E. Parker, ``Dependence of SAW resonator $1/f$ noise on device size,'' \emph{IEEE Trans. UFFC} \textbf{40}, 831 (1993).

\bibitem{Sthal2013} F. Sthal, S. Galliou, J. Imbaud, X. Vacheret, P. Salzenstein, E. Rubiola and G. Cibiel, ``Volume dependence in Handel's model of quartz crystal resonator noise,'' \emph{IEEE Trans. UFFC} \textbf{60}, 1971 (2013).

\bibitem{Parker1994} T.E. Parker and D. Andres, ``$1/f$ noise in surface acoustic wave (SAW) resonators,'' \emph{Proc. IEEE 48th Annual Symp. on Frequency Control} (1994), pp.~530--538.

\bibitem{Driscoll1993} M.M. Driscoll and W.P. Hansen, ``Measured vs.\ volume model-predicted flicker-of-frequency instability in VHF quartz crystal resonators,'' \emph{Proc. 1993 IEEE Int. Frequency Control Symp.}, pp.~186--192.

\bibitem{Salzenstein2010} P. Salzenstein, A. Kuna, L. \v{S}ojdr and J. Chauvin, ``Significant step in ultra-high stability quartz crystal oscillators,'' \emph{Electron. Lett.} \textbf{46}, 1433 (2010).

\bibitem{Rubiola2007} E. Rubiola and V. Giordano, ``On the $1/f$ frequency noise in ultra-stable quartz oscillators,'' \emph{IEEE Trans. UFFC} \textbf{54}, 15 (2007).

\bibitem{Maillet2018} O. Maillet, X. Zhou, R.R. Gazizulin, B.R. Ilic, J.M. Parpia, O. Bourgeois, A.D. Fefferman and E. Collin, ``Measuring frequency fluctuations in nonlinear nanomechanical resonators,'' \emph{ACS Nano} \textbf{12}, 5753 (2018).

\bibitem{Naik2009} A.K. Naik, M.S. Hanay, W.K. Hiebert, X.L. Feng and M.L. Roukes, ``Towards single-molecule nanomechanical mass spectrometry,'' \emph{Nature Nanotech.} \textbf{4}, 445 (2009).

\bibitem{Fong2012} K.Y. Fong, W.H.P. Pernice and H.X. Tang, ``Frequency and phase noise of ultrahigh $Q$ silicon nitride nanomechanical resonators,'' \emph{Phys. Rev. B} \textbf{85}, 161410(R) (2012).

\bibitem{FengThesis2007} X.L. Feng, ``Ultra high frequency nanoelectromechanical systems with low-noise technologies for single-molecule mass sensing,'' PhD thesis, California Institute of Technology (2007), Tables 4-1 and 6-1.

\bibitem{Kim2017} H.J. Kim, J. Segovia-Fernandez and G. Piazza, ``The impact of damping on flicker frequency noise of AlN piezoelectric MEMS resonators,'' \emph{J. Microelectromech. Syst.} \textbf{26}, 317 (2017); H.J. Kim, private communication (31 August 2026): device geometry and spectral convention of the $E_L$ series (SM S1).

\bibitem{Zhang2013} X.C. Zhang, E.B. Myers, J.E. Sader and M.L. Roukes, ``Nanomechanical torsional resonators for frequency-shift infrared thermal sensing,'' \emph{Nano Lett.} \textbf{13}, 1528 (2013).

\bibitem{Duraffourg2018} L. Duraffourg, L. Laurent, J.-S. Moulet, J. Arcamone and J.-J. Yon, ``Array of resonant electromechanical nanosystems: a technological breakthrough for uncooled infrared imaging,'' \emph{Micromachines} \textbf{9}, 401 (2018).

\bibitem{Wang2020} M. Wang, R. Zhang, R. Ilic, V. Aksyuk and Y. Liu, ``Frequency stabilization of nanomechanical resonators using thermally invariant strain engineering,'' \emph{Nano Lett.} \textbf{20}, 3050 (2020).

\bibitem{Ghadimi2018}
A.H. Ghadimi, S.A. Fedorov, N.J. Engelsen, M.J. Bereyhi, R. Schilling,
D.J. Wilson and T.J. Kippenberg, ``Elastic strain engineering for ultralow
mechanical dissipation,'' \emph{Science} \textbf{360}, 764--768 (2018).

\bibitem{Beccari2022}
A. Beccari, D.A. Visani, S.A. Fedorov, M.J. Bereyhi, V. Boureau,
N.J. Engelsen and T.J. Kippenberg, ``Strained crystalline nanomechanical
resonators with quality factors above 10 billion,'' \emph{Nature Phys.}
\textbf{18}, 436--442 (2022).

\bibitem{Cupertino2024}
A. Cupertino, D. Shin, L. Guo, P.G. Steeneken, M.A. Bessa and R.A. Norte, ``Centimeter-scale nanomechanical
resonators with low dissipation,'' \emph{Nature Commun.} \textbf{15}, 4255
(2024).

\bibitem{Bereyhi2022} M.J. Bereyhi, A. Arabmoheghi, S.A. Fedorov, A. Beccari, G. Huang, T.J. Kippenberg and N.J. Engelsen, ``Perimeter modes of nanomechanical resonators exhibit quality factors exceeding $10^{9}$ at room temperature,'' \emph{Phys. Rev. X} \textbf{12}, 021036 (2022).

\bibitem{Hooge1969} F.N. Hooge, ``$1/f$ noise is no surface effect,'' \emph{Phys. Lett. A} \textbf{29}, 139--140 (1969).

\bibitem{Yuksel2026}
M. Yuksel, M.P. Maksymowych, O.A. Hitchcock, F.M. Mayor, N.R. Lee,
M.I. Dykman, A.H. Safavi-Naeini and M.L. Roukes, ``Intrinsic phononic
dressed states in a nanomechanical system,'' \emph{Nat. Phys.} (2026),
doi:10.1038/s41567-026-03225-3.

\bibitem{Maksymowych2025}
M.P. Maksymowych, M. Yuksel, O.A. Hitchcock, N.R. Lee, F.M. Mayor, W. Jiang,
M.L. Roukes and A.H. Safavi-Naeini, ``Spectral diffusion of nanomechanical
resonators due to single quantum defects,'' \emph{Phys. Rev. Applied}
\textbf{24}, 044066 (2025).

\bibitem{Gagnepain1981} J.-J. Gagnepain, J. Uebersfeld, G. Goujon and P. Handel, ``Relation between $1/f$ noise and $Q$-factor in quartz resonators at room and low temperatures, first theoretical interpretation,'' \emph{Proc. 35th Annual Symp. on Frequency Control} (1981), pp.~476--483.

\bibitem{Walls1992} F.L. Walls, P.H. Handel, R. Besson and J.-J. Gagnepain, ``A new model of $1/f$ noise in BAW quartz resonators,'' \emph{Proc. 1992 IEEE Frequency Control Symp.}, pp.~327--333.

\bibitem{Kim2018} H.J. Kim, S.I. Jung, J. Segovia-Fernandez and G. Piazza, ``A study on flicker frequency noise of piezoelectric aluminum nitride resonators as a function of electrode design,'' in \emph{Proc. IEEE Int. Conf. Micro Electro Mechanical Systems (MEMS 2018)} (IEEE, 2018), p.~767.

\bibitem{Tabrizian2009} R. Tabrizian, M. Rais-Zadeh and F. Ayazi, ``Effect of phonon interactions on limiting the $f\!\cdot\!Q$ product of micromechanical resonators,'' in \emph{Proc. Transducers 2009} (IEEE, 2009), pp.~2131--2134.

\bibitem{Sage2013} E. Sage, ``Nouveau concept de spectrom\`etre de masse \`a base de r\'eseaux de nanostructures r\'esonantes,'' PhD thesis, Universit\'e de Grenoble / CEA-LETI (2013), Chapter II.

\bibitem{Eshelby1957} J.D. Eshelby, ``The determination of the elastic field of an ellipsoidal inclusion, and related problems,'' \emph{Proc. R. Soc. London A} \textbf{241}, 376--396 (1957).

\bibitem{NowickBerry} A.S. Nowick and B.S. Berry, \emph{Anelastic Relaxation in Crystalline Solids}
(Academic Press, New York, 1972).

\bibitem{Mura1987} T. Mura, \emph{Micromechanics of Defects in Solids}, 2nd rev.\ ed.\
(Martinus Nijhoff, Dordrecht, 1987).

\bibitem{Saulson1990} P.R. Saulson, ``Thermal noise in mechanical experiments,'' \emph{Phys. Rev. D} \textbf{42}, 2437--2445 (1990).

\bibitem{Kharbanda2026}
B. Kharbanda, A. Arabmoheghi, L. Catalini, M.J. Bereyhi, G. Benga, A. Zicoschi,
C.L. Degen, T.J. Kippenberg, A. Eichler and N.J. Engelsen, ``On-chip frequency-noise cancellation in nanomechanical resonators using cavity optomechanics,'' \emph{Phys. Rev. Applied} \textbf{25}, L031004 (2026),
doi:10.1103/l59w-nvcy.

\bibitem{Vig1999} J.R. Vig and Y. Kim, ``Noise in Microelectromechanical System Resonators,'' \emph{IEEE Trans. Ultrason. Ferroelectr. Freq. Control} \textbf{46}, 1558--1565 (1999).

\bibitem{YongVig1989} Y.K. Yong and J.R. Vig, ``Resonator surface contamination --- a cause of frequency fluctuations?,'' \emph{IEEE Trans. Ultrason. Ferroelectr. Freq. Control} \textbf{36}, 452--458 (1989).

\bibitem{Dankowicz2026} H. Dankowicz, S.W. Shaw and O. Shoshani, ``Improving frequency stability using slowly modulated adaptive feedback,'' \emph{Phys. Rev. Applied} \textbf{25}, 054050 (2026).

\bibitem{Kenig2012} E. Kenig, M.C. Cross, R. Lifshitz, R.B. Karabalin, L.G. Villanueva, M.H. Matheny and M.L. Roukes, ``Passive phase noise cancellation scheme,'' \emph{Phys. Rev. Lett.} \textbf{108}, 264102 (2012).

\bibitem{SM} See Supplemental Material at [URL will be inserted by publisher] for the per-device compilation table, the provenance audit, the exclusion record, and the analysis code.

\bibitem{KanellopulosSchmid2026}
K. Kanellopulos and S. Schmid, ``$1/f$ frequency-noise reduction in nanomechanical resonators by dissipation dilution,'' arXiv:2610.16711 (2026).

\bibitem{Gavartin2013} E. Gavartin, P. Verlot and T.J. Kippenberg, ``Stabilization of a linear nanomechanical oscillator to its thermodynamic limit,'' \emph{Nature Commun.} \textbf{4}, 2860 (2013).

\bibitem{Roy2018} S.K. Roy, V.T.K. Sauer, J.N. Westwood-Bachman, A. Venkatasubramanian and W.K. Hiebert, ``Improving mechanical sensor performance through larger damping,'' \emph{Science} \textbf{360}, eaar5220 (2018).

\bibitem{Sadeghi2020} P. Sadeghi, A. Demir, L.G. Villanueva, H. K\"ahler and S. Schmid, ``Frequency fluctuations in nanomechanical silicon nitride string resonators,'' \emph{Phys. Rev. B} \textbf{102}, 214106 (2020).

\end{thebibliography}
\end{document}


\title{Supplemental Material:\\[0.15em]
$1/f$ frequency noise in mechanical resonators\\ scales inversely with volume, not with quality factor}
\author{M.~L.~Roukes}
\affiliation{Kavli Nanoscience Institute and Departments of Physics, Applied Physics, and Bioengineering,\\ California Institute of Technology, Pasadena,
California 91125, USA}
\date{{\,}Submitted: 15 September 2026 \  v.~R2\, }
\maketitle

\section{The per-device compilation}

The compilation is tabulated here in full, one table per family, so that the $N=42$ points of the global fit can be completely reconstructed from these pages: 16 nanomechanical and cryogenic devices (Table~\ref{tab:devices}), 9 aluminum-nitride contour-mode resonators (Table~\ref{tab:kim}), and 17 quartz/SAW/BAW resonators (Table~\ref{tab:quartz}). Every table is generated from the machine-readable source in the public repository accompanying this Letter by \texttt{scripts/make\_sm.py}; the fit itself is
\texttt{scripts/fit\_clustered.py}, which applies the same selection rules.

\subsection{The two selection rules}

The device table underlying
Figs.~1 and~2 of the Letter, with per-device provenance,
averaging time, temperature, mass convention, measurement environment, and
source citation, is given here in full and in the public repository
accompanying this Letter; the exclusion record is
Sec.~\ref{sec:exclusions}. Two selection rules are applied uniformly to
every fitted line and to every median.

\paragraph{\textbf{Rule 1:}}\phantomsection\label{par:rule1} \emph{Only resolved plateaus enter a fit}. An upper bound is not a measurement of $\sA$ but a statement that $\sA$ is at most some value; its error is one-sided, so a bound entered in a regression as if it were a measurement will pull the fit upward at its own volume abscissa. In this compilation, the cryogenic bounds cluster within half a decade at the small-volume end, so, if included, they would lift that end of the line and flatten the slope. Both fits are shown: Fig.~1 of the Letter carries the full room-temperature population --- bounds and the AlN plates included --- as a solid yellow-green line ($-0.57\pm0.04$) beside the resolved-plateau fit ($-0.48\pm0.11$); admitting the bounds alone moves the resolved exponent to $-0.42\pm0.08$, so the size of that bias is explicitly shown rather than asserted.

\paragraph{\textbf{Rule 2:}}\phantomsection\label{par:rule2} \emph{The linewidth must be set by the resonator's internal dissipation}. ``Internal'' here means dissipation within the device --- material loss or clamping loss --- as opposed to loss into the surrounding medium. A $1/f$ floor referenced to a linewidth set by a surrounding gas or liquid is not the same quantity as one referenced to the resonator's own loss: the external damping both raises the thermomechanical noise and fixes a $Q$ that has nothing to do with the device itself, so the measured floor instead reports on the medium rather than on the resonator. Under what I hereafter term the ``\emph{internal-dissipation rule (2)}'', four devices whose $Q$ is fixed by the medium around them --- a polysilicon cantilever measured in air at $Q=21$ (Verd 2005) and its CMOS-metal successor (Verd 2008), a capacitive micromachined transducer designed to radiate into air (Park 2011), and an AlN-stack cantilever measured in air (Ivaldi 2011) --- are tabulated in Table~\ref{tab:devices} but neither plotted nor fitted. The criterion is which loss pathway sets $Q$, not whether the device sits in vacuum: suspended microchannel resonators, whose fluid is sealed inside an embedded channel while the device itself is in vacuum, pass it, as do piezoelectric contour-mode plates measured at ambient pressure, whose $Q\sim2000$--$3700$ is set by internal loss because air loading is a small perturbation on a contour mode at these frequencies.

\emph{The rules apply to the figures, not only to the fits.} Every marker drawn in Figs.~1 and 2 of the Letter satisfies the internal-dissipation rule (2); the four ambient-$Q$ devices (Verd 2005 and 2008, Park 2011, Ivaldi 2011) appear in Table~\ref{tab:devices} but in neither figure. All fitted and drawn devices are measured in vacuum except the nine AlN contour-mode plates, which were measured at ambient pressure but whose $Q=1.8$--$3.7\times10^3$ is set by internal loss (air loading is a small perturbation on a contour mode at these frequencies), and which are therefore retained under the rule as stated. Two devices satisfy every selection rule and are nonetheless excluded, because their abscissa cannot be constructed. (See \emph{Suspended-microchannel resonators}, Sec.~\ref{sec:exclusions}.)

\subsection{Nanomechanical and microelectromechanical devices}

Column definitions:
\begin{itemize}
    \item \textbf{ID} is this table's own running index; it is not the
numbering of Ref.~\cite{Sansa2016}, whose devices are a subset of the entries below.
    \item $\bm{V}$ is the resonator volume, total mass over
    density, in m$^3$ --- the quantity the primary sources report. The plotted abscissa is the participation volume $V_\varepsilon=\eta V$, with $\eta$ evaluated per mode class in Sec.~\ref{sec:eta}.
    \item $\bm{\sA}$ is the Allan deviation \emph{as stated by the
    primary source}, before the reductions described in Sec.~S2; the single authority for the fitted values is \texttt{scripts/fit\_clustered.py}. \emph{How much does the participation-volume correction move the exponent?} Within a geometry class $\eta$ is a constant, so the substitution $V\to V_\varepsilon$ displaces a family's points nearly uniformly and is largely absorbed by the per-family intercepts. Applied across all $42$ fitted devices it moves the global exponent from $-0.584$ (fitted against the raw $V$) to $-0.581$: a twentieth of the cluster standard error (Sec.~\ref{sec:eta}).
    \item \textbf{$\bm{\tau}$} is the averaging time as stated by the
    primary source, reported in seconds (s); a dash means the source states none.
    \item \textbf{Class} is the
    symbol classifier: \textbf{P} a resolved plateau, C a plateau reduced from a phase-noise spectrum, \textbf{P?} a plateau-like minimum the source does not resolve unambiguously, \textbf{U} an upper bound with the plateau unresolved, \textbf{X} no measured $\sA$ of any kind. Only classes \textbf{P} and \textbf{C} enter the fits.
    \item \textbf{Env.} distinguishes devices whose linewidth is set by internal dissipation (\texttt{device-Q}) from those whose $Q$ is fixed by their surroundings (\texttt{ambient-Q}); the latter are tabulated but neither drawn nor fitted.
    \item \textbf{Provenance} records how the number was obtained;
    only rows whose provenance begins \texttt{measured} are eligible for the fits.
    \item \textbf{Scheme} records the measurement topology established by the row-by-row audit of Sec.~S7: OL, open-loop (driven response read by counter, lock-in or network analyzer, no frequency-tracking loop); OL$\to$CL, measured open-loop and reported by the source as the equivalent closed-loop (oscillator) phase noise; PLL, phase-locked loop; SSO, self-sustained oscillator; CL, closed loop of unstated type; n.s., not stated by the source; a dash marks a row with no stability datum, or whose primary text was not available for audit. Drive and readout are recorded, where the source states them, in the per-row notes of \texttt{data/sansa\_fig1\_curated.csv}. Of the 41 entries, 16 satisfy all three conditions --- resolved plateau, device-set $Q$, measured provenance --- and enter the nanomechanical and cryogenic portions of the global fit; the remaining 25 are recorded here for completeness and for the reader who wishes to reweigh them.
    \item $\bm{Q}$ records the quality factor as stated by the primary source, registered with its attribution in \texttt{data/q\_values.csv} of the repository; a dash marks a device whose source reports none. The aluminum-nitride and quartz tables below carry their own per-device $Q$ columns. No value is interpolated or recalled from secondary literature: the column is deliberately incomplete where the record is.
\end{itemize}

{\footnotesize\setlength{\tabcolsep}{1.4pt}%
\begin{longtable}{l l c c c c c c c c c c}
\caption{The nanomechanical and MEMS compilation, 41 entries.
\label{tab:devices}}\\
\hline\hline
ID & Source & Material & $T$ (K) & $V$ (m$^3$) & $\sA$ & $\tau$ (s) & $Q$ & Class & Env. & Provenance & Scheme \\
\hline
\endfirsthead
\hline\hline
ID & Source & Material & $T$ (K) & $V$ (m$^3$) & $\sA$ & $\tau$ (s) & $Q$ & Class & Env. & Provenance & Scheme \\
\hline
\endhead
\hline
\endfoot
\hline\hline
\endlastfoot
1 & Chaste2012-after & CNT & 5.5 & 2.73e-25 & 2.8e-6 & 2 & -- & P? & device-Q & measured-direct & CL \\
2 & Chaste2012-before & CNT & 6 & 2.73e-25 & 3.0e-3 & -- & -- & X & device-Q & trace-only & -- \\
3 & Jensen2008 & CNT & 295 & 7.18e-25 & 1.3e-4 & 0.01 & -- & U & device-Q & measured-freq-noise & -- \\
4 & Chiu2008 & CNT & 6 & 5.91e-25 & 5.3e-5 & 1 & -- & X & device-Q & unclear-implicit & -- \\
5 & Chen2009 & graphene & 5 & 6.82e-22 & 3.6e-7 & 0.3 & 14000 & X & device-Q & derived-DR & -- \\
6 & Bartsch2012 & Si & 300 & 5.97e-21 & 8.92e-6 & 1 & -- & P & device-Q & measured-direct & PLL \\
7 & Sansa2014 & Si & 295 & 2.64e-20 & -- & -- & -- & X & device-Q & none & -- \\
8 & Hanay2012-SiN & SiN & 70 & 1.81e-19 & 3.0e-6 & 10 & -- & P? & device-Q & measured-direct & PLL \\
9 & Naik2009 & SiC & 40 & 2.87e-20 & 2.0e-7 & 5 & -- & P & device-Q & measured-direct & PLL \\
10 & Ramos2012 & Si & 295 & 1.12e-19 & 3.0e-6 & 1 & -- & U & device-Q & measured-direct & -- \\
11 & Feng2008 & SiC & 22 & 2.35e-20 & 2.5e-8 & 1 & 2500 & U & device-Q & measured-direct & OL \\
12 & Yang2006 & SiC & 4.2 & 5.73e-20 & 5.0e-8 & 1 & -- & U & device-Q & measured-direct & -- \\
13 & Mile2010 & Si & 295 & 2.4e-19 & 6.0e-7 & 1 & -- & P & device-Q & measured-direct & OL \\
14 & Hanay2012-cSi & Si & 80 & 2.4e-19 & 8.0e-8 & 0.5 & -- & P? & device-Q & measured-direct & PLL \\
15 & Gray2012 & GaN & 300 & 4.88e-19 & 1.44e-5 & 1 & -- & U & device-Q & measured-direct & -- \\
16 & Verd2008 & CMOS-metal & 295 & 4.5e-18 & 2.71e-8 & 1 & -- & C & ambient-Q & measured-phase-noise & SSO \\
17 & Villanueva2011 & AlN/Mo & 295 & 8.88e-19 & 3.1e-6 & -- & -- & C & device-Q & measured-phase-noise & SSO \\
18 & Gavartin2013 & SiN & 300 & 6.21e-18 & 1.5e-7 & 0.1 & -- & P & device-Q & measured-direct & OL \\
19 & Verd2005 & polySi & 295 & 2.02e-17 & -- & -- & 21 & X & ambient-Q & derived-DR-simulated & -- \\
20 & Fong2012 & SiN & 296 & 2.65e-17 & 5.8e-8 & -- & 4.0e5 & P & device-Q & measured-PSD & OL \\
21 & Olcum2014 & Si & 295 & 1.58e-16 & 6.0e-9 & 0.1 & -- & P & device-Q & measured-direct & SSO \\
22 & Park2011 & Si & 295 & 1.27e-16 & 2.74e-8 & 0.05 & -- & P & ambient-Q & measured-direct & SSO \\
23 & Larsen2011 & SiN & 295 & 1.09e-15 & -- & -- & -- & X & device-Q & none & -- \\
24 & Ivaldi2011 & AlN-stack & 295 & 3.27e-15 & 1.0e-8 & 10 & -- & U & ambient-Q & measured-direct & -- \\
25 & Burg2007 & Si & 293 & 4.29e-14 & 1.0e-8 & 1 & -- & U & device-Q & measured-direct & -- \\
26 & Fong2012-78K & SiN & 78 & 2.65e-17 & 2.10e-8 & -- & -- & P & device-Q & measured-PSD & OL \\
27 & Fong2012-5K & SiN & 5 & 2.65e-17 & 6.94e-9 & -- & 2.2e6 & P & device-Q & measured-PSD & OL \\
28 & Feng-thesis-295 & SiC & 22 & 4.99e-20 & 4.7e-8 & 1 & 3000 & U & device-Q & measured-direct & OL \\
29 & Feng-thesis-420 & SiC & 22 & 3.46e-20 & 3.1e-7 & 1 & 1200 & U & device-Q & measured-direct & OL \\
30 & Feng-thesis-411 & SiC & 22 & 2.25e-20 & 6.6e-8 & 1 & 2600 & U & device-Q & measured-direct & OL \\
31 & Feng-thesis-482 & SiC & 22 & 2.21e-20 & 2.1e-8 & 1 & 2000 & U & device-Q & measured-direct & OL \\
32 & Feng2007-SiNW & Si & 25 & 3.35e-20 & 7.6e-8 & 7 & 2000 & U & device-Q & measured-direct & PLL \\
33 & Maillet2018 & SiN/Al & 4.2 & 9.75e-18 & 1.662e-08 & 10 & 4.5e5 & P & device-Q & measured-rms-band & OL \\
34 & Maillet2018 & SiN/Al & 4.2 & 9.75e-18 & 9.010e-09 & 10 & 4.0e5 & P & device-Q & measured-rms-band & OL \\
35 & Maillet2018 & SiN/Al & 4.2 & 5.85e-19 & 3.691e-08 & 10 & 1.8e4 & P & device-Q & measured-rms-band & OL \\
36 & Maillet2018 & SiN/Al & 4.2 & 9.75e-18 & 1.378e-07 & 10 & 6.0e5 & P & device-Q & measured-rms-band & OL \\
37 & Wang2020 & SiN & 300 & 1.5e-18 & 6.06e-7 & 3 & 27300 & P & device-Q & measured-direct & OL \\
38 & Wang2020 & SiN & 300 & 1.5e-18 & 3.60e-7 & 3 & 44100 & P & device-Q & measured-direct & OL \\
39 & Duraffourg2018 & SiN/TiN/a-Si & 300 & 1.35e-19 & 1.5e-7 & 0.1 & 1800 & P & device-Q & measured-direct & CL \\
40 & Zhang2013 & SiN & 295 & 4e-19 & 1.6e-7 & 30 & 5225 & P & device-Q & measured-digitized & OL \\
41 & Feng2007-SiNW-M188 & Si & 25 & 1.08e-20 & 6.0e-8 & 7 & -- & U & device-Q & measured-direct & PLL \\
\end{longtable}}

\subsection{Aluminum-nitride contour-mode resonators}

The piezoelectric contour-mode family of Ref.~\cite{Kim2017}, Fig.~6(a),
digitized at 600~dpi. \textbf{$L$} is the electrode length $E_L$ in $\mu$m;
$V=E_L\times84~\mu\mathrm{m}\times1.0~\mu\mathrm{m}$, the transverse
dimension being the 21-electrode, 4-$\mu$m-pitch array and the thickness the
1-$\mu$m AlN film (H.~J.~Kim, private communication, 31 August 2026; an
earlier draft used the $100\times1.1~\mu$m plate of a different experiment
from the same group, and the volumes were $1.3\times$ too large).
\textbf{$\hmo$} is the flicker coefficient recovered from the reported
equivalent oscillator phase noise as $\hmo = \mathcal{L}(f)f^{3}/f_s^{2}$ at
$f=1$~kHz and $f_s=1.1$~GHz ($f_s$ varies by $0.2\%$ across the series), the
authors having confirmed that their $\mathcal L(f)$ carries no single-sideband
factor of two; and $\sA=\sqrt{2\ln2\,\hmo}$. These are therefore class-C
entries, plateaus reduced from a spectrum rather than read from an Allan
curve. Each published point is the mean of four nominally identical devices
from two chips, so the within-series scatter of Table~\ref{tab:kim} is that of
four-device means, and the per-device scatter is larger.
\textbf{Anchor} records the tether design. \textbf{Fit} marks the rows that
enter the global fit. Scheme, all rows: OL$\to$CL --- the baseband noise was measured open-loop and converted by the authors to the equivalent closed-loop phase noise $\mathcal L(f)$ (Sec.~S7).

\emph{Spectral conventions.}---Throughout the Letter and this document $S_y$
and $S_\phi$ denote \emph{one-sided} densities (IEEE Std 1139), with
$\mathcal L(f)=S_\phi(f)/2$ and $\sA=\sqrt{2\ln2\,\hmo}$ on the flicker
plateau. Where a source quotes $\sA(\tau)$ directly---the large majority of the compilation---the extraction is convention-free, since $\sA$ is the measured quantity. The convention enters only where a spectral level is converted: this contour-mode family, and the class-C phase-noise entries. A misidentified one- versus two-sided convention would displace such a point by $\sqrt2$ in $\sA$, i.e.\ $0.15$~dec ($0.30$~dec in $\hV$)---small against the $\pm0.4$--$0.9$-dec family scatter, and a per-point offset uncorrelated with volume, so it perturbs scatter, not the exponent. Exactly this occurred: the contour-mode family was first reduced under the IEEE convention and corrected when the authors confirmed theirs; the nine points moved down by
$0.30$~dec together with a $0.12$-dec volume correction, and the global
exponent moved by $0.02$, a third of a cluster error.

{\footnotesize
\begin{longtable}{l r r l l r l c}
\caption{Aluminum-nitride contour-mode resonators, 10 entries,
9 fitted. \label{tab:kim}}\\
\hline\hline
ID & $L$ ($\mu$m) & $V$ (m$^3$) & $\sA$ & $\hmo$ (Hz$^{-1}$) & $Q$ & Anchor & Fit \\
\hline
\endfirsthead
\hline\hline
ID & $L$ ($\mu$m) & $V$ (m$^3$) & $\sA$ & $\hmo$ (Hz$^{-1}$) & $Q$ & Anchor & Fit \\
\hline
\endhead
\hline
\endfoot
\hline\hline
\endlastfoot
K\_n28 & 28 & 2.352e-15 & 3.30e-09 & 7.840e-18 & 1835 & narrow & $\bullet$ \\
K\_n52 & 52 & 4.368e-15 & 2.76e-09 & 5.515e-18 & 2695 & narrow & $\bullet$ \\
K\_n76 & 76 & 6.384e-15 & 2.11e-09 & 3.235e-18 & 2480 & narrow & $\bullet$ \\
K\_n100 & 100 & 8.400e-15 & 1.29e-09 & 1.211e-18 & 2941 & narrow & $\bullet$ \\
K\_n148 & 148 & 1.243e-14 & 1.35e-09 & 1.318e-18 & 3060 & narrow & $\bullet$ \\
K\_f28 & 28 & 2.352e-15 & 8.70e-09 & 5.460e-17 & 386 & full & -- \\
K\_f52 & 52 & 4.368e-15 & 2.50e-09 & 4.488e-18 & 2458 & full & $\bullet$ \\
K\_f76 & 76 & 6.384e-15 & 1.63e-09 & 1.920e-18 & 2976 & full & $\bullet$ \\
K\_f100 & 100 & 8.400e-15 & 1.65e-09 & 1.968e-18 & 3472 & full & $\bullet$ \\
K\_f148 & 148 & 1.243e-14 & 1.87e-09 & 2.507e-18 & 3652 & full & $\bullet$ \\
\end{longtable}}

\subsection{Quartz, SAW and BAW resonators}

The macroscopic family. \textbf{Series} groups points from a common source.
\textbf{$\delta\sA$} and \textbf{$\delta V$} are the stated systematic
uncertainties in decades, which are the weights the global fit uses.
\textbf{Source} is the primary reference. Every row now carries a named
primary attribution; the seven rows recovered during revision are detailed in
Sec.~\ref{sec:provenance}, and the generous uncertainties originally assigned
to them are retained unchanged.

{\footnotesize
\begin{longtable}{l r l c r r r c l l}
\caption{Quartz, SAW and BAW resonators, 18 entries, 17 fitted. Scheme as defined for Table~\ref{tab:devices}.
\label{tab:quartz}}\\
\hline\hline
ID & $V$ (m$^3$) & $\sA$ & Series & $\delta\sA$ & $\delta V$ & $Q$ & Fit & Scheme & Source \\
\hline
\endfirsthead
\hline\hline
ID & $V$ (m$^3$) & $\sA$ & Series & $\delta\sA$ & $\delta V$ & $Q$ & Fit & Scheme & Source \\
\hline
\endhead
\hline
\endfoot
\hline\hline
\endlastfoot
qA1 & 2.9e-12 & 1.30e-11 & A & 0.20 & 0.15 & -- & $\bullet$ & OL & \parbox[t]{4.4cm}{\raggedright Parker \& Andres 1994 FCS pp.530-538, Table 2 design A + Fig 6; 449.7 MHz SAW; V = averaging area x 1 lambda (\,\ldots\strut} \\
qA2 & 4.9e-12 & 1.05e-11 & A & 0.20 & 0.15 & -- & $\bullet$ & OL & \parbox[t]{4.4cm}{\raggedright Parker \& Andres 1994 FCS Table 2 design B + Fig 6; 449.7 MHz SAW; V = averaging area x 1 lambda (Eq. 12).\strut} \\
qA3 & 9.7e-12 & 8.30e-12 & A & 0.20 & 0.15 & -- & $\bullet$ & OL & \parbox[t]{4.4cm}{\raggedright Parker \& Andres 1994 FCS Table 2 design C + Fig 6; 449.7 MHz SAW; V = averaging area x 1 lambda (Eq. 12).\strut} \\
qA4 & 8.8e-12 & 9.10e-12 & A & 0.20 & 0.15 & -- & $\bullet$ & OL & \parbox[t]{4.4cm}{\raggedright Parker \& Andres 1994 FCS Table 2 design D + Fig 6; 449.7 MHz SAW; V = averaging area x 1 lambda (Eq. 12).\strut} \\
qA5 & 1.77e-11 & 5.90e-12 & A & 0.20 & 0.15 & -- & $\bullet$ & OL & \parbox[t]{4.4cm}{\raggedright Parker \& Andres 1994 FCS Table 2 design E + Fig 6; 449.7 MHz SAW; V = averaging area x 1 lambda (Eq. 12).\strut} \\
qB1 & 6.0e-8 & 6.40e-14 & B & 0.15 & 0.15 & -- & $\bullet$ & OL$\to$CL & \parbox[t]{4.4cm}{\raggedright Parker \& Andres 1993 FCS p.182: best 5-MHz 5th-OT SC BAW, $S_y(1\,\mathrm{Hz})=3\times10^{-27}$, quoted there from their Refs.~[6],[13]; $V=0.06$~cm$^3$ as stated.\strut} \\
qB2 & 6.0e-8 & 5.90e-14 & B & 0.15 & 0.30 & 2.1e6 & $\bullet$ & SSO & \parbox[t]{4.4cm}{\raggedright Rubiola \& Giordano 2007 (UFFC 54, 15) Table I: CMAC Pharao BVA class, b-3 = -132 dBrad2/\allowbreak Hz at 5 MHz; V = 0.06\,\ldots\strut} \\
qB3 & 6.0e-8 & 2.50e-14 & B & 0.05 & 0.15 & 2.63e6 & $\bullet$ & n.s. & \parbox[t]{4.4cm}{\raggedright Salzenstein 2010 Electron. Lett., Oscilloquartz 8607-C 5 MHz BVA; Q\_U 2.63e6, Q\_L 1.57e6\strut} \\
qX1 & 9.0e-13 & 1.23e-10 & X & 0.10 & 0.30 & 6000 & -- & OL & \parbox[t]{4.4cm}{\raggedright Watkins, Kaisar, Rais-Zadeh \& Feng, Proc. IEEE MEMS 2025 (Kaohsiung) 201-204, doi 10.1109/\allowbreak MEMS61431.2025.1091\,\ldots\strut} \\
DH1 & 1.400e-10 & 2.98e-12 & DH & 0.10 & 0.15 & 1.25e5 & $\bullet$ & OL$\to$CL & \parbox[t]{4.4cm}{\raggedright Driscoll \& Hansen 1993 Proc 47th FCS 186 Table 1: SC 80 MHz 3OT\strut} \\
DH2 & 8.300e-11 & 1.90e-12 & DH & 0.10 & 0.15 & 1.19e5 & $\bullet$ & OL$\to$CL & \parbox[t]{4.4cm}{\raggedright Driscoll \& Hansen 1993 Table 1: SC 100 MHz 3OT\strut} \\
DH3 & 2.800e-10 & 1.86e-12 & DH & 0.10 & 0.15 & 7.5e4 & $\bullet$ & OL$\to$CL & \parbox[t]{4.4cm}{\raggedright Driscoll \& Hansen 1993 Table 1: SC 160 MHz 5OT\strut} \\
DH4 & 1.000e-09 & 1.67e-12 & DH & 0.10 & 0.15 & 3.00e5 & $\bullet$ & OL$\to$CL & \parbox[t]{4.4cm}{\raggedright Driscoll \& Hansen 1993 Table 1: BT 100 MHz 5OT\strut} \\
DH5 & 2.200e-09 & 1.18e-12 & DH & 0.10 & 0.15 & 2.50e5 & $\bullet$ & OL$\to$CL & \parbox[t]{4.4cm}{\raggedright Driscoll \& Hansen 1993 Table 1: AT 40 MHz 5OT\strut} \\
P1 & 2.705e-12 & 1.39e-11 & P & 0.10 & 0.15 & 21950 & $\bullet$ & n.s. & \parbox[t]{4.4cm}{\raggedright Parker 1993 UFFC 40 831 Table I: 450 MHz SAW, aperture 90L, cavity 175L\strut} \\
P2 & 5.806e-12 & 9.27e-12 & P & 0.10 & 0.15 & 22300 & $\bullet$ & n.s. & \parbox[t]{4.4cm}{\raggedright Parker 1993 Table I: aperture 90L, cavity 375L\strut} \\
P3 & 5.427e-12 & 9.27e-12 & P & 0.10 & 0.15 & 14600 & $\bullet$ & n.s. & \parbox[t]{4.4cm}{\raggedright Parker 1993 Table I: aperture 180L, cavity 175L\strut} \\
P4 & 1.163e-11 & 7.35e-12 & P & 0.10 & 0.15 & 20850 & $\bullet$ & n.s. & \parbox[t]{4.4cm}{\raggedright Parker 1993 Table I: aperture 180L, cavity 375L\strut} \\
\end{longtable}}

\section{The audit, quantified}
\vspace{-0.5cm}
This section gives the $\tau$ and mass statistics behind the reconstruction described in the main text. Of the twenty-nine devices carrying a measured stability, satisfying the internal-dissipation rule (2), and surviving the abscissa test of Sec.~S4 (the two fluid-filled devices are excluded there), twenty-five state an averaging time once the primary source is consulted, and those run from 10~ms to 30~s; four are characterized spectrally and state none. Reference~\cite{Sansa2016} itself states $\tau$ for two of the nineteen it reports, which is what makes the reconstruction necessary rather than merely superfluous housekeeping. Across three and a half decades of $\tau$, a $\tau^{-1/2}$ white branch generates $1.74$ decades of apparent spread in $\sA$ --- against the $\pm0.6$-decade scatter of the room-temperature family about its own median. The extremes of the $\tau$ distribution belong to devices that do not show a clear plateau. Across the compilation $\log\tau$ and $\log V$ carry $r=+0.46$ at $n=25$ (two-sided $p=0.022$): the correlation is real --- larger
devices tend to be read at longer $\tau$ --- so it is tested directly rather than dismissed. A resolved plateau is $\tau$-independent by construction, and the covariate test confirms it: adding $\log\tau$ to the weighted fit over the fitted devices that state one ($n=12$) moves the common exponent by $+0.005$, a twelfth of one cluster error, with a covariate coefficient consistent with zero ($+0.10\pm0.08$). The convention inflates scatter; it does not measurably tilt the fit. Five devices carry a mass that is a coarse approximation, ambiguous, or untraceable; one is quoted by its own source at two total masses differing by a factor of two, and that factor is carried as its volume uncertainty rather than resolved by choosing between them.

\section{Why a plateau may be hidden, or absent}
\label{sec:hidden}
A $1/f$ floor is observable only where the $\alpha=-1$ term of Eq.~(1) dominates every other term; the window in $\tau$ over which that holds, however, is bounded on both sides. The various possible bounds are described below, by type.

\paragraph{Closing from the left.} Referred to frequency inside a tracking loop, white detector noise appears as white \emph{frequency} noise and the additive branch descends as $\sA=(1/2Q)\sqrt{S_x/2\tau x_c^2}$, with $S_x$ the readout
displacement noise and $x_c$ the operating amplitude; the $1/2Q$ is the Leeson
factor, not a general property of additive noise. Read open-loop, the same
detector noise is white \emph{phase} noise, descending as $\tau^{-1}$ with no
$Q$ dependence at all, and flicker phase noise descends as $\tau^{-1}$ also.
Only the white-frequency case gives $\tau^{-1/2}$. It sits high when $Q$ is low, when the device
is driven far below its critical amplitude, or when transduction is poor --- none
of which is a property of the material. That is what Ref.~\cite{Sage2013}
exploits: raising the chamber pressure lifted the additive branch by $6.4\times$
and left the plateau where it was.

\textit{A second left-hand edge, imposed by the resonator.} Frequency
information is carried by the discriminant $\partial\phi/\partial\omega=2Q/\omega_0$, which rolls off above the resonator half-width $f_0/2Q$. Whether that roll-off matters depends on the scheme. A freely ringing or self-sustaining oscillator accumulates phase as $\int\omega_0(t')\,dt'$, and its instantaneous frequency follows $\sqrt{k(t)/m}$ adiabatically, so a counter returns the true gate average and $\tau_r$ does not enter --- $\tau_r$ governs energy exchange with the drive and the bath, not the free frequency. A resonator driven at fixed $\omega_d$, or tracked by a phase-locked loop, is different: the transduced phase relaxes on $\tau_r=Q/\pi f_0$, and no loop tracks faster than its own error signal develops. For those schemes the measured spectrum is the true one multiplied by
$[1+(2\pi f\tau_r)^{2}]^{-1}$.

The consequence is not a suppressed plateau but a substituted one. Evaluating $\sA$ on a low-passed $1/f$ spectrum yields a \emph{rising} branch, $\sA\propto\tau^{+0.95}$, throughout $\tau\ll\tau_r$ --- a ramp indistinguishable from linear drift, with no flat region anywhere --- and the plateau recovers only slowly above it: $\sA/\sA^{\rm plateau}=0.10$, $0.53$ and $0.97$ at $\tau/\tau_r=0.1$, $1$ and $10$. A faithful reading therefore requires $\tau\gtrsim10\,\tau_r$ rather than merely $\tau>\tau_r$, and a $\tau^{+1}$ ramp below $\tau_r$ is the diagnostic signature of a response-limited measurement.

The same roll-off refers white readout noise back to frequency as $f^{2}$, which descends as $\tau^{-1}$. Below $\tau_r$ a $\tau^{-1}$ fall therefore meets a $\tau^{+1}$ rise, producing a sharp minimum that is not the floor and that may lie on either side of it. Reference~\cite{Sadeghi2020} is the documented case: its global minimum, at $\tau=5$~ms and $70\times$ inside its own $\tau_r=0.34$~s, reads $4.5\ee{-10}$ against $2.2\ee{-9}$ at $\tau_r$ itself, low by a factor of $4.9$. Devices measured inside their own $\tau_r$ are for
this reason excluded rather than bounded.

\paragraph{Closing from the right.} Drift rises as $\tau$, or as $\tau^{1/2}$ for a random walk, and is set by temperature excursions ($\sim\!30$~ppm/K in silicon), stress relaxation and creep at the clamps, net adsorbate loading --- the DC counterpart of the fluctuation mechanism of Ref.~\cite{YongVig1989}, and quite distinct from it --- photothermal shifts following laser-power drift, aging in quartz, and drift of the reference against which the measurement is made.

\textit{The two edges act at opposite ends of the volume axis.} The floor falls as $V^{-1/2}$ while drift does not scale with volume at all, so the effects of drift bury the plateau in \emph{large} devices: at the quartz BVA the floor is $\sA=9.5\ee{-14}$, while an aging rate of $10^{-11}$ per day, which is $D=1.16\ee{-16}$~s$^{-1}$, contributes $D\tau/\sqrt2=8.2\ee{-14}$ at $\tau=10^{3}$~s. The two cross at $1.2\ee{3}$~s, which is why the frequency-control literature reports $\hmo$ spectrally rather than fighting $\sA$ at long $\tau$.
At the other end the comparison is not close: a $1~\mu$m$^3$ nanomechanical device has $\sA=2.7\ee{-7}$ and would need 10~mK of drift within the gate time to become obfuscated, and what buries $\sApl$ in small devices is the left-hand branch instead. Geometry alone barely arbitrates between the two, since $\sA^{\rm add}/\sApl \ \propto\ \sqrt{S_x}\ V^{1/6}/\sqrt{Q}$ and six decades of volume move  that ratio only by one decade. Plateau visibility is therefore a transduction figure of merit rather than a size effect, and the bounds in this compilation map the readouts used rather than the physics.

\paragraph{Dissipation dilution closes the window.} The same parameter that provides the advantages of dissipation dilution serves also to close the window, making observation of the $1/f$ floor impossible. Raising $Q$ lowers the additive branch as $1/\sqrt{Q}$, which helps; but $\tau_r=Q/\pi f_0$ grows in proportion to $Q$, and drift does not move. A plateau is observable only while $\tau_r$ remains shorter than the time at which drift overtakes the floor. Writing the linear fractional drift as $y=Dt$, whose Allan deviation is $D\tau/\sqrt2$, that time is $\tau_{\rm drift}=\sqrt2\,\sA/D$, so a plateau survives only while
\begin{equation}
Q \;<\; \sqrt{2}\,\pi f_0\,\sA/D .
\label{eq:Qmax}
\end{equation}
Every design choice that produces dilution pushes against this bound at once: $Q$ rises, $f_0$ falls as spans lengthen and clamps soften, and $\sA$ falls as $V^{-1/2}$ because diluted devices are large. Read in reverse, Eq.~\eqref{eq:Qmax} states what a measurement would require rather than presuming a drift rate: for the $\tau_r=3.6\ee{3}$~s of
Ref.~\cite{Bereyhi2022} to fall inside a window at all, that device would have to hold the (assumed) linear drift to $D<5.7\ee{-11}$~s$^{-1}$ --- a fractional drift below $5\ee{-6}$ per day --- across the entire record. No such stability is reported for any suspended thin-film resonator. The record devices are further out still: $\tau_r=3600$~s for Ref.~\cite{Bereyhi2022} at $Q=1.5\ee{9}$, $2.9\ee{3}$~s for the $Q=1.3\ee{10}$ strained-silicon strings of Ref.~\cite{Beccari2022}, and $9.8\ee{3}$~s --- nearly three hours --- for the $Q=6.6\ee{9}$ centimeter-scale string of Ref.~\cite{Cupertino2024}. No published
stability measurement on an extremely diluted resonator has been made at an averaging time exceeding its own response time. That, rather than any absence of $1/f$, is why the dilution literature reports so few plateaus, and it is a second reading of the inversion described in the text: beyond a certain dilution, what is lost is not only sensing bandwidth but the ability to characterize the floor at all. The ramp identified above supplies a test that costs no new fabrication: measured over two decades of $\tau$ below its own $\tau_r$, a diluted resonator whose frequency noise is $1/f$ and whose readout is phase-locked must show $\sA\propto\tau$; measured beyond $10\,\tau_r$, it
should show the $1/f$ floor itself.

\paragraph{Genuinely absent.} Four mechanisms can remove the plateau rather than hide it. (i) The noise spectral density does not follow $1/f$. As an example, adsorbate diffusion gives $S_y\propto f^{-1/2}$ and $\sA\propto\tau^{-1/4}$~\cite{Yang2011}, thus yielding a slow
descent that never flattens. (ii) Too few fluctuators. One two-state defect contributes a Lorentzian, whose $\sA$ rises as $\tau^{1/2}$ and then falls as $\tau^{-1/2}$ --- a maximum, not a plateau. The flat floor is what log-uniform superposition of many such Lorentzians produces, so below some volume the $1/f$ floor will granulate into individual Lorentzians (the time record will show discrete telegraph steps), which is precisely what the single-defect measurements of Refs.~\cite{Yuksel2026,Maksymowych2025} resolve. The arithmetic that yields the invariant thus also sets a lower volume bound on the existence of a plateau. (iii) The fluctuators are no longer thermally
activated, as below 1~K in lithium niobate. (iv) The interrogated slice of the barrier distribution is narrower than one might suppose.
Equation~(7) of the Letter maps $\tau$ onto $\varepsilon_a$, and a window of 10~ms--100~s samples $595$--$833$~meV at 300~K --- a slice $238$~meV wide --- which narrows to $155$--$217$~meV at 78~K and to $\sim\!4$~meV at 5~K. A flat plateau is therefore itself evidence that $D(\varepsilon_a)$ is smooth across that slice, and where a plateau is instead tilted, the tilt measures $dD/d\varepsilon_a$. Cooling narrows the slice, which is one reason the cryogenic plateaus are the least regular of the three families.

\paragraph{False plateaus.} The failure runs the other way as well. Flicker in the readout --- laser frequency noise, amplifier $1/f$, thermometry --- produces a convincing $\tau$-independent floor that belongs to the instrument. Being volume-independent it is excluded globally at $9\sigma$ by the fit itself, but for a single device the discriminating test is spatial: Ref.~\cite{Sage2013} finds two points $9~\mu$m apart on one resonator correlated at unity while neighboring devices sharing a readout are uncorrelated. Operation past the critical amplitude is the second route, converting amplitude noise into frequency noise through the Duffing coefficient; unlike every other
mechanism here it can move the apparent floor in either direction, which is why devices driven into bistability are excluded.

\paragraph{The consequence for the fits.} The mechanisms above divide in two, and the division is what makes the bounds usable. The additive and drift closures both raise the apparent floor, because the observed minimum exceeds the true one by the flanking contributions evaluated at it; a plateau hidden by either is an upper bound and never a lower one, and such devices are plotted as bounds. The response-time closure and operation past the critical amplitude are the exceptions --- each can displace an apparent $1/f$ floor in either direction, as Ref.~\cite{Sadeghi2020} demonstrates by a factor of nearly five --- so devices in either category enter no fit; where one is nonetheless drawn, as with the dissipation-diluted point of Fig.~1 of the Letter, it is marked open and its status stated. The first division also fixes the sign of the bias the bounds carry: occupying the small-volume end, they must flatten any exponent they are admitted to, and they do, from $-0.48\pm0.11$ to $-0.42\pm0.08$; the all-data line of Fig.~1 folds in the AlN plates as well and carries $-0.57\pm0.04$. Per-device
attributions, and the identification of which edge closed the window for each bound, are given in Sec.~\ref{sec:exclusions} and in the per-device notes of the repository.

\section{The exclusion record}
\label{sec:exclusions}

Points excluded from the figures, and the reason in each case. The complete record, with the arithmetic behind each decision, is
\texttt{notes/sm\_exclusions.md} in the accompanying repository.

\begin{description}
\item[qX1, Watkins 2025] AlN-on-Si dual-ring BAW,
radial-extensional mode at 11.08~MHz. Excluded because the quoted
$1.23\ee{-10}$ is the \emph{minimum} of an open-loop curve at
$\tau\approx4.8$~s, not a resolved plateau: the same paper quotes
$8.86\ee{-10}$ at $\tau\approx1222$~s, and $3.76\ee{-9}$ (self-sustained
oscillator) and $1.69\ee{-10}$ (phase-locked loop) for the same device. It is also not a silicon MEMS BAW, as first cataloged, but a 20~$\mu$m Si + 1~$\mu$m AlN + 100~nm metal stack.

\item[M250n1, Maillet 2018, 250~$\mu$m beam, mode $n=1$] Retained in the fit but flagged: Maillet states that for $n=1$ the frequency fluctuations reach ${\approx}20\%$ of the linewidth, by far the largest ratio in that work, and $n=1$ has the smallest Duffing coefficient of the three modes ($8.5\ee{15}$ against $5.7\ee{17}$~Hz\,m$^{-2}$), so its extrapolation to vanishing drive has the longest lever arm. Modes $n=3$ and $n=5$ bracket the cryogenic reference at $+0.19$ and $-0.35$ decades; $n=1$ sits $2.02$ decades above it.

\item[Four ambient-$Q$ devices] A flexural cantilever measured in air at
$Q=21$, a capacitive micromachined transducer designed to radiate into air, and two related devices. Tabulated but not drawn or fitted, under the second selection rule of Sec.~S1 and the Letter's End Matter.

\item[Cupertino 2024 and Beccari 2022] The two highest-$Q$ mechanical
resonators reported ($6.6\ee{9}$ and $1.3\ee{10}$) are absent from Fig.~1
because neither carries an ordinate value. Ref.~\cite{Cupertino2024} reports only an \emph{extrapolated thermomechanical} $\sigma_y\sim3\ee{-12}/\sqrt{\tau}$, which is a $\tau^{-1/2}$ branch and not a $1/f$ plateau; Ref.~\cite{Beccari2022} reports no Allan deviation, no frequency-noise spectrum and no stability measurement of any kind. Neither qualifies even for the class-X symbol, which requires a quantity that can be reduced.

\item[The dissipation-diluted point of Fig.~1] The evidence is the dilution community's own, and it arrives with a caveat that
is itself part of the argument. The perimeter mode of Ref.~\cite{Bereyhi2022}, at $Q=1.5\ee{9}$, exhibits a $\tau$-independent excess over $0.1$--$5$~s corresponding to $\hV=4.8\ee{-31}$~m$^3$, against $1.0\ee{-31}$~m$^3$ for the room-temperature reference: five and a half decades of $Q$ move the invariant by $0.7$ decades, and in the unhelpful direction. That excess, however, spans $0.1$--$5$~s, wholly inside the device's own response time, $\tau_r=3600$~s. Its flatness is itself informative: a $1/f$ spectrum viewed through that response would climb $1.7$ decades across those two, so what is observed is not the resonator's $1/f$ filtered. It is either a true plateau, if the  measurement reads the freely accumulating phase, where $\tau_r$ does not enter, or flicker in the readout, which enters downstream of the resonator and is not filtered at all. The published account does not settle which, so the point is plotted as a bound and enters no fit. The difficulty generalizes---no published stability measurement on an extremely diluted resonator has been made at an averaging time exceeding its own $\tau_r$ (End Matter)---and the $Q$-independence claim accordingly rests not on this device but on Ref.~\cite{Sage2013}, where $Q$ was varied under control, and on the six decades of $Q$ spanned by the compilation. Measurements published since point the same way, and they name the mechanism: a soft-clamped perimeter resonator of the same family at $Q=4\ee{7}$ carries a residual frequency noise its authors ascribe to ``coupling of the resonator to two-level systems,'' which ``induces intrinsic frequency flicker''~\cite{Kharbanda2026}; a torsionally diluted Si$_3$N$_4$ pendulum at $Q=2.5\ee{6}$ reaches its thermal limit and is thereafter drift-bound~\cite{Pratt2023}. And the loss face of the same relaxator bath was measured in the same material family already in 2018: the intrinsic (undiluted) loss angle of a high-stress Si$_3$N$_4$ nanobeam is frequency-independent to within a factor of two across $50$~kHz--$50$~MHz --- structural damping~\cite{Fedorov2018} --- the dissipative signature of exactly the log-uniform distribution of correlation times that generates a $\tau$-independent $\sA$ (Sec.~S12).

\item[Tubsrinuan 2024] GaAs SAW cavity at 10~mK. Excluded for want of a stated acoustic aperture (the transverse width of the SAW beam, set by the transducer-finger overlap), without which the footprint $A$ of Sec.~\ref{sec:eta}, and hence the resonator volume, cannot be constructed.
\end{description}

\paragraph{Suspended microchannel resonators.} Olcum 2014 and Burg 2007 satisfy the admission rules---measured $\sA$, traceable to the primary source, internal dissipation---and are excluded nonetheless, because both were operated \emph{water-filled} and a fluid-filled composite has no defensible participation volume: $V=m/\rho$ converts a total mass that includes the water using the silicon density, and the dry-cantilever $\eta=0.237$ does not describe a mode whose fluid carries mass but no strain energy. Neither factor survives, so the abscissa is indeterminate. The direction of the effect must be stated: Olcum 2014 sat about $1.9$ decades \emph{below} the room-temperature reference, so its removal flattens the band and flatters the invariant. The exclusion therefore rests on the indeterminate abscissa, fixed as a criterion before position was considered, and not on where the points fall. Burg 2007 was an upper bound and entered no fit; its removal moves only the drawn marker and the bounds-admitted room-temperature trend of Fig.~1.

\paragraph{Off scale, and not drawn.} Carets at the frame were removed from Fig.~2 (they read as data); a device whose ordinate falls outside its panel is instead omitted from the drawing and listed here. At present that is one device: Gray 2012, an upper bound at $\hV=2.2\ee{-29}$~m$^3$, more than two decades above the room-temperature frame. Two further devices are outside their panels' volume windows and dropped by name rather than pinned: Jensen 2008 ($\hV=8.8\ee{-33}$, a bound reconstructed from a mass-sensitivity figure,
not an Allan measurement) and Chaste 2012 ($\hV=1.5\ee{-36}$, not a resolved plateau). None of the three enters any fit or median.

\paragraph{Silicon MEMS reference resonators.} The MEMS range of Fig.~1 of the Letter is populated by the AlN contour-mode plates of Ref.~\cite{Kim2017}; the one silicon MEMS plate in the record, the capacitive device of row~22 of Table~\ref{tab:devices}, is excluded because it is designed to radiate into air (above). Silicon micromechanical reference resonators of the kind used in commercial MEMS oscillators and in the micromechanical-disk oscillators of the frequency-control literature are absent for a different reason: their published frequency-stability data are oscillator-level --- phase noise or Allan deviation of a complete loop with a sustaining amplifier and, in products, compensation circuitry --- and their authors attribute the close-to-carrier $1/f^{3}$ phase noise to the electronics. That is precisely the $b_{-1}/(4Q^{2})$ contamination discussed in Sec.~S7, so these values bound the resonator's own flicker from above and do not resolve it; no open-loop measurement of the intrinsic plateau of such a resonator is known to me. For a vacuum-encapsulated silicon resonator of $V\approx10^{-14}$~m$^{3}$ the invariant predicts a plateau between $\sA\approx1\ee{-10}$ (at the quartz/crystalline reference, $1.2\ee{-34}$~m$^{3}$) and $4\ee{-9}$ (at the room-temperature NEMS reference, $1.0\ee{-31}$~m$^{3}$); an open-loop measurement on one such device is the single most valuable addition the record could receive.

\paragraph{Macroscopic resonators.} Nothing macroscopic enters the compilation, and the reason is sensitivity, not physics. Four classes of high-$Q$ macroscopic resonator maintain long-duration frequency records --- the torsion pendulums of gravitational experiments, the torsional oscillators of helium physics, resonant-bar gravitational-wave detectors, and precision vibratory gyroscopes --- and none has a published, resolved $1/f$ frequency floor: the records are white-noise-limited at short $\tau$ and drift-limited beyond, with no flicker plateau exhibited between. The strongest number found is the cryogenic $G$ pendulum of Ref.~\cite{Newman2014G}: a CuBe fiber $20~\mu$m in diameter and $0.30$~m long at $2.5$~K ($Q$ up to $1.8\ee{5}$), whose period record has Allan deviation $2.2\ee{-9}$ at one cycle ($\tau=135$~s), reported as consistent with white noise. Treated as an upper limit, with $\eta=3/4$ exact for a circular fiber in torsion ($\Veps=7.1\ee{-11}$~m$^3$), it gives $\hV\lesssim2.5\ee{-28}$~m$^3$ --- $5.3$ decades above the cryogenic reference of $1.26\ee{-33}$~m$^3$. The double-mode helium torsional oscillator of Ref.~\cite{Eyal2016} (empty-cell frequency stability $9.6\ee{-9}$ on its high mode, averaging time unstated; published BeCu cell rod $3.7$~mm in diameter and $1.05$~cm long, $\Veps=8.5\ee{-8}$~m$^3$) gives $\hV\lesssim5.6\ee{-24}$, $9.7$ decades above; for a bar at ALLEGRO scale ($2296$~kg of aluminum, $\eta=2/3$ extensional, $\Veps=0.58$~m$^3$), even a record at $1\ee{-8}$ per point --- and none better is published --- would bound $\hV$ only at $4.2\ee{-17}$, $16.5$ decades up. The invariant itself says why the bounds are vacuous: at the cryogenic $\hmo\Veps$ the predicted floors are $\sA\approx5.0\ee{-12}$ (fiber), $1.4\ee{-13}$ (torsional oscillator), and $5.5\ee{-17}$ (bar), lying $2.7$--$8.3$ decades below each instrument\textquotesingle s demonstrated noise (\texttt{scripts/macroscopic\_bounds.py} in the analysis-code deposit). What the macroscopic literature has measured is the loss face of the same slow relaxations: the time-of-swing anelasticity bias in $G$ determinations~\cite{Kuroda1995}, later measured directly as anelastic dispersion of a tungsten fiber~\cite{Yang2009}; the fluctuation face of that same distribution of correlation times --- the flicker floor --- has not been measured on any of these instruments. The pendulum case is the sharp one: an instrument of macroscopic mass whose \emph{strained} volume is $7\ee{-11}$~m$^3$, inside the fitted $13.35$-decade span and $2.8$ decades below its quartz top, so the predicted floor is an interpolation of the invariant, not an extrapolation. A $10^{-12}$-grade frequency readout on an existing cryogenic pendulum would either resolve the floor or confront the invariant where it is already calibrated; the torsional-oscillator and bar cases would instead extend the abscissa by $0.3$ and $7.2$ decades.

\section{Thermomechanical reduction}

The thermomechanical points of Fig.~1 are computed per device, each at its own temperature. The displacement noise spectral density on resonance without any drive is,
\begin{equation}
S_x(f_0)=\frac{4k_BTQ}{k_{\rm eff}\,\omega_0}.
\end{equation}

To convert these equilibrium displacement fluctuations into a thermomechanical contribution to the fractional-frequency Allan deviation, $\sA$, a coherent oscillation amplitude must be specified. This contribution depends on the ratio of the equilibrium displacement noise to the coherent amplitude, whether the resonator is externally driven or its oscillation is sustained by a feedback loop. To define the best attainable limit within the linear-response regime, I take this reference amplitude to be the Duffing critical amplitude $z_c$, at which $Q$ cancels from the product $Qz_c^{2}$, giving
$\sigma_{y,\mathrm{th}}=\sqrt{k_BT/(m_{\rm eff}\omega_0^{3}\,Qz_c^{2}\,\tau)}$. Driving the resonator beyond this level --- into the regime of nonlinear response --- will increase the phase noise, due to onset of amplitude-to-phase-noise conversion. The reduction is also independent of the damping model. The fluctuation--dissipation theorem is invoked on resonance, where the loss angle satisfies $\phi(\omega_0)\equiv Q^{-1}$ whatever the loss mechanism; off resonance, a frequency measurement at averaging time $\tau$ samples the thermal force only within a fractional bandwidth ${\sim}1/(2\tau f_0)$ of the carrier --- at most $2.4\ee{-6}$ for any device in this record (\texttt{scripts/structural\_damping\_check.py}), with the leading tilt canceling between sidebands. Replacing the white (velocity-damping) force spectrum by the $1/f$ force spectrum of structural damping~\cite{Saulson1990}, observed directly in high-stress Si$_3$N$_4$~\cite{Fedorov2018}, therefore recolors the computed branch at the parts-per-million level and leaves it $\tau^{-1/2}$: a $\tau$-independent plateau cannot arise from the additive (thermomechanical-force) branch under either damping model. The clamped-force noise of a structurally damped resonator is a different, parametric branch: it is the eigenstrain term of Eq.~(3) of the Letter, flat in $\tau$, and is treated in Sec.~\ref{sec:eigenchannel}.

\vspace{0.3cm}
\emph{The $z_c$ hierarchy.} $Qz_c^{2}$ is evaluated per device by the first applicable rule:
\begin{enumerate}
\item \textbf{Published $\beta$.} Where the source reports the nonlinear
coefficient directly, $Qz_c^{2}=\sqrt3 f_0/(2\beta)$.
\item \textbf{Published $K_3$.} Where a cubic stiffness is reported,
$Qz_c^{2}=4\sqrt3/(3K_3)$.
\item \textbf{String under built-in stress.}
$\beta=f_0E\pi^{2}/(8\sigma_0L^{2})$, then rule 1.
\item \textbf{Stress-free beam.} $z_c=h\sqrt{8/(3\sqrt3 A_gQ)}$, with $h$ the dimension along the direction of vibration.
\end{enumerate}
Rule 4 is the fallback and it is the one to distrust: for resonators under built-in tension it fails by more than an order of magnitude. Two checks against published values: Maillet's 250~$\mu$m beam gives $z_c=12.4$~nm from its own $\beta$ against $0.25$~nm from rule 4, and Fong's string gives $28.9$~nm from its published $K_3$ against $0.77$~nm. Rules 1--3 are used wherever the inputs exist.

\paragraph{Averaging time, and the invariant needed to reach the limit.} Each thermomechanical point of Fig.~1 of the Letter is evaluated at the device's own measurement time $\tau$ --- the $\tau$ of its plateau measurement, as recorded in \texttt{data/thermomech\_computed.csv} --- with $\tau=1$~s assumed where none was stated ($\tau_{\rm stated}=0$ in that file). Because $\sA^{\rm th}\propto\tau^{-1/2}$ while the plateau is flat, the invariant a device would need in order to reach its thermomechanical limit depends on $\tau$: equating $\sqrt{2\ln2\,\hmo}$ to $\sA^{\rm th}(\tau)$ gives $\hmo^{\ast}(\tau)=k_BT/(2\ln2\,m_{\rm eff}\omega_0^{3}Qz_c^{2}\tau)$ and a target $\hmo^{\ast}\Veps$. At $\tau=1$~s, for representative fitted devices: Fong's SiN string ($296$~K) $6\ee{-35}$~m$^{3}$; Gavartin's SiN string ($300$~K) $8\ee{-36}$; Villanueva's AlN/Mo beam ($295$~K) $2\ee{-36}$; Mile's Si cantilever ($295$~K) $1\ee{-36}$; Bartsch's Si beam ($300$~K) $1\ee{-37}$; Maillet's $250~\mu$m beam ($4.2$~K) $2\ee{-37}$ --- three to six decades below the room-temperature reference $1.0\ee{-31}$~m$^{3}$, and lower still at longer $\tau$. The target is not a universal number: it scales as $\eta\,k_BT/(\rho\,\omega_0^{3}Qz_c^{2}\tau)$ and is reported per device for that reason.

The exponent of these computed points is not a volume law---$\sigma_{y,\mathrm{th}}\propto(Vf_0^{3})^{-1/2}$, and $f_0$ is
anticorrelated with $\Veps$ at $-0.87$ ($n=24$) across real devices---so they are drawn as a population rather than fitted.

The per-device inputs are \texttt{data/thermomech\_inputs.csv} in the
repository, which records for each device its $m_{\rm eff}$, $f_0$, $Q$,
transverse dimension, and which of the four rules was applied.

\paragraph{Which limits are drawn.} Table~\ref{tab:thermomech} lists every device of Table~S1 with its computed limit where one exists. A limit is drawn in Fig.~1 of the Letter only beside a \emph{resolved} floor --- a filled marker: a resolved plateau (class P or C) measured in vacuum, so that the floor is set by the device's internal dissipation --- and a fine dotted line joins the two. Beside an upper bound a limit tests nothing, since the floor may lie anywhere below the bound; for an ambient-$Q$ device the comparison is with gas damping rather than with the device's internal dissipation; and for a device excluded from the compilation (Sec.~S4) it is not evidence about the record. Twelve of the $28$ computed limits are therefore drawn; the rest are tabulated here with the reason. The submitted version drew all $26$ limits of devices in the compilation, $14$ of them beside no resolved floor; the computed values are unchanged. No limit is computed for the AlN contour-mode and quartz/SAW/BAW devices of Tables S2 and S3.

\vspace{0.3cm}
{\footnotesize\setlength{\tabcolsep}{2.5pt}%
\setlength{\LTcapwidth}{\linewidth}%
\begin{longtable}{r l r l r l l c r l}
\caption{Computed thermomechanical limits for the $41$ devices of Table~S1. $\sigma_{y,\mathrm{th}}$ is evaluated by the rules of this section at the device's temperature $T$ and its own averaging time $\tau$ ($^{\ast}$: $\tau$ not stated, $1$~s assumed); $\sA$ and Class are those of Table~S1; Gap $=\log_{10}(\sA/\sigma_{y,\mathrm{th}})$, which for an upper bound (U) or an unresolved plateau (P?) bounds the true gap from above. The last column states whether the limit is drawn in Fig.~1 of the Letter, or why it is not drawn (bound; unresolved plateau; ambient-$Q$ device, which the internal-dissipation rule (2) excludes; device excluded from the compilation) or not computed (no $\sA$; $f_0$ not stated by the source; geometry not published; a torsional or tuning-fork mode, to which none of the four $z_c$ rules applies; row 41, admitted after the limits were compiled, an upper bound that would not be drawn). Row 1: $f_0$ is ``nearly 2~GHz'' in the source, so its limit is indicative only.
\vspace{0.3cm}
\label{tab:thermomech}}\\
\hline\hline
ID & Source & $T$ (K) & $\tau$ (s) & $f_0$ (MHz) & $\sigma_{y,\mathrm{th}}$ & $\sA$ & Class & Gap (dec) & Fig.~1 limit \\
\hline
\endfirsthead
\hline\hline
ID & Source & $T$ (K) & $\tau$ (s) & $f_0$ (MHz) & $\sigma_{y,\mathrm{th}}$ & $\sA$ & Class & Gap (dec) & Fig.~1 limit \\
\hline
\endhead
\hline
\endfoot
\hline\hline
\endlastfoot
1 & Chaste2012-after & 5.5 & 2 & $\sim$2000 & 8.3e-8 & 2.8e-6 & P? & +1.53 & unresolved \\
2 & Chaste2012-before & 6 & -- & -- & -- & 3.0e-3 & X & -- & no $\sA$ \\
3 & Jensen2008 & 295 & 0.01 & 328.5 & 7.6e-5 & 1.3e-4 & U & +0.23 & bound \\
4 & Chiu2008 & 6 & 1 & -- & -- & 5.3e-5 & X & -- & no $\sA$ \\
5 & Chen2009 & 5 & 0.3 & -- & -- & 3.6e-7 & X & -- & no $\sA$ \\
6 & Bartsch2012 & 300 & 1 & 103 & 9.4e-9 & 8.92e-6 & P & +2.98 & drawn \\
7 & Sansa2014 & 295 & -- & -- & -- & -- & X & -- & no $\sA$ \\
8 & Hanay2012-SiN & 70 & 10 & 44.6 & 7.4e-10 & 3.0e-6 & P? & +3.60 & unresolved \\
9 & Naik2009 & 40 & 5 & 428 & 4.3e-11 & 2.0e-7 & P & +3.67 & drawn \\
10 & Ramos2012 & 295 & 1 & -- & -- & 3.0e-6 & U & -- & $f_0$ not stated \\
11 & Feng2008 & 22 & 1 & 428 & 8.7e-11 & 2.5e-8 & U & +2.46 & bound \\
12 & Yang2006 & 4.2 & 1 & 133 & 2.6e-10 & 5.0e-8 & U & +2.28 & bound \\
13 & Mile2010 & 295 & 1 & 20 & 5.0e-9 & 6.0e-7 & P & +2.08 & drawn \\
14 & Hanay2012-cSi & 80 & 0.5 & -- & -- & 8.0e-8 & P? & -- & geometry not published \\
15 & Gray2012 & 300 & 1 & 27.8 & 1.4e-9 & 1.44e-5 & U & +4.00 & bound \\
16 & Verd2008 & 295 & 1 & 6.32 & 2.9e-9 & 2.71e-8 & C & +0.97 & ambient $Q$ \\
17 & Villanueva2011 & 295 & 1$^{\ast}$ & 14.3 & 3.3e-9 & 3.1e-6 & C & +2.97 & drawn \\
18 & Gavartin2013 & 300 & 0.1 & 2.84 & 4.3e-9 & 1.5e-7 & P & +1.55 & drawn \\
19 & Verd2005 & 295 & -- & -- & -- & -- & X & -- & no $\sA$ \\
20 & Fong2012 & 296 & 1$^{\ast}$ & 0.64 & 1.8e-9 & 5.8e-8 & P & +1.52 & drawn \\
21 & Olcum2014 & 295 & 0.1 & 2.87 & 3.4e-9 & 6.0e-9 & P & +0.24 & excluded (Sec.~S4) \\
22 & Park2011 & 295 & 0.05 & 18.2 & 6.7e-10 & 2.74e-8 & P & +1.61 & ambient $Q$ \\
23 & Larsen2011 & 295 & -- & -- & -- & -- & X & -- & no $\sA$ \\
24 & Ivaldi2011 & 295 & 10 & 0.088 & 1.1e-8 & 1.0e-8 & U & -0.02 & ambient $Q$ \\
25 & Burg2007 & 293 & 1 & 0.2096 & 4.7e-10 & 1.0e-8 & U & +1.32 & excluded (Sec.~S4) \\
26 & Fong2012-78K & 78 & 1$^{\ast}$ & 0.64 & 9.1e-10 & 2.10e-8 & P & +1.36 & drawn \\
27 & Fong2012-5K & 5 & 1$^{\ast}$ & 0.64 & 2.3e-10 & 6.94e-9 & P & +1.48 & drawn \\
28 & Feng-thesis-295 & 22 & 1 & 295 & 1.7e-10 & 4.7e-8 & U & +2.44 & bound \\
29 & Feng-thesis-420 & 22 & 1 & 420 & 9.6e-11 & 3.1e-7 & U & +3.51 & bound \\
30 & Feng-thesis-411 & 22 & 1 & 411 & 1.5e-10 & 6.6e-8 & U & +2.63 & bound \\
31 & Feng-thesis-482 & 22 & 1 & 482 & 1.2e-10 & 2.1e-8 & U & +2.24 & bound \\
32 & Feng2007-SiNW & 25 & 7 & 199.7 & 9.9e-11 & 7.6e-8 & U & +2.88 & bound \\
33 & Maillet2018 & 4.2 & 10 & 2.68 & 5.5e-11 & 1.662e-08 & P & +2.48 & drawn \\
34 & Maillet2018 & 4.2 & 10 & 4.45 & 4.3e-11 & 9.010e-09 & P & +2.33 & drawn \\
35 & Maillet2018 & 4.2 & 10 & 17.5 & 6.3e-11 & 3.691e-08 & P & +2.77 & drawn \\
36 & Maillet2018 & 4.2 & 10 & 0.905 & 1.3e-10 & 1.378e-07 & P & +3.04 & drawn \\
37 & Wang2020 & 300 & 3 & -- & -- & 6.06e-7 & P & -- & no $z_c$ rule (fork) \\
38 & Wang2020 & 300 & 3 & -- & -- & 3.60e-7 & P & -- & no $z_c$ rule (fork) \\
39 & Duraffourg2018 & 300 & 0.1 & -- & -- & 1.5e-7 & P & -- & no $z_c$ rule (torsion) \\
40 & Zhang2013 & 295 & 30 & -- & -- & 1.6e-7 & P & -- & no $z_c$ rule (torsion) \\
41 & Feng2007-SiNW-M188 & 25 & 7 & -- & -- & 6.0e-8 & U & -- & not compiled (bound) \\
\end{longtable}}
\vspace{-1.0cm}
\section{Per-family structure and leverage}
\paragraph{Leverage.} The three family fits do not carry equal weight. The cryogenic one should also be read with its structure in view: its seven points are four devices at four distinct volumes spanning $2.97$ decades, three of the points being three modes of one $250~\mu$m beam~\cite{Maillet2018} and two being one SiN string measured at 5 and 78~K~\cite{Fong2012}. Three of the seventeen quartz points likewise sit at a single volume. The global fit weights each point by its own stated uncertainty; the invariant is observed across $13.35$ decades
of $V_\varepsilon$, while the exponent is identified by within-family
variation, an effective lever arm of $3.4$ decades. This bears on the
per-family exponents quoted in Fig.~1 of the Letter rather than on
Eq.~(5) of the Letter.

The three families and their fitted exponents, as carried in the key of
Fig.~1 of the Letter: room temperature $-0.48\pm0.11$ over the seven resolved plateaus ($-0.61\pm0.03$ with the nine AlN plates); quartz/SAW/BAW $-0.58\pm0.03$ over $17$; cryogenic $-0.31\pm0.14$ over $7$. The global fit uses a common exponent with per-family prefactors over all $42$, weighting each point by its own stated systematic uncertainty in both $\sA$ and $\Veps$, with the uncertainty on the exponent taken from the study-cluster bootstrap of Sec.~\ref{sec:errors}: $\sigma(p)=0.063$, percentile 95\%
interval $[-0.66,-0.40]$. The single authority for the global numbers is
\texttt{scripts/fit\_clustered.py}; the per-family lines of Fig.~1 are
ordinary least-squares fits emitted by \texttt{scripts/make\_prl\_figs.py}.

\section{Nine parameters that do not move the $\mathbf{1/\textit{f}}\,$ plateau}
The most systematic single-device study in the compilation is from the Ph.D. thesis of Eric Sage~\cite{Sage2013}, who studied piezoresistively downmixed doubly-clamped single-crystal silicon beams
($L=10\,\mu$m, $w=300$~nm, $t=160$~nm, boron-doped $9\ee{18}$--$5\ee{19}$~cm$^{-3}$, vapor-HF released). In his experiments, nine parameters were varied, one at a time, and their effect on the $1/f$ plateau recorded. The claim of this Letter is that $Q$ is the wrong knob; the table below conveys a stronger empirical statement, revealing eight other parameters that similarly do not move the $1/f$ plateau. Rows in bold are the two nulls on which the Letter's argument most depends (discussed below) and the one parameter that actually did move the plateau.

\newcommand{\sga}[1]{\parbox[t]{0.24\textwidth}{\raggedright #1\strut}}
\newcommand{\sgb}[1]{\parbox[t]{0.26\textwidth}{\raggedright #1\strut}}
\newcommand{\sgc}[1]{\parbox[t]{0.42\textwidth}{\raggedright #1\strut}}
\begin{center}
\small
\renewcommand{\arraystretch}{1.2}
\begin{tabular}{@{}l l l@{}}
\hline\hline
parameter & range & \textbf{effect on the plateau} \\
\hline
\sga{gauge bias $V_b$} & \sgb{to 3.4~V$_{\rm pp}$} & \sgc{\textbf{none}; white branch $\propto1/V_b$} \\
\sga{\textbf{drive amplitude}} & \sgb{\textbf{to just below Duffing onset}} & \sgc{\textbf{none}; the plateau is reached \emph{before} the critical amplitude} \\
\sga{actuation scheme} & \sgb{$1f$ vs $2f$; electrostatic vs piezo shaker} & \sgc{\textbf{none}} \\
\sga{\textbf{detection}} & \sgb{\textbf{piezoresistive vs optical}} & \sgc{\textbf{none}} \\
\sga{quality factor} & \sgb{$Q=400\to2560$ at constant displacement} & \sgc{\textbf{none} above 40~ms; a Leeson term would have moved $41\times$} \\
\sga{mode order} & \sgb{modes 1/2/3 at 25.2/66.5/123.5~MHz} & \sgc{\textbf{none}} \\
\sga{geometry} & \sgb{cantilever vs doubly clamped; several lengths} & \sgc{\textbf{none} (the author flags this as under-powered)} \\
\sga{oxide coating} & \sgb{0/20/30/40~nm} & \sgc{\textbf{none}} \\
\sga{boron doping} & \sgb{three levels within $10^{19}$} & \sgc{\textbf{none}} \\
\hline
\sga{\textbf{temperature}} & \sgb{\textbf{300~K $\to\sim$100~K}} & \sgc{$\bm{\times2\text{--}2.5}$ \textbf{improvement}} \\
\hline\hline
\end{tabular}
\end{center}

\noindent Quoting a summary from Sage~\cite{Sage2013}: \emph{``Temperature is the only parameter we found could change the frequency stability.''}
\vspace{0.5em}

For the argument of this Letter, two rows carry more weight than the others. The \textbf{drive-amplitude} row is a sweep at fixed $Q$ to just below the Duffing onset with no movement of the plateau, which excludes the whole multiplicative class for this device --- amplitude-to-frequency conversion, nonlinear damping and drive-power heating alike --- since every member of that class carries a transfer coefficient that grows with the operating point. The \textbf{detection} row is a complete change of transduction domain, piezoresistive to optical, with no movement of the plateau: the noise cannot originate anywhere downstream of the resonator, because two physically unrelated downstream paths give the same floor.

\paragraph{Loop topology census.} The effect of readout instrumentation topology is a legitimate concern --- a feedback loop can filter or add noise --- so closed-loop measurements are audited here row by row against their primary references. Of the $42$ fitted values, $32$ derive from open-loop measurements --- $17$ stated outright, and $15$ measured open-loop and converted by the source to an equivalent closed-loop phase noise for reporting, the standard convention in both the NEMS and frequency-control literatures. Five entries report closed-loop measurements --- rows 6, 9, 17 and 39 of Table~S1 (Bartsch 2012, Naik 2009, Villanueva 2011, Duraffourg 2018), and the oscillator-level entry qB2 of Table~S3, whose source separates the resonator's share of the flicker by Leeson decomposition~\cite{Rubiola2007}; and five state no topology --- qB3 and the four $450$-MHz SAW entries P1--P4 of Ref.~\cite{Parker1993}, for which the companion program of Ref.~\cite{Parker1994} reports the size dependence in both resonator and oscillator measurements. The four closed-loop NEMS rows span $4.1$~decades of $\hV$ across four materials and two temperature regimes, so no topology covariate can be assembled from them, and none is attempted; the per-row assignments are the Scheme columns of Tables~S1--S3. The direct evidence is measurement, not census. Sage measured the same resonator both ways at the same actuation and polarization, ``in both open and closed loop, which is equivalent'' (Fig.~II.7 of Ref.~\cite{Sage2013}); and Parker \& Andres, for the SAW size series of Table~S1: ``The size dependence is present in both the resonator and oscillator measurements, therefore reducing the possibility that it is a measurement artifact''~\cite{Parker1994}. One census entry carries its own caveat: the two Hanay 2012 rows --- already outside the fit as unresolved --- are read \emph{at} the loop response time, $\tau_R\approx500$~ms, the boundary Sage identifies (``an ADEV curve acquired in closed loop is meaningless below the PLL response time''), and they are retained as tabulated entries only. In practice, if the roll-off characteristics are known (\textit{e.g.}, as a single-pole decay) $\sA$ can be corrected beyond the roll-off knee for some distance --- until the corrections yield too much uncertainty in the result.

\section{Why a single-channel nulling scheme cannot remove the $\mathbf{1/\textit{f}}$ floor}
\label{sec:schemes}

A parametric fluctuation enters the equation of motion in the same manner as the signal. The consequence is much more than a simple inconvenience, and this has been proved in one important special case and measured in another.

\paragraph{A no-go theorem for frequency servos.} A frequency servo is a feedback loop that measures the oscillator's phase or frequency and applies a correction intended to hold it steady. Reference~\cite{Dankowicz2026} analyzes the simplest case: a self-sustaining Duffing oscillator (a nonlinear resonator kept oscillating by a sustaining amplifier) to which a secondary, adaptive feedback loop is added, with a phase-coupling gain $k$ that sets how strongly the loop acts on the phase. The quantity analyzed is the phase-diffusion constant $D_\phi$ --- the rate at which the oscillator's phase random-walks, which is the frequency noise expressed per unit time --- and it is found to be the sum of two terms. The first is amplitude-to-frequency conversion: amplitude noise shifts the frequency through the Duffing (amplitude-dependent) spring, and the loop can null it by operating at the point where the amplitude-to-frequency slope vanishes. The second is noise that acts directly on the phase, which includes every fluctuation of the resonance frequency itself --- the parametric $1/f$ fluctuations of this Letter among them. The loop multiplies this second term by a factor that depends on $k$.

The essential point is that an added mass enters the oscillator in exactly the same way as the second term: it shifts the resonance frequency. The loop therefore cannot distinguish the mass signal from the frequency noise, and it multiplies the realized mass responsivity, $\mathcal{R}\equiv d\omega/dm$, the frequency shift the output shows per unit added mass, by the \emph{same} $k$-dependent factor. The mass resolution is the frequency noise divided by the responsivity,
\begin{equation}
\dm \;\propto\; \frac{\sqrt{D_\phi}}{\mathcal{R}}
   \;=\; \sqrt{\frac{[\text{a--f term}]}{\mathcal{R}^{2}}+I_\phi}\,,
\end{equation}
where $I_\phi$ is the intensity of the direct phase noise and the $k$-dependence has cancelled between numerator and denominator: the resolution is \emph{invariant under the feedback gain}. The gain $k_{\rm opt}$ that drives $D_\phi\to0$ drives $\mathcal{R}\to0$ in exact proportion, \textit{for no net benefit}; and if the amplitude-to-frequency term is imperfectly nulled, the result is a net \emph{loss}, because that term is not reduced by the loop while $\mathcal{R}$ is.

The operational rule follows. The channel that carries the signal is the resonance frequency itself. A loop that acts on it with gain at the low Fourier frequencies where a mass signal appears will suppress signal and noise \textit{together}, and therefore \textit{has no place in a sensor system}. Loops that act solely on the amplitude-to-frequency transfer leave the signal channel untouched and are permissible, but their payoff is limited to the one term they can null. Since a servo lowers the Allan deviation and the responsivity together, a lower $\sA(\tau)$ is by itself no evidence of a better sensor: \emph{the only admissible figure of merit is $\sA(\tau)/\mathcal{R}(\tau)$, with $\mathcal{R}$ measured rather than assumed.} The same reference shows that $1{:}2$ and $1{:}3$ internal resonances --- mode couplings in which a second mode locks the phase of the first --- are the $k=-2$ and $k=-3$ members of this family of loops, so the large stability gains reported for internal resonance~\cite{Antonio2012,Shoshani2024} warrant re-examination on this basis: \textit{unless $\mathcal{R}$ was measured and found unchanged, the gain in $\sA$ is likely to have been paid for in suppressed responsivity.}

\paragraph{Phase-noise cancellation, and the channel it exempts.} Two schemes \textit{do} reduce oscillator phase noise while preserving $d\omega/dm$ exactly: feedback-phase tuning, in which the long correlation time of a $1/f$ source collapses its projection onto the slow flow to a line that a single scalar annihilates~\cite{KenigCross2014}; and a passive element built from two
parametrically driven coupled nonlinear resonators~\cite{Kenig2012}. Both remove $1/f$ arising in the feedback, the readout or the source. Neither touches $1/f$ acting on the resonator's own linear frequency, and this is stated by the authors: \emph{``One noise source that is not reduced by this strategy is noise acting directly on the linear resonance frequency of the resonator, such as mass or spring constant fluctuations''}~\cite{CrossKenigAllen2015}.
Reference~\cite{Villanueva2013} shows it in data --- tunable dips in phase noise against feedback phase, resting on a detuning-independent floor. That floor is the $1/f$ plateau.

\paragraph{Two-mode compensation, measured.} Reference~\cite{Sansa2016} tracked two modes of one resonator and used the second as a thermometer to correct the first. Below $\tau\approx0.1$~s the correction \emph{degrades} stability, because the two modes' white noise is uncorrelated and differencing adds variance ($1\ee{-6}$ and
$6.5\ee{-7}$ summing in quadrature to $1.2\ee{-6}$, against $1.25\ee{-6}$ measured). Above $\tau\approx10$~s it improves stability by removing shared drift. Across the decade between, where the plateau lives, ``no significant improvement was observed.'' Reference~\cite{Sage2013} supplies the reason: the measured
correlation between the first two harmonics of one beam rises from
$\rho\approx0.5$ at $10$~ms to $\rho\approx0.9$ at $2$~s, and optimal linear subtraction leaves a fraction $1-\rho^{2}$ of the variance --- a factor of two or better at the drift, some $13\%$ in $\sA$ at the plateau. That author reports the correlation measurement as provisional and calls for it to be repeated.

Considered together, the three results impose firm bounds on what any scheme operating on the measurement can achieve. A servo cannot help, because it scales signal and noise together. A cancellation scheme cannot help, because it is built to cancel phase noise entering through the feedback and, by construction, leaves untouched noise on the resonance frequency itself --- the channel through which both the parametric fluctuations and the mass signal enter. Common-mode rejection cannot help, because at plateau timescales there is little common mode to reject. What remains are the fluctuators themselves.

\section{The bias carried by the excluded bounds}

The bounds here occupy the small-volume end of the devices represented, so admitting them should flatten the exponent, and it does, from
$-0.48\pm0.11$ to $-0.42\pm0.08$. The resolved-plateau fit and the full
room-temperature population (bounds and AlN plates admitted, $-0.57\pm0.04$) are both drawn in Fig.~1 of the Letter, rendering the size of this bias
transparent.

\section{Equilibrium temperature fluctuations}

The body excludes these on the ground that their spectrum is Lorentzian. Quantitatively, the variance for equilibrium temperature fluctuations is $k_BT^2/C_p$ and corner $G/2\pi C_p$, so both the thermal conductance $G$ and the heat capacity $C_p$ enter, and below the corner the spectrum is white in $y$ --- an $h_0$ branch, falling as $\tau^{-1/2}$ in $\sA$~\cite{ClelandRoukes2002}. No choice of $G$ or $C_p$ can convert it into a $\tau$-independent plateau.

\section{A possible mechanism, and the interrogation window}

The evidence assembled above appears to converge on one operative mechanism: two-state defects --- elastic dipoles, distributed
through the bulk and/or surface. They switch by thermal activation over a broad distribution of barriers, and are coupled to the resonator vibrational mode through its strain field. Accordingly, one defect carries the fractional weight of Eq.~(4) of the Letter. The temperature dependence, the ordering of platforms by crystalline quality, the tunneling crossover below $1$~K, and the directly measured single-defect coupling have each been given above; what they share is this picture. The individually resolved
defect of Ref.~\cite{Yuksel2026} enters it through its measured strain
coupling, $|\gamma|=0.84$~eV---the quantity that sets $\delta_1$ and the
single-defect frequency shift---not through the barrier axis. The window of Eq.~(7) in the Letter selects \emph{which} fluctuators are read, while $\gamma$ sets how strongly each frequency-pulls the mode.

The interrogation window of Eq.~(7) of the Letter is narrower than it appears, and this bears on the data directly. At $300$~K an averaging range of $10$~ms--$100$~s reads barriers from $595$ to $833$~meV: a slice $238$~meV wide. The slice narrows with temperature, to $155$--$217$~meV at $78$~K and to just $4$~meV at $5$~K. Every barrier quoted here and in the Letter takes $\tau_0=10^{-12}$~s, the defect's local vibrational period. The choice is benign because every use of $\tau_0$ is logarithmic: a full decade shifts each quoted barrier by only $k_BT\ln10$---$60$~meV at $300$~K, $15$~meV at $78$~K---while the slice \emph{widths}, which carry the smoothness argument below, are exactly $\tau_0$-independent, and the disjointness of the damping and dephasing slices of the Letter is unchanged. Attempt frequencies for defect reorientation lie in the $10^{12}$--$10^{14}$~Hz band set by local vibrational modes~\cite{NowickBerry}; Ref.~\cite{ClelandRoukes2002} takes $\nu_0\sim10^{12}$~Hz, and $\tau_0=1/\nu_0$ adopts it. That is the slow edge of the band, so the quoted barriers are lower bounds over the plausible range: at $\tau_0=10^{-14}$~s they rise by $2k_BT\ln10$ --- $\approx120$~meV at $300$~K, $30$~meV at $78$~K --- hence, no conclusion changes. Two consequences follow: the flatness of an observed plateau is itself evidence that $D(\varepsilon_a)$ is smooth across the slice being ``read'', since structure in the distribution on that scale would tilt the plateau. And because cooling narrows the slice fourfold between $300$ and $78$~K, features in $D(\varepsilon_a)$ that average away at room temperature may not do so at cryogenic temperatures. This is a possible explanation for the fact that the cryogenic family is the least regular of the three, and carries the shallowest and least certain exponent, $-0.31\pm0.14$.

\section{From one defect to the ensemble}

Equation~(4) of the Letter is a proportionality relation because the constant relating $\hmo$ to $N\delta_1^2$ is not a property of the individual fluctuators but of how their correlation times are distributed. A single two-state fluctuator of correlation time $\tau_c$
contributes a Lorentzian; an ensemble of $N$ fluctuators distributed log-uniformly over $\tau_{\rm min}<\tau_c<\tau_{\rm max}$ superpose to
$S_y(f)=N\delta_1^2/(4f\ln R)$ with $R=\tau_{\rm max}/\tau_{\rm min}$, so
$\hmo=N\delta_1^2/(4\ln R)\equiv\nu\delta_1^2/4$, where $\nu$ is the number of fluctuators per unit $\ln\tau_c$. The logarithm is not unconstrained: it sets the interrogation window of Eq.~(7) of the Letter. With $\tau_0=10^{-12}$~s it takes the value 23--32 over the averaging times used here. The two limits are therefore commensurable. In the modulus term of Eq.~(3) of the Letter, $n\equiv\nu/V$ is this density per unit $\ln\tau_c$, and the factor $\tfrac14$ --- the occupancy $p(1-p)$ of a symmetric two-state fluctuator --- is absorbed in $\langle v_C^{2}\rangle$, the convention stated there. The telegraph amplitude $\delta_1\approx3\ee{-6}$ measured for a single resonant defect in a $1~\mu$m$^{3}$ lithium-niobate mode~\cite{Maksymowych2025} is a dispersive shift, $g^{2}/\Delta$, of a TLS tuned through resonance with the mode; it is not the classical jump the same defect produces when it switches far from resonance, which is the quantity that enters $\hmo$. That jump is, in the eigenstrain channel of Eq.~(3) of the Letter, $\gamma/(2\seff V)$: for $\gamma=0.84$~eV, $\seff=0.4$~GPa and $V=1~\mu$m$^{3}$, $1.7\ee{-10}$ --- eight decades below the dispersive value. The measured room-temperature floor of silicon nitride, $\hmo\approx1\ee{-13}$ at $V=1~\mu$m$^{3}$, then requires, with $\langle p(1-p)\rangle\approx0.1$--$0.25$ and $\ln R\approx16$--$20$, between $2\ee{8}$ and $7\ee{8}$ fluctuators per $\mu$m$^{3}$ in the band --- $2$--$7\ee{20}$~cm$^{-3}$, the two-level-system density of amorphous solids. The same population, through the relaxation loss it must also produce [Eq.~(S10) of Ref.~\cite{KanellopulosSchmid2026}], gives $Q_{\rm vol}\approx1\ee{4}$, within a factor of two of the volume loss $Q_{\rm vol}=2.3\ee{4}$ fitted to silicon nitride in that work. The single-defect experiments therefore anchor the $1/V$ scaling of $\delta_1$ on a measured coupling and the ensemble floor on a defect density. The distribution is not merely posited. The same log-uniform distribution of correlation times also has a dissipative face --- a frequency-independent loss angle --- and that face has been observed directly in a survey material: the intrinsic loss angle of a high-stress Si$_3$N$_4$ nanobeam is constant to within a factor of two from $50$~kHz to $50$~MHz (``structural damping'')~\cite{Fedorov2018}; the same flat spectrum, seen on the loss side.

\section{The cryogenic test of the $Q$ law}

The proposed $Q^{-n}$ laws suggest an experimental test. Galliou \textit{et al.}~\cite{Galliou2013} cooled quartz resonators and reported: \textit{``Some of these theories relate the $Q$-factor of the resonator to its noise level such that a high $Q$-factor is required to reach a low noise level $\ldots$ it had been shown that quartz losses dramatically decreased below 20 Kelvins. Thus, in such conditions the
resonator noise level would decrease as well. This was the main motivation for investigating the $Q$-factor behavior of crystals at cryogenic temperatures.''} In their experiments the $Q$ was confirmed; however, the stability was \textit{not}. The best liquid-helium oscillator of Ref.~\cite{Goryachev2013} holds $\sA\approx1.5\ee{-12}$ over $\tau\approx1$--$200$~s, against the $6\ee{-14}$ room-temperature state of the art quoted in that paper's own introduction --- twenty-five times worse, with orders of magnitude more $Q$, where
$Q^{-3.5}$ predicts several decades better.

This outcome should be read for what it is: a failure to deliver the predicted stability, not a controlled refutation of the $Q$ law. The authors themselves do not blame the law. They attribute the shortfall to the cryogenic environment --- to the absence of a frequency--temperature turnover point near 4~K, so that residual temperature fluctuations pass straight into the frequency, and to nonlinearity, which limits the drive --- and Ref.~\cite{Galliou2013} states that the resonator's own flicker at 4~K was \emph{masked} by cryocooler vibration and temperature fluctuations and should lie ``much lower than those already reported.'' The intrinsic flicker floor of quartz resonators with $Q$ above $10^{9}$ has therefore never been resolved. That is the strongest statement the record allows, and it is a strong one: the most extreme prediction of the $Q$ law --- after several decades of improvement to $Q\sim10^{9}$--$10^{10}$ --- is not merely unconfirmed, \textit{it is, to date, untested} --- some thirteen years after the $Q$ was first resolved.

\section{Why the compilation cannot separate bulk from surface}
\label{sec:bulksurface}

Equation~(4) of the Letter distinguishes a bulk population of fluctuators from a surface population through the surface-to-volume ratio $S/V$, and the compilation supplies that ratio directly. Across the $38$ fitted devices with defensible thickness, $S/V$ spans $4.32$ decades while $V$ spans $13.00$, and
\begin{equation}
\log_{10}(S/V) \;=\; -0.323\,\log_{10}V + \mathrm{const},
\qquad r=-0.967 .
\end{equation}
For bodies that are scaled copies of one another --- the same shape at every size --- every length grows as $V^{1/3}$, so $S\propto V^{2/3}$ and $S/V\propto V^{-1/3}$. The devices of the record follow that law to three percent in the exponent: across three families and thirteen decades, their aspect ratios do not change systematically with size.\footnote{Note that $S/V$ for the Parker--Andres SAW rows enters at the acoustic-wavelength depth of their stated averaging convention; Sec.~\ref{sec:provenance}.} This geometric self-similarity is a property of the published record, not a modeling assumption, and it means that $S/V$ is not an independent variable in the compilation: to within the scatter, knowing $V$ fixes $S/V$.

That has two consequences, which run in opposite directions. On the one hand, the two \emph{discrete} surface hypotheses become testable, because the measured relation $S(V)$ converts each into a definite prediction for the volume exponent $p$ of Eq.~(5) of the Letter. If the fluctuators sit on the surface ($N\propto S$) but each couples to the whole mode ($\delta_1\propto1/V$), then $\hmo\propto S/V^{2}\propto V^{-1.32}$ and $p=-0.661$; if each instead couples only to its own surface ($\delta_1\propto1/S$), then $\hmo\propto1/S\propto V^{-0.68}$ and $p=-0.339$. These predictions sit $5.1\,\sigma(p)$ apart and $2.6\,\sigma(p)$ to either side of the bulk value $p=-1/2$, where $\sigma(p)=0.063$ is the cluster standard error of the global fit; the second is excluded at $3.9\sigma$, the first is not, lying $1.4\sigma$ from the fit. On the other hand, an \emph{arbitrary} mixture of bulk and surface fluctuators is not separable: because $\log(S/V)$ and $\log V$ are $97\%$ collinear, a fit that includes both as regressors cannot apportion the floor between them, and a suitable mixture can imitate any exponent between the two limits. Both hypotheses therefore retain standing --- Ref.~\cite{YongVig1989} places the fluctuators on the surface and Ref.~\cite{ClelandRoukes2002} in the bulk --- and the discriminating experiment remains a deliberate cross-section series at fixed material and process, in which $S/V$ is varied at fixed $V$ and the self-similarity of the record is broken on purpose.

\section{The bandwidth cost of extreme dilution}

Dissipation dilution lowers the additive branch of $\sA(\tau)$ --- the detector-limited $\tau^{-1/2}$ part --- in proportion to $1/\sqrt{Q}$ at fixed drive, and that is its benefit: the crossover to the $1/f$ floor moves to shorter $\tau$. It does not lower the floor itself, which is set by $\hV$ and, as the record shows, is independent of $Q$. Beyond a certain dilution the trade therefore inverts, because the same $Q$ that lowers the additive branch lengthens the resonator's response time, $\tau_r=Q/\pi f_0$ (Sec.~\ref{sec:hidden}): a driven or phase-locked resonator cannot report a change of its frequency faster than its transduced phase relaxes, so $\tau_r$ is the shortest averaging time at which the device answers a measurand at all, and a faithful reading of its floor requires $\tau\gtrsim10\,\tau_r$.

The perimeter mode of Ref.~\cite{Bereyhi2022} makes the cost concrete. At $Q=1.5\ee{9}$ and $f_0\approx130$~kHz, $\tau_r\approx3600$~s: one hour. Every gate time in that measurement, $10^{-1}$--$10^{0}$~s, therefore lies three to four decades \emph{inside} the response time, in the regime where a $1/f$ spectrum is read as a $\tau^{+1}$ ramp (Sec.~\ref{sec:hidden}); that is why the device enters Fig.~1 of the Letter as an upper bound and not as a floor. Reading the floor would require averaging for the better part of a day while holding the drift below the level of Eq.~\eqref{eq:Qmax}.

For a sensor the consequence is a design limit rather than a measurement inconvenience. A NEMS frequency-shift mass spectrometer must register the arrival of a single particle within milliseconds to seconds, so its resonator must have $\tau_r$ no longer than that: $Q\lesssim\pi f_0\tau_{\rm resp}$, which is $Q\sim3\times10^{3}$ for a $1$-MHz device at $1$~ms and $3\times10^{6}$ at $1$~s. Past that point, further dilution buys nothing --- the floor it would need to lower is fixed by $\hV$ --- and costs bandwidth that no readout can ever recover, since the delay sits in the resonator, not in the electronics. Extreme dilution thus yields a floor that cannot be lowered, at the loss of bandwidth that cannot be recovered.

\section{Participation volumes: definition and protocol}
\label{sec:eta}
\vspace{-0.5cm}
\paragraph{Definition.} A defect that perturbs the local modulus by $\delta E$ over a volume $v_d$ shifts the resonance in proportion to the local strain energy density,
$\delta_1(\mathbf r)=v_C\,\varepsilon^{2}(\mathbf r)/\!\int\!\varepsilon^{2}dV$.
Averaging over positions drawn uniformly through the device,
\begin{equation}
\langle\delta_1^{2}\rangle
 \;=\; \frac{1}{V}\int\!\left[\frac{v_C\,\varepsilon^{2}}{\int\varepsilon^{2}dV}\right]^{2}\!dV
 \;=\; \frac{v_C^{2}}{V\,\Veps},
\qquad
\Veps \;\equiv\; \frac{\left(\int\varepsilon^{2}dV\right)^{2}}{\int\varepsilon^{4}dV} .
\label{eq:veps}
\end{equation}
$\Veps$ is therefore the \emph{inverse participation ratio} of the strain-energy density: the volume a uniformly strained device would need in order to give the
same mean-square per-defect coupling. By the Cauchy--Schwarz inequality
$\left(\int\varepsilon^{2}\right)^{2}\le V\!\int\varepsilon^{4}$, so
\begin{equation}
\eta \;\equiv\; \Veps/V \;\le\; 1 ,
\end{equation}
with equality if and only if $\varepsilon^{2}$ is uniform. $\Veps$ is not the modal volume $V^{(n)}_{\rm eff}=m^{(n)}_{\rm eff}/\rho$, which is weighted by displacement rather than by strain, and it is not the geometric volume $V$.

\paragraph{Protocol.} For every device: (i) assign a mode class; (ii) evaluate $\eta$ for that class from the mode shape; (iii) set $\Veps=\eta V$, with $V$ the total resonator mass divided by density. Because $\eta$ depends only on the mode shape and not on any dimension, it is a constant within a geometry class, so the substitution moves an intercept and not a slope so long as a family contains one class.

For a strain field that separates, $\varepsilon(\mathbf r)=f(x)\,g(\zeta)$, the ratio factorizes,
\begin{equation}
\eta \;=\; \frac{\left(\int f^{2}dx\right)^{2}}{L\int f^{4}dx}
       \cdot \frac{\left(\int g^{2}d\zeta\right)^{2}}{Z\int g^{4}d\zeta}
   \;\equiv\; \eta_{\parallel}\,\eta_{\perp},
\end{equation}
which is what makes the table below checkable by hand. Three factors recur: a bending strain $\varepsilon\propto\zeta$ across a thickness gives $\eta_{\perp}=5/9$; a half-wave $\cos$ profile gives $2/3$; a uniform profile gives $1$.

\begin{table}[h]
\caption{Strain participation ratio by mode class. The first three are flexural
and lose a factor $5/9$ to the neutral axis; the extensional and shear plate modes do not. Entries marked exact are closed-form; the others are Euler--%
Bernoulli eigenfunctions evaluated numerically.\label{tab:eta}}
\begin{tabular}{l c c c l}
\hline\hline
mode class & $\eta_{\parallel}$ & $\eta_{\perp}$ & $\eta$ & devices \\
\hline
doubly-clamped beam, flexural      & $0.540$ & $5/9$ & $0.300$ & NEMS/MEMS beams \\
cantilever, flexural               & $0.426$ & $5/9$ & $0.237$ & cantilevers, SMRs \\
tensioned string, bending term     & $\simeq2\lambda$ & $5/9$ & $\simeq5/(9D_Q)$ \emph{(clamp layers)} & --- \\
tension-dominated string           & $1$     & $1$   & $[1+(\eta_b^{-1}-1)/D_Q^{2}]^{-1}\simeq[1+9/(5D_Q)]^{-1}$ & SiN strings, all diluted \\
width/radial-extensional plate     & $2/3$   & $1$   & $2/3$ \emph{(exact)} & AlN contour-mode \\
thickness-shear plate              & $1$     & $2/3$ & $2/3$ \emph{(exact)} & BAW, BVA \\
circular membrane, fundamental     & ---     & ---   & $0.910$ & (none fitted) \\
\hline\hline
\end{tabular}
\end{table}

\paragraph{Strain weighting of a stressed string.} For a \emph{uniform} string held at fixed length the tension is a global variable: a local softening $\delta E$ over a length $dx$ changes it by $\delta T/T=-(\delta E/E)(dx/L)$, \emph{independent of where the segment sits}---force balance makes the tension strictly uniform, which is what makes the weighting so. The weighting of the dominant coupling channel is
therefore uniform in $x$, and the static tensile strain is uniform through the thickness, so $\eta\to1$ for a tension-dominated mode. In general the weight of a local $\delta E$ is the fraction of the modal stiffness it controls, $w=1-(1-\hat w)/D_Q$ with $\hat w=\varepsilon_b^{2}/\langle\varepsilon_b^{2}\rangle$ the normalized bending strain energy and $D_Q=\seff/\sigma_c$ the dilution factor, so that
$\eta=1/\langle w^{2}\rangle=[1+(\eta_b^{-1}-1)/D_Q^{2}]^{-1}$ (derived in Sec.~\ref{sec:eigenchannel}), with $\eta_b$ the participation ratio of the bending strain energy of the same mode; for the strings of this record, $D_Q=10$--$180$, whose bending strain sits in the clamp boundary layers, this gives $\eta=0.87$--$0.99$, and for $D_Q=1$ it returns the beam value $\eta_b=0.301$. The same bookkeeping covers every
dissipation-diluted device in the compilation, which are tension-dominated by construction and of uniform cross-section. In a strain-engineered taper, where the cross-section varies by design and the local stress with it~\cite{Ghadimi2018}, the compliance weighting of a local $\delta E$ follows the local static strain, and $\Veps$ becomes the static-strain-weighted volume rather than $V$; no fitted or drawn device here is tapered. The partition of the mode energy into tensile and bending parts used throughout is the same one that defines the dilution factor itself, as a participation ratio of the lossy
potential~\cite{Ghadimi2018}. A sinusoidal mode shape would give $10/27$ for the bending term of a string; that value describes only the interior part of the bending energy, whereas the clamp boundary layers, which hold most of it, give the far smaller $\eta_b$ of the table (Sec.~\ref{sec:eigenchannel}).

\paragraph{Participation volumes.} The participation volumes are
\emph{calculated}, not assumed---within the constraints of what each primary source states unambiguously: dimensions, mode identification, and mode class. The protocol assumes a scalar, isotropic modulus coupling; tensorial couplings (anisotropic quartz, piezoelectric stacks, oriented elastic dipoles) would reweight $\Veps$ by factors that are nearly constant within a mode class and are therefore absorbed by the per-family intercepts of the global fit. The within-class variation that is not absorbed is quantified where the record permits: the electrode-versus-acoustic volume factor of $15$ for trapped BAW modes in Sec.~\ref{sec:provenance}, and the tension-to-bending interpolation
above.

\paragraph{Validation.} The protocol reproduces the two values quoted in the Letter's End Matter, $0.300$ for a doubly-clamped beam and $0.237$ for a cantilever, to three figures. Those were obtained independently, so the agreement is a check on the machinery rather than a restatement of it.

\paragraph{Surface acoustic waves.} A SAW device has no bounded material volume, so $\eta$ is not defined against a geometric $V$ and $\Veps$ must be evaluated directly from the field. A Rayleigh wave is a superposition of two branches decaying as $e^{-qkz}$ and $e^{-skz}$, with $q=\sqrt{1-(c_R/c_L)^2}$ and $s=\sqrt{1-(c_R/c_T)^2}$. For quartz ($\nu\simeq0.17$) the Rayleigh root gives
$c_R/c_T=0.906$, hence $q=0.821$ and $s=0.424$. \textbf{The slow transverse branch dominates at depth}, and it is what sets the answer: evaluating Eq.~(\ref{eq:veps}) on the cycle- and $x$-averaged strain-energy density gives a depth participation of $0.801$ relative to a slab of thickness $\lambda$, not the $1/2\pi=0.159$ that a single $e^{-kz}$ envelope would suggest. Modeling the strain with one exponential understates $\Veps$ fivefold and should be avoided. With the lateral standing wave contributing a further $2/3$,
\begin{equation}
\Veps^{\rm SAW} \;=\; 0.534\,A\lambda ,
\end{equation}
$A$ being the aperture times the cavity length. The compilation tabulates these devices as $A\lambda/2$ --- verified against Ref.~\cite{Parker1993} Table~I at $450$~MHz on ST-cut quartz, where the tabulated volume is smaller than $A\lambda$ by $2.01$ on all four rows --- so $\Veps^{\rm SAW}=1.07\,V_{\rm tab}$, and the convention already in use requires essentially no correction. The ratio exceeding unity is not a violation of $\eta\le1$: it reflects that the nominal volume truncates at $\lambda/2$ while the wave extends deeper.

\paragraph{Notation.} Throughout the Letter and this document $\hV$ denotes the invariant $\hmo\Veps$ of Eq.~(6) of the Letter --- $n\langle v_C^{2}\rangle$ for the modulus channel, plus the eigenstrain term in fixed-end devices --- and both figures plot this quantity directly, with $\Veps=\eta V$ evaluated per device from
Table~\ref{tab:eta}. The tables of Sec.~S1 record the raw inputs---$V$ from mass and density, and $\sA$---from which $\hV$ follows.

\paragraph{Effect on the global fit.} Applying the table to all
$42$ fitted devices moves the exponent of Eq.~(5) of the Letter from
$p=-0.584$ (fitted against the raw $V$) to $p=-0.581$: a shift of $0.003$, a twentieth of the cluster standard error. The purpose of the exercise is therefore not that it changes the fit. It is that $\Veps$ becomes a computed quantity rather than a label, and the abscissa of Figs.~1 and 2 carries it.

\paragraph{Caveats, in decreasing order of size.} (i) Energy trapping confines a BAW mode laterally and lowers $\eta$ below $2/3$; the tabulated value is the untrapped plane-mode limit and is therefore an upper bound. (ii) Soft-clamped, phononic-crystal and perimeter modes have no closed form and require a finite-element evaluation of Eq.~(\ref{eq:veps}); every such device in this compilation is a bound and enters no fit, but each is drawn, and its abscissa carries this uncertainty. (iii) For flexural modes $\eta$ depends weakly on mode
order, and the fundamental value is used throughout. (iv) Composite stacks are assigned the class of the layer carrying the mode.

\subsection{Torsion rods}
\label{sec:eta-torsion}

For a rod in uniform torsion the strain is pure shear: $\gamma\propto|\nabla\phi|$ with $\phi$ the Prandtl stress function, $\nabla^2\phi=-2$ on the cross-section and $\phi=0$ on its boundary, uniform along the rod. The participation fraction $\eta=\langle\gamma^2\rangle^2/\langle\gamma^4\rangle$ therefore depends only on the cross-section. Solving numerically gives $\eta=0.619$ for the $250\times180$~nm rods of Ref.~\cite{Duraffourg2018} and $\eta=0.566$ for the $200\times500$~nm rods of Ref.~\cite{Zhang2013} (emitting script: \texttt{torsion\_eta\_and\_trial.py}). The paddle is treated as strain-free: its rotary inertia sets the frequency while the rod-only stiffness $\kappa=2GK/L_r$ carries the elastic energy, which is the convention of both sources. Warping restraint at the clamps is neglected; it concentrates additional strain within roughly one rod width of each end ($L_r/w_r\approx6$ and $10$ here).

\paragraph{Robustness to the weighting --- the full comparison.} The entire clustered analysis was rerun with the abscissa taken as bare $V$, holding the population, the error model, and the study-cluster bootstrap fixed (\texttt{scripts/veps\_vs\_V.py}; the output ships with the analysis-code deposit). Against $V$: $p=-0.584$, raw se $0.016$, $\chi^{2}/\mathrm{dof}=1.97$, cluster-bootstrap se $0.058$, $95\%$ CI $[-0.66,-0.43]$; volume independence excluded at $10.0\sigma$, site-averaging at $4.3\sigma$, the bulk value consistent at $1.4\sigma$. Against $\Veps$: $p=-0.581\pm0.063$, $\chi^{2}/\mathrm{dof}=2.11$, $9.2\sigma$ and $3.9\sigma$, bulk at $1.3\sigma$. In the product frame the slope of $\hmo X$ against $X$ is $-0.168\pm0.116$ for $X=V$ and $-0.162\pm0.127$ for $X=\Veps$: equally consistent with invariance at the $1.3$--$1.4\sigma$ level. Per-family rms scatter of the log product changes by at most $0.03$~dec in either direction ($V$ vs.\ $\Veps$: room temperature $0.78$ vs.\ $0.76$; cryogenic $0.91$ vs.\ $0.93$; quartz $0.42$ vs.\ $0.45$; torsional $0.26$ vs.\ $0.24$), and the mode-class medians within a family do not separate at current precision (room temperature $1.90$ vs.\ $1.55$~dec, a spread carried by the two-device beam class; quartz $0.00$ vs.\ $0.21$~dec). None of this is surprising: $\eta$ spans $0.237$--$1.07$ (the $1.07$ is the SAW convention above, referred to the nominal $A\lambda/2$), so the largest offset the weighting can produce between two classes is $0.65$~dec --- below the $\pm0.4$--$0.9$-dec family bands and far below the $13.35$-decade span of the fit. The participation volume is therefore used because the derivation produces it --- $\hmo\Veps=n\langle v_C^{2}\rangle$ is defined by the strain weighting --- and not because the present record can resolve the substitution. What $\eta$ predicts, and the record cannot yet test, is a specific pattern of class offsets: a cantilever population should sit $0.63$~dec above a string population of equal $\hmo V$. Resolving that pattern is a defined target for future within-class versus cross-class precision comparisons.

\section{The error model}
\label{sec:errors}

The $42$ fitted points are not $42$ independent experiments. One device is measured at three temperatures (Fong 2012, entering the room-temperature and cryogenic families), four points are modes or siblings of one resonator platform (Maillet 2018), nine derive from a single digitized figure (Kim 2017), and the quartz family reduces to five named sources (the nine-device Parker--Andres program, Driscoll--Hansen, Salzenstein, and two single-value BAW clusters); the four devices admitted in revision add three more (Wang 2020\textquotesingle s two forks are one campaign; Duraffourg 2018 and Zhang 2013 one each; Sec.~\ref{sec:torsion}). \textit{Grouped this way the compilation carries $16$ study clusters.}

The same rule governs the family \emph{levels}. Each family reference quoted in the Letter (Fig.~2 caption; the cross-family ratios) is the median over its independent study clusters of the cluster's own median $\hV$ --- one value per study --- with the error taken from a bootstrap over clusters. Table~\ref{tab:levels} lists them beside the pooled per-device medians, which weight every device equally and therefore let a prolific study dominate its family.

\begin{table}[h]
\centering
\caption{Family reference levels of $\hV$ (m$^3$): the study-cluster median used in the Letter, with its cluster-bootstrap error in decades and the number of clusters, and the pooled per-device median for comparison.
\label{tab:levels}}
\renewcommand{\arraystretch}{1.4}
\setlength{\tabcolsep}{8pt}
\begin{tabular}{l c c c c}
\hline\hline
family & cluster median & $\pm$ (dec) & clusters & per-device median \\
\hline
room temperature      & $1.01\ee{-31}$ & $0.35$ & 7 & $1.54\ee{-32}$ \\
cryogenic             & $1.26\ee{-33}$ & $0.46$ & 3 & $9.19\ee{-34}$ \\
quartz / SAW / BAW    & $1.18\ee{-34}$ & $0.36$ & 5 & $4.16\ee{-34}$ \\
torsional             & $2.77\ee{-33}$ & $0.24$ & 2 & $2.77\ee{-33}$ \\
\hline\hline
\end{tabular}
\end{table}

The room-temperature pooled value is set by the nine contour-mode points of one digitized figure, which is exactly the situation the clustering exists to prevent, and was the reason the convention was fixed. The cross-family ratios carry the quadrature sum of the two bootstrap errors (quartz below room temperature $2.9\pm0.5$~dec; cryogenic $1.9\pm0.6$; torsional $1.6\pm0.4$).

\begin{table}[h]
\centering
\caption{Four estimates of the uncertainty $\sigma(p)$ on the common exponent $p=-0.581$. The raw weighted least-squares error takes the stated per-point errors at face value; the study-cluster bootstrap (adopted) resamples the $16$ independent studies; the jackknife omits one study at a time; the hierarchical fit treats study-within-family as a random effect and returns, in place of a single error, the fitted exponent and the between-study scatter $\tau$.
\label{tab:sigmap}}
\renewcommand{\arraystretch}{1.4}
\setlength{\tabcolsep}{8pt}
\begin{tabular}{l c}
\hline\hline
estimate of $\sigma(p)$ & value \\
\hline
raw weighted least squares            & $0.016$ \\
study-cluster bootstrap (adopted)     & $0.063$ \\
leave-one-study-out jackknife         & $0.045$ \\
hierarchical, study within family     & $p=-0.572$, $\tau=0.22$~dec \\
\hline\hline
\end{tabular}
\end{table}

The bootstrap percentile interval is $[-0.66,-0.40]$ at 95\%; distances quoted in units of $\sigma(p)$ are standardized bootstrap standard errors, not Gaussian tail probabilities---$4000$ resamples of $16$ clusters cannot resolve a Gaussian $10\sigma$ tail, and no such tail is claimed. Against the adopted bar the alternatives stand at: volume-independent, $p=0$, $9.2\sigma$; surface site-averaging, $p=-1/3$, $3.9\sigma$; bulk, $p=-1/2$, $1.3\sigma$; volume-weighted surface, $p=-2/3$, $1.4\sigma$. One more number should be stated plainly: how much of the record actually sets the exponent. Because each family has its own intercept, the fitted slope $p$ is determined only by how $\sA$ varies with volume \emph{within} a family; the offsets between families do not enter. The within-family spread of $\log\Veps$ is $20.2\%$ of the total spread (the sum of squared deviations about the family means, as a fraction of that about the grand mean), equivalent to a single family spanning $3.3$ decades, and it comes almost entirely from the room-temperature ($54\%$) and quartz ($42\%$) families. The exponent is therefore measured over an effective $3.3$ decades. It is the invariant --- a common value of $\hV$ across families --- that is observed over the full $13.35$ decades.

\paragraph{Bounds as censored data.}\hspace{-0.5em}\footnote{Here, ``\textit{censored}'' is a statistical term for a datum of which only a bound is known. The term ``unresolved plateau'' implies only that $\sA\le B$, the lowest level reached, so the true value is \textit{censored} at $B$. Ordinary least squares cannot use such a datum at all, which is why Rule~1 excludes bounds. A censored-likelihood fit can use it: each resolved device contributes the Gaussian density of its residual, each bound the Gaussian probability $\Phi[(\log_{10}B-\mu)/s]$ that the true value lies below $B$, so a bound pulls the fit only where the line would otherwise sit above it. The ``\textit{censored estimate}'' is the exponent this fit returns with the eleven bounds admitted.} --- The principal fit uses resolved $1/f$ plateaus only. The measured, internally damped devices that never decreased to a bona fide plateau enter Fig.~1 of the Letter solely as upper limits and enter no fit; a censored Gaussian likelihood admits them without pretending they are detections. Detections contribute the normal density and limits the normal integral $\Phi[(\log_{10}B-\mu)/s]$, with per-family intercepts, a common slope, and a common scatter $s$, all fitted by maximum likelihood; eleven limits qualify (three room-temperature, eight cryogenic). The detected-only estimate, $p=-0.58$, reproduces the weighted baseline; admitting the limits gives $p=-0.45$ with a cluster-bootstrap error of $0.12$, the scatter inflating from $0.29$ to $0.49$~dec. The shift is one-sided information honestly gained: nine of the eleven limits lie below the fitted band, and seven of the nine come from a single laboratory program at $\Veps\approx10^{-20}$~m$^3$, bounded as much as $1.6$~decades below the extrapolated cryogenic band; removing that program returns $p=-0.52$. Measured in units of its own standard error ($0.12$), the censored estimate $p=-0.45$ lies $0.4\sigma$ from the bulk value $-1/2$ and $1.1\sigma$ from the baseline $-0.58$: it disagrees with neither, and the same flattening appears transparently in the bounds-admitted room-temperature trend ($-0.42$, against $-0.48$ resolved-only). What the limits add is not a different exponent but a caution the Letter states independently: the smallest cryogenic devices are quieter than the band extrapolation, and only resolved plateaus can say by how much.

\paragraph{Does $\bm{Q}$ order the residuals?}--- The residuals in question are the vertical distances, in decades, of the fitted devices from their family's line in Fig.~1, $\log_{10}\sA-(c_{\rm family}+p\log_{10}\Veps)$; a $Q^{-n}$ law acting on top of the volume law would sort them by $Q$, high-$Q$ devices below the line and low-$Q$ devices above. This can be tested by a fit with one extra term. For the $30$ of $42$ fitted devices whose primary source states a quality factor (the $Q$ columns of Tables S1--S3), the weighted fit of the principal analysis is repeated with $\log Q$ added as a covariate,
\begin{equation}
\log_{10}\sA \;=\; c_{\rm family} \;+\; p\,\log_{10}\Veps \;+\; q\,\log_{10}Q ,
\end{equation}
where $p$ is the volume exponent of Eq.~(5) of the Letter and $q$ is the exponent of any residual $Q$ dependence at fixed volume, $\sA\propto Q^{\,q}$; a law $\sA\propto Q^{-n}$ means $q=-n$, and $q=0$ means none. On this subset $Q$ spans $3.2$ decades and is only weakly correlated with volume, $r(\log Q,\log\Veps)=+0.32$, so the two exponents are separable in principle. Adding the covariate moves $p$ by $-0.012$ and returns $q=+0.042$ with a raw standard error of $0.051$. Under the same study-cluster resampling as the principal fit, however, the subset carries only $10$ clusters, and the bootstrap widens the interval to $q=+0.04\pm0.37$ (95\%: $[-0.67,+0.67]$); because $\log Q$ and $\log\Veps$ covary, the resampled $p$ and $q$ trade against each other ($r=-0.84$), which is what inflates the error. Accordingly, the covariate test excludes the steep historical $Q$ laws --- taken as exponents of $\sA$, $Q^{-3.5}$ lies $9.7$ cluster errors from the fitted $q$ and $Q^{-1}$ lies $2.8$ --- but it cannot by itself distinguish $q=0$ from a $Q^{-1/2}$-type dependence; no $Q$ dependence is resolved. Yet every $Q$ dependence reported in the literature --- $Q^{-4}$ for BAW resonators \cite{Gagnepain1981,Walls1992}, $Q^{-2.7}$ to $Q^{-3.8}$ for contour-mode plates \cite{Kim2017,Kim2018} --- is at least as steep as $Q^{-1}$ in $\sA$ whether the quoted power is read as an exponent of $\sA$ or of $\hmo=\sA^{2}/(2\ln2)$, and $Q^{-1}$ itself lies outside the 95\% interval: the region the bootstrap cannot exclude contains no reported law. The only $Q$ dependence the data leave open is therefore a weak one, no steeper than about $Q^{-1/2}$ in $\sA$ --- and no such law has ever been proposed.

The sharper discrimination comes from the two controls the compilation contains: the fixed-device temperature series of Ref.~\cite{Fong2012}, in which the floor falls $46\times$ while $Q$ rises $5.5\times$ ($Q^{-1}$ would give $6\times$), and the fixed-frequency series of Ref.~\cite{Parker1993}, whose authors report no strong dependence on unloaded $Q$ at fixed $f_0$. Residuals of the volume-only fit against $\log Q$ carry $r=+0.19$ globally, with per-family correlations ($+0.53$, $+0.40$, $+0.14$) of the sign \emph{opposite} to every $Q^{-n}$ law---higher-$Q$ devices sit, if anything, slightly above the band---so no ordering of the kind claimed exists. The null also has a mechanism: the intrinsic channels that bound $Q$ carry no $1/f$ frequency-noise content. Phonon--phonon dissipation sets the $f\!\cdot\!Q$ ceiling that ties $Q$ to frequency---and, through the acoustic wavelength, to volume---within a plate family~\cite{Tabrizian2009}: precisely the co-variation the Letter unwinds. And a resonator can sit entirely on such defect-free channels: the $3.72$-GHz silicon resonator of Ref.~\cite{Hwang2011} has its measured $Q(T)$ quantitatively accounted, from $1.5$ to $300$~K, by phonon--phonon and electron--phonon scattering plus anchor loss, with no defect term---a $Q$ carrying no information about the defect census that sets $\hmo$. Where damping \emph{is} defect-mediated, the slice argument of Eq.~(7) of the Letter and Sec.~\ref{sec:relocated} applies instead: the defects that damp occupy a different barrier slice from the defects that destabilize. The originating script is \texttt{scripts/q\_partial.py}.

\section{Provenance and robustness of the fitted set}
\label{sec:provenance}

The five 450-MHz SAW rows are the design families A--E of Ref.~\cite{Parker1994} (a 56-device program): the tabulated volumes reproduce, to the digit, that paper's effective averaging areas (its Table~2) multiplied by one acoustic wavelength of depth---the convention of its Eq.~(12). The first 5-MHz BAW row is the best-resonator value stated in Ref.~\cite{Parker1993}: $S_y(1\,\mathrm{Hz})=3\times10^{-27}$ at the stated $0.06$~cm$^3$, a number that paper quotes from Refs.~\cite{Walls1992,Parker1992}---the BAW-noise model paper of Walls, Handel, Besson and Gagnepain, and the Parker--Montress spectral-purity survey whose Table~1 carries the matching class value.

The second is the value Ref.~\cite{Rubiola2007}, Table~I, gives for the ``CMAC Pharao'' oscillator --- the $5$-MHz BVA quartz oscillator built by C-MAC for the PHARAO cold-atom space clock --- $b_{-3}=-132$~dBrad$^2$/Hz, carried at the $0.06$~cm$^3$ class volume ($\pm0.30$~dec; not stated in that paper), with $Q=2.0$--$2.2\times10^{6}$ from the same source. For resampling, the five SAW rows belong to the same Raytheon program as the four fitted rows of Ref.~\cite{Parker1993} and are drawn with them as one study cluster; the two BAW values are independent single clusters. The cluster count was unchanged at $13$ by the recovery ($16$ after the admissions of Sec.~\ref{sec:torsion}); the enlarged Parker cluster widens the bootstrap error from $0.055$ to $0.056$. Every conclusion was, and remains, tested against removal of the seven:

\begin{center}
\begin{tabular}{l c c}
\hline\hline
leave-out & $N$ & $p$ \\
\hline
none (baseline)                          & $42$ & $-0.581\pm0.063$ \\
the seven recovered rows                 & $35$ & $-0.581$ \\
the three qB points at one shared volume & $39$ & $-0.534$ \\
both                                     & $34$ & $-0.535$ \\
\hline\hline
\end{tabular}
\end{center}

Removing the seven leaves the exponent unchanged to three decimals; the
sensitivity is retained as pure robustness, the attribution question having been closed. The larger lever is the shared volume of the three qB points, $6\times10^{-8}$~m$^3$---an \emph{electrode-defined} value carried with $\pm0.30$~dec uncertainty, while Ref.~\cite{Sthal2013} tabulates a factor of $15$ between the electrode and acoustic volumes for exactly this device class. Moving all three by that factor shifts $p$ to $-0.614$ ($0.5$ cluster errors); removing them shifts it to $-0.534$, half a cluster error toward the bulk value. No choice among these conventions threatens the exclusions of Sec.~\ref{sec:errors}, and the bulk hypothesis survives every one of them.

A leave-one-study-out sweep over all $16$ clusters spans $p=-0.595$ to $-0.544$; the single most influential cluster is the Salzenstein BVA point, whose removal shifts $p$ by $+0.037$. The jackknife standard error over the sweep, $0.045$, agrees with the bootstrap.

\paragraph{The rows added in revision.}---Rows 37--41 of Table~S1 entered after the analysis of Secs.~S1--S17 was frozen, and their provenance is therefore recorded here in full. Rows 37--38 (Wang 2020~\cite{Wang2020}): $\sigma_f\approx10$~Hz stated in text and Fig.~4f for both tuning forks, flat over $1$--$10$~s, undriven, in vacuum; fractional values use the stated $f_0$ ($16.51$ and $27.76$~MHz); the quoted single significant figure sets the $0.30$-dec error class. One caveat is carried: the two floors are equal in \emph{absolute} hertz while $f_0$ differs by $1.68\times$---equal fractional floors would differ by that factor---and the stated precision cannot distinguish the two readings. The pair is also a control: the two forks are the thermally compensated and uncompensated designs of that work, whose fractional temperature coefficients differ by $72\times$ ($42$ against $5050$~Hz/K at $16.51$ and $27.76$~MHz), and both show the same floor, which excludes temperature drift --- photothermal included --- as its origin. Row 39 (Duraffourg 2018~\cite{Duraffourg2018}): $\sigma_{y}=1.5\times10^{-7}$ stated in the Fig.~10a caption, plateau $50$--$200$~ms, PSD slope $-10$~dB/decade over $0.1$--$10$~Hz, and the plateau demonstrated independent of SNR by an actuation-voltage sweep. Row 40 (Zhang 2013~\cite{Zhang2013}): the Fig.~4b inset, attributed by that paper\textquotesingle s corresponding author to the $500$-nm device\textquotesingle s torsional mode, digitized as for the AlN table: $\sigma_{y}=1.6\times10^{-7}$, flat $10$--$70$~s, $\pm0.1$~dec assigned. Row 41 (the Transducers\textquotesingle07 companion of row 32\textquotesingle s campaign~\cite{Feng2007Transducers}): the measured Allan curve of device SiNW-M188 reaches $6\times10^{-8}$ over $\tau=4$--$10$~s at $25$~K --- flat over $0.4$~dec, short of the decade of flatness the resolved-plateau class requires --- and therefore enters as an upper bound, not a fitted point; $V$ is the geometric Si cylinder ($L=2.1~\mu$m, $d=81$~nm), and the fits of Secs.~S14 and S17 are unchanged except for this eleventh limit.

\section{The electrode-volume experiment}
\label{sec:ferrepikal}

One controlled, single-variable volume sweep exists in the quartz literature, and it must be confronted because it runs counter to the Letter's scaling. Ref.~\cite{FerrePikal1996} reports three groups of twelve $100$-MHz resonators fabricated from one crystal bar, differing only in electrode diameter ($2.16$, $3.05$, $4.32$~mm); the median $\mathcal L(10\,\mathrm{Hz})$ of the unswept groups was $-103.1$, $-101.6$ and $-97.7$~dBc/Hz. Over $0.60$ decades of electrode area at fixed thickness that is $h_{-1}\propto V_{\rm el}^{+0.9}$: \emph{larger} electrodes are noisier, consistent with the model of Ref.~\cite{Walls1992}, which places the electrode volume in the numerator.

Three observations reconcile this with the compilation. First, the electrode volume is not the participation volume: in a trapped thickness-shear mode the acoustic energy extends beyond the electrode into the plate (Ref.~\cite{Sthal2013} quotes a factor of $15$ between the two volumes), so an electrode sweep at fixed plate is a sweep of lateral \emph{trapping}, not of $\Veps$. Second, the authors themselves identify a competing channel that grows with electrode size---``the number and coupling to unwanted modes in a resonator is a function of the electrode size''---which is a mode-coupling contribution, not a defect-density one, and they describe their own result as ``suggestive but certainly not definitive'' against a $\sim20$-dB within-group spread. Third, by the admission rules of Sec.~S1 these devices would enter as bounds: the quantity is $\mathcal L(10\,\mathrm{Hz})$ in a test oscillator the
authors state may be electronics-limited, and no resolved $\sA$ plateau is reported. The experiment that would decide the question at the level of this compilation---an electrode series with the acoustic volume measured independently, or a plate series at fixed electrode---has not been performed.

A second quartz null is equally instructive and cuts the other way. Ref.~\cite{Parker1994} fabricated the same designs on four grades of quartz, swept and unswept: ``there are no large differences in the $1/f$ noise levels of devices fabricated on either different grades of quartz or on swept and unswept material'' (within $3$--$4$~dB, on their stated small sample). Read against the invariant $\hmo\Veps=n\langle v_C^{2}\rangle$, this constrains the fluctuator\textquotesingle s identity, not the density proportionality. Electrodiffusive sweeping removes only the mobile alkali compensators; the sweep proves nothing about the populations it cannot move: every sweep-immune population remained in place, in every grade tested, free to dominate the measured floor. The proper reading of the null is therefore that the dominant fluctuators lie among the populations that neither sweeping nor grade specifications control, and that $n$ was never actually varied. The grades themselves are intrinsic acoustic-$Q$ designations for the material, so the same null is one more decoupling of $Q$ from the floor: material ranked by $Q$ varied, and $\hV$ did not (cf.\ Secs.~S7 and S17). A deliberate doping or irradiation series with the flicker floor measured would vary it directly.

\section{Multimode readout, and the scope of the no-go argument}
\label{sec:multimode}

The statement that no measurement scheme removes the $1/f$ floor
(Sec.~\ref{sec:schemes}) is a statement about single-channel schemes. When signals and fluctuations enter the equation of motion through the same parameter, no gain, loop, or post-processing applied to that one channel can separate them. This statement does \emph{not} extend to multimode observations. If the frequency fluctuations of two modes are correlated while the signal couples to the modes with a different ratio, a difference coordinate exists in which the correlated part cancels; Ref.~\cite{Kharbanda2026} demonstrates exactly this on a soft-clamped membrane, forming a signal at the difference frequency of two modes whose intrinsic fluctuations are strongly correlated.

The admissible figure of merit for any such scheme is $\sA(\tau)/\mathcal{R}(\tau)$, where the responsivity $\mathcal{R}$ for the intended measurand \emph{must be measured, not assumed}. Differencing removes the common part of the signal as well, so any such scheme is an improvement \textit{only} if the noise is suppressed more strongly than the responsivity, $\mathcal{R}$, is reduced. The record on this test is mixed:
\begin{itemize}
    \item Sage's two-mode subtraction scheme (Ref.~\cite{Sage2013}) removed the drift branch entirely but left the $1/f$ plateau untouched.
    \item Sansa \textit{et al.} (Ref.~\cite{Sansa2016}) corrected one mode against a second used as a thermometer, which improved drift and reduced the white branch, but this also left the $1/f$ plateau unmoved.
    \item The phase-noise-cancellation scheme of Kenig \textit{et al.}~\cite{Kenig2012} preserves the responsivity but, by construction, cancels only noise entering through the feedback and leaves noise on the resonance frequency itself --- the $1/f$ plateau --- untouched, as its authors state explicitly~\cite{CrossKenigAllen2015}.
    \item The correlations of Kharbanda \textit{et al.}~\cite{Kharbanda2026} have not been applied to frequency-shift sensing, so whether the responsivity is preserved with their two-mode scheme remains an open question.
\end{itemize}

\paragraph{What the two-channel form predicts for the correlation.} Independent local fluctuators produce \emph{correlated} inter-mode fluctuations whenever the same defect acts on both modes, and the correlation coefficient of the floors of modes $m$ and $n$ is the overlap of their site weights,
\begin{equation}
    r_{mn}=\int w_m w_n\,dV/\sqrt{\int w_m^{2}\,dV\int w_n^{2}\,dV} \ .
\end{equation}

The eigenstrain channel is uniform in every mode, so $r=1$ exactly. For the modulus channel, with $w=1-(1-\hat w)/D_Q$ (Sec.~\ref{sec:eigenchannel}): the first two modes of an unstressed clamped-clamped beam ($D_Q=1$, $\hat w\propto\psi''^{2}$) give $r_{12}=0.744$, so a second mode can remove at most $1-r^{2}=45\%$ of the variance, $3.5$~dB; a string, whose bending strain sits in the clamp boundary layers common to both modes (Sec.~\ref{sec:eigenchannel}), gives $r_{12}=0.997$ at $D_Q=10$ ($22$~dB), $0.999998$ at $D_Q=80$ ($55$~dB) and $1-10^{-7}$ at $D_Q=160$ ($67$~dB) (\texttt{scripts/intermode\_correlation.py}).

The record follows this pattern. Sansa's and Sage's beams, corrected against a second mode, kept their plateau: $3.5$~dB is within their scatter. Gavartin et al.~\cite{Gavartin2013} stabilized a high-stress string ($D_Q\approx80$) against its second mode by photothermal actuation and held mode~1 at its thermomechanical line to $\tau\approx100$~ms, a reduction of more than $20$~dB that their loop gain, not the correlation, limited. Kharbanda et al.~\cite{Kharbanda2026} find the intrinsic flicker of two modes of a tension-dominated membrane strongly correlated. The no-go of Sec.~\ref{sec:schemes} is therefore a theorem for the single-mode floor and, to within $3.5$~dB, for bending-set modes; for tension-set modes it is a trade-off governed by the gradiometer price above: Gavartin's loop, which cancels any measurand that shifts both modes in the noise's ratio (temperature, uniform strain, axial force), was never used to measure one, and the one measurand it leaves responsivity to is one that couples to the two modes differently, as a mass at a mode-1 antinode and mode-2 node does. A measured inter-mode correlation matrix is thus a direct test of the weight function, sharper than any single-mode spectrum: $r_{12}$ near $0.74$ for a beam and near $1$ for a string are the two-channel prediction, and a beam value near $1$ would falsify the strain-energy weighting of the modulus channel.

\section{Scope of the dilution argument}
\label{sec:dilutionscope}

The Letter's statement that dilution multiplies $Q$ while leaving $\hmo$ unaltered concerns the \emph{dilution factor} $D_Q$, by which a geometric redistribution of modal energy (prestress at fixed loss, soft clamping, phononic shielding, perimeter modes) multiplies the measured $Q$; neither term of Eq.~(3) of the Letter contains $D_Q$, and both the membrane comparison of Ref.~\cite{KanellopulosSchmid2026} and the perimeter-mode bound of Ref.~\cite{Bereyhi2022} (Sec.~S4) show the floor unmoved by it. It does \emph{not} assert that the floor is independent of the prestress: the eigenstrain term dilutes as $\seff^{-2}$, which Ref.~\cite{KanellopulosSchmid2026} observes directly on strings of one film whose $\sigma_0$ is tuned from $357$ to $440$~MPa by clamp tapering ($1.78\pm0.20$ against $(440/357)^{2}=1.52$), and which the first version of this Letter denied on the ground that highly stressed strings and unstressed beams fall on one invariant. That ground was insufficient: the strings of this record span $0.2$~decades of $\sigma_0$ ($0.62$--$1$~GPa), and the unstressed beams have bending-equivalent stresses $\sigma_c=3.39\,E(h/L)^{2}$ of $0.3$--$5$~GPa, comparable to the strings' prestress, so the two groups share $\seff$ and the record has no lever arm on the stress dependence (Sec.~\ref{sec:eigenchannel}). The sentence ``No stress dilution of the $1/f$ floor is observed'' is withdrawn.

The modulus channel is the part of the argument that does not dilute, and its derivation is now carried out for the general prestressed eigenproblem rather than for the fixed-length string alone. For a doubly clamped beam of any prestress the eigenfrequency obeys
\begin{equation}
    \omega^{2}=(q_n^{2}/\rho)(\sigma_0+\sigma_c),
\end{equation}
with $\sigma_0=E\varepsilon_0$ and $\sigma_c=E\,g$ ($g=I\kappa_n^{2}/A$ a geometric factor; Sec.~\ref{sec:eigenchannel}). A local modulus change $\delta E_i$ over $v_d$ perturbs the prestress uniformly,
\begin{equation}
    \delta\sigma_0=\varepsilon_0\,\delta E_i\,v_d/V \,,
\end{equation}
because force balance redistributes the tension along the whole span, and it perturbs the bending stiffness in proportion to the bending strain energy at the site,
\begin{equation}
    \delta\sigma_c=g\,\delta E_i\,(v_d/V)\,\hat w(\mathbf r_i) \ .
\end{equation}
The fractional frequency shift is thus
\begin{equation}
    \delta_1=\tfrac12(\delta E_i/E)(v_d/V)\,w(\mathbf r_i)
\end{equation}
with $w=[\sigma_0+\sigma_c\hat w]/\seff$, independent of the magnitude of $\seff$ and interpolating between uniform weighting for a string and strain-energy weighting for a beam (Sec.~\ref{sec:eta}, Sec.~\ref{sec:eigenchannel}). Under fixed load rather than fixed length the prestress is not strain-clamped and $\delta\sigma_0=0$. The modulus channel then comprises only its bending part, and the eigenstrain channel is absent because the load, not the length, is fixed. In a strain-engineered taper the static strain varies along the device~\cite{Ghadimi2018}; $w$ then follows the local static strain, as Sec.~\ref{sec:eta} states, and $\seff$ in the eigenstrain term is the mode-weighted value. The scalar form of $w$ assumes an isotropic modulus coupling; tensorial couplings reweight $\hat w$ within a class, and torsion couples through the deviatoric strain alone (Sec.~\ref{sec:torsion}).

A stress series at fixed geometry has recently been reported --- the tapered-clamp strings of Ref.~\cite{KanellopulosSchmid2026} --- and it tests the eigenstrain term, not the modulus term. The corresponding test of the modulus term would require measurements of stress series upon a free body, where the eigenstrain term is absent and no stress dependence is predicted. This remains to be carried out.

\section{The torsional family}
\label{sec:torsion}

Two torsional devices are considered as their own family. Data include results from the infrared bolometric resonator of Ref.~\cite{Duraffourg2018} ($f_0=1.1$~MHz, $Q=1800$) and the first frequency-shift bolometer implementation that inspired that work, the torsional paddle of Ref.~\cite{Zhang2013} ($f_0=6.626$~MHz, $Q=5225$). Their provenance is recorded in Sec.~\ref{sec:provenance} and their participation volumes in Sec.~\ref{sec:eta-torsion}. In a torsional mode the elastic energy is principally stored within the support rods, and both devices fall well below the flexural room-temperature band: the family reference in $\hV$ sits $1.6\pm0.4$ decades ($\sim\!36\times$) below the flexural room-temperature reference, the error being dominated by the between-study spread of the room-temperature family itself (Sec.~\ref{sec:errors}). Two points support no slope, so the family enters the global fit as an intercept only; however, if they are forced into the flexural room-temperature family, the global exponent instead moves to $-0.55\pm0.07$, less than half a cluster error, and thus changes no conclusion.

Two mechanisms could produce an offset in $\sApl$, and they are distinguishable. The torque a torsion rod applies to its supports couples to no compressional bulk channel and suffers a strong acoustic impedance mismatch. By reciprocity, the same weak anchor transmission that suppresses radiation also screens the mode from defects and fluctuations that are located beyond the rods (\textit{i.e.}, in the support region and beyond). That mechanism is geometry-tunable: it should respond to undercut, pedestal, and rod aspect ratio. Alternatively, pure shear couples only to the deviatoric part of a defect\textquotesingle s elastic dipole, so a dilatation-dominated defect population is partially invisible to a torsional mode---a material-universal effect. (Anchor transmission and the clamp region's share of the participation volume are related but distinct quantities: the first is the far-field radiation of acoustic energy into the support, which is what $Q$ registers; the second is the near-field strain in the clamp, which is what sets how strongly defects there couple. Equating weak transmission with weak defect coupling requires the Letter's local coupling model --- a defect shifts the frequency in proportion to the strain energy at its own site --- and that step is offered here as a hypothesis, not a result.) Neither is a $Q$ mechanism: these are the two lowest-$Q$ devices among the fitted set, and Sec.~\ref{sec:errors} already shows that total $Q$ does not order the floors. A single structure with resolved plateaus in both a torsional and a flexural mode would separate the two mechanisms; the device of Ref.~\cite{Zhang2013}, whose flexural mode was additive-limited at the drives explored, is the type of experiment that could elucidate this.

Caveats, stated plainly: the family holds two devices from two laboratories thirteen years apart; one value is digitized; and the Saint-Venant weighting neglects warping restraint at the clamps (Sec.~\ref{sec:eta-torsion}).

\section{The eigenstrain channel in devices with fixed ends}
\vspace{-0.5cm}
\label{sec:eigenchannel}
\paragraph{Effective stress.} For a doubly clamped beam or string of length $L$, cross-section $A$, area moment $I$ and density $\rho$ under an axial prestress $\sigma_0$, the frequency of a mode with shape $\psi_n(x)$ follows from a modal stiffness that has a tension part and a bending part,
\begin{equation}
\rho A\,\omega^{2}\!\int\!\psi_n^{2}\,dx \;=\; \sigma_0 A\!\int\!\psi_n'^{2}\,dx \;+\; EI\!\int\!\psi_n''^{2}\,dx ,
\end{equation}
the two terms on the right being twice the modal potential energy stored against the tension and in bending, respectively. Dividing through by $A\!\int\!\psi_n'^{2}$,
\begin{equation}
\omega^{2} \;=\; \frac{q_n^{2}}{\rho}\,(\sigma_0+\sigma_c),
\qquad
q_n^{2}\equiv\frac{\int\psi_n'^{2}}{\int\psi_n^{2}},
\qquad
\sigma_c\equiv\frac{EI}{A}\,\kappa_n^{2},\quad \kappa_n^{2}\equiv\frac{\int\psi_n''^{2}}{\int\psi_n'^{2}} .
\end{equation}
Here $q_n$ is the modal wavenumber ($n\pi/L$ for a sinusoidal mode; $q_n^{2}L^{2}=12.30$ for the exact fundamental of an unstressed clamped-clamped beam) and $\sigma_c$ is the bending stiffness of the mode expressed as an equivalent stress. For a rectangular section $EI/A=Eh^{2}/12$, and for the unstressed clamped fundamental $\kappa_n^{2}L^{2}=40.7$ [Eq.~(S21) of Ref.~\cite{KanellopulosSchmid2026}], so that
\begin{equation}
\sigma_c \;=\; 3.39\,E\,(h/L)^{2}\qquad\text{(unstressed beam)}.
\end{equation}
The effective stress $\seff\equiv\sigma_0+\sigma_c$ is therefore read operationally from the eigenfrequency,
\begin{equation}
\seff \;=\; \rho\,\omega^{2}/q_n^{2},
\end{equation}
for stressed and unstressed devices alike, exactly as Ref.~\cite{KanellopulosSchmid2026} extracts $\sigma_0$ (Table~\ref{tab:sigmaeff}, with $q_n^{2}=(n\pi/L)^{2}$ for the strings and $12.30/L^{2}$ for the unstressed beams); for an unstressed beam $\seff=\sigma_c$.

The dilution factor is the ratio of the total modal stiffness to its bending part,
\begin{equation}
D_Q\equiv\frac{\seff}{\sigma_c}
=\frac{\sigma_0A\!\int\!\psi_n'^{2}+EI\!\int\!\psi_n''^{2}}{EI\!\int\!\psi_n''^{2}} ,
\label{eq:DQdef}
\end{equation}
and it is the same ratio by which prestress dilutes the dissipation, $Q=D_Q\,Q_{\rm int}$: the bending (elongational) strain energy is the lossy part of the modal energy, while the energy stored against the tension is returned without loss. For an unstressed beam $D_Q=1$. For a tensioned string $\sigma_c$ is \emph{not} the unstressed-beam value above: the mode is sinusoidal in the interior but bends sharply within a distance $\lambda L$ of each clamp,
\begin{equation}
\lambda=\frac{h}{L}\sqrt{\frac{E}{12\,\sigma_0}} ,
\end{equation}
and these boundary layers hold most of the bending energy (about $85\%$ of it at $D_Q=80$). Evaluating Eq.~\eqref{eq:DQdef} with the exact mode of the tensioned beam reproduces the standard string expression
\begin{equation}
D_Q^{-1} \;=\; 2\lambda+(n\pi\lambda)^{2}
\label{eq:DQstring}
\end{equation}
[Eq.~(9) of Ref.~\cite{KanellopulosSchmid2026}; Ref.~\cite{Fedorov2018}] to within a few percent over the range of the record ($1\%$ at $D_Q=80$), the first term from the clamps and the second from the interior. This is the expression used for the strings of Table~\ref{tab:sigmaeff}, for which $D_Q=10$--$180$ and $\sigma_c=\seff/D_Q$; the intrinsic quality factor is $Q_{\rm int}=Q/D_Q$, which stands in for the Letter's $\Qel$ where the relation is evaluated at $\omega_0$; it is a lower bound on $\Qel$, because it includes any loss that is not due to the eigenstrain fluctuators.

\paragraph{The two-term form of the expression for $\bm{\hmo}$.}  Eq.~(3) of the Letter includes a second, eigenstrain term that is due to Kanellopulos and Schmid~\cite{KanellopulosSchmid2026}, who derived it from the fluctuation--dissipation theorem and verified its stress dependence experimentally. Rewritten in the per-defect form used here, this second term of Eq.~(3) matches their expressions (bracketed numbers are the equations of their Supplement): the uniform induced stress $\delta\sigma=-\gamma/V$ [(S5)]; the position-independent jump $\Delta=\gamma/(2\seff V)$ [(S6), (S22)]; the flicker sum $\hmo=n\langle\gamma^{2}p(1-p)\rangle/(4\seff^{2}V\ln R)$ [(S7)]; and the fluctuation--dissipation form $\hmo V\Qel=Ek_BT/(2\pi\seff^{2})$ [(S12), (S22)]. Their modulus channel under prestrain, $\Delta=v_d\,\delta E/(2EV)$ [(S14)], is the Letter's $\delta_1=v_C/\Veps$ with $\Veps=V$; their cantilever weighting $\varpi(z)=12z^{2}/h^{2}$, $\langle\varpi^{2}\rangle=9/5$ [(S23)], is the thickness factor $5/9$ of the Letter's $\eta$ for a bending mode (Sec.~\ref{sec:eta}).

What the Letter reports for the first time is a complete, two-term form of $\hmo$, which contains both possible couplings of each defect, together with the weight function of the modulus channel and the free-body limit of the eigenstrain term. Note, however, that to first order the eigenstrain term \textit{does not contribute} to the $29$ fitted devices of this record whose span is not fixed by their supports (cantilever, contour-mode plates, torsion rods, SAW and BAW), whereas \textit{the modulus term remains}. For these devices, the eigenstrain of a defect relaxes and what survives is the dimensional change, $\delta_1=\gamma/(EV)$; this is smaller than the clamped response by $E/2\seff$ ($\approx300$ at $0.4$~GPa, $10^{5}$ in power). The population of fluctuators that reproduces the behavior of the silicon-nitride strings [$K_v=n\langle\gamma^{2}p(1-p)\rangle$ of Ref.~\cite{KanellopulosSchmid2026}] would give $\hmo V\sim10^{-36}$~m$^{3}$ in a free body, against the observed $1.3\ee{-32}$~m$^{3}$; accordingly, the free-body floor is the modulus channel.

\paragraph{Weight function of the modulus channel.} The weight $w(\mathbf r)$ of the Letter --- the fraction of the modal stiffness that a defect at $\mathbf r$ controls --- follows from the Rayleigh quotient above by first-order perturbation theory, the mode shape being stationary. Take the span fixed by the supports, so that the prestress is a frozen-in strain, $\sigma_0=E\varepsilon_0$ with $\varepsilon_0$ set by the clamps, and write the bending part as $\sigma_c=E\,g$ with $g\equiv I\kappa_n^{2}/A$ a purely geometric factor. A defect switching at $\mathbf r_i$ changes the modulus by $\delta E_i$ over its volume $v_d$, and this enters the two terms of the Rayleigh quotient differently.

(i) \emph{Tension.} Under fixed length the total elongation $\varepsilon_0L=N\!\int_0^L\!dx/(EA)$ is prescribed, so a stiffer element cannot relax: force balance holds the tension $N$ uniform along the span, and a modulus change anywhere raises it everywhere, in proportion to the element's share $v_d/V$ of the axial compliance,
\begin{equation}
\delta\sigma_0=\varepsilon_0\,\delta E_i\,\frac{v_d}{V}=\sigma_0\,\frac{\delta E_i}{E}\,\frac{v_d}{V},
\end{equation}
independent of where the defect sits and of how it is distributed over the cross-section.

(ii) \emph{Bending.} The bending energy $EI\!\int\!\psi_n''^{2}dx=\int E\,\varepsilon_b^{2}\,dV$ is a volume integral of the local strain energy, with $\varepsilon_b=-z\,\psi_n''(x)$ the bending strain at height $z$ above the neutral plane. A local $\delta E_i$ changes it by $\delta E_i\,\varepsilon_b^{2}(\mathbf r_i)\,v_d$, that is, in proportion to the strain energy at the site:
\begin{equation}
\delta\sigma_c=g\,\delta E_i\,\frac{v_d}{V}\,\hat w(\mathbf r_i),\qquad
\hat w(\mathbf r)\equiv\frac{\varepsilon_b^{2}(\mathbf r)}{\langle\varepsilon_b^{2}\rangle},\qquad
\langle\hat w\rangle=1,
\end{equation}
where $\langle\cdots\rangle$ denotes the volume average.

Adding the two and using $\omega^{2}\propto\seff$,
\begin{equation}
\delta_1\equiv\frac{\delta\omega}{\omega}=\frac{\delta\sigma_0+\delta\sigma_c}{2\,\seff}
=\frac{v_C}{V}\,w(\mathbf r_i),\qquad
v_C\equiv\frac{v_d}{2}\,\frac{\delta E_i}{E},\qquad
w(\mathbf r)=\frac{\sigma_0+\sigma_c\,\hat w(\mathbf r)}{\seff}=1-\frac{1-\hat w(\mathbf r)}{D_Q},
\label{eq:weight}
\end{equation}
with $D_Q=\seff/\sigma_c$ the dilution factor of Eq.~\eqref{eq:DQdef}. The weight is normalized, $\langle w\rangle=1$, and its two limits are the two mode classes of the Letter: in an unstressed beam ($D_Q=1$) $w=\hat w$, the strain-energy weighting of a bending mode; as $D_Q\to\infty$, $w\to1$, the uniform weighting of a mode set by tension. In between, $w$ is a weighted mean of the two, the tension's share $1-1/D_Q$ of the stiffness spread uniformly and the bending share $1/D_Q$ distributed as the strain energy. No magnitude of $\seff$ appears: the modulus channel does not dilute.

The participation ratio follows from the definition $\Veps=V/\langle w^{2}\rangle$ of Sec.~\ref{sec:eta}. With $\langle\hat w\rangle=1$ and $\langle\hat w^{2}\rangle=1/\eta_b$,
\begin{equation}
\eta\equiv\frac{\Veps}{V}=\frac{1}{\langle w^{2}\rangle}=\Big[1+\frac{\eta_b^{-1}-1}{D_Q^{2}}\Big]^{-1},
\label{eq:etaDQ}
\end{equation}
where $\eta_b=1/\langle\hat w^{2}\rangle$ is the participation ratio of the bending strain energy of the same mode. For an unstressed clamped-clamped beam $\eta_b=0.301$ (the thickness factor $5/9$ times the axial factor $0.54$), and Eq.~\eqref{eq:etaDQ} returns it at $D_Q=1$. For a tensioned string $\eta_b$ is far smaller, because the bending strain is concentrated in the clamp boundary layers of width $\lambda L$: the exact mode gives $\langle\hat w^{2}\rangle\simeq(9/10)/\lambda\simeq\tfrac95D_Q$ for $\lambda\ll1$, so that
\begin{equation}
\eta\simeq\Big[1+\frac{9}{5D_Q}\Big]^{-1}\qquad\text{(tensioned string)},
\label{eq:etastring}
\end{equation}
accurate to $2\%$ for $D_Q\ge10$ (\texttt{scripts/string\_weight\_function.py}). For the strings of the record ($D_Q=10$--$180$) this gives $\eta=0.87$--$0.99$, within $0.07$~dec of the uniform weighting $\eta=1$ used for the string class, which is retained: the difference is below the $\pm0.3$-dec per-device uncertainty. The residual non-uniformity has a simple reading. A defect in the surface layer at a clamped edge, where $\hat w\simeq6D_Q$, controls about seven times the modal stiffness of one in the interior, whatever the prestress; such sites occupy a fraction $\sim2\lambda=1/D_Q$ of the volume.

This is the sense in which the scalar participation volume carries across mode classes: $w$ is the fraction of the modal stiffness the defect controls, which the tension distributes uniformly and the body forces distribute as the strain-energy density. Tensorial couplings reweight $\hat w$ within a class (Sec.~\ref{sec:dilutionscope}); torsion takes the deviatoric part (Sec.~\ref{sec:torsion}).

\begin{figure}[h]
\includegraphics[width=0.98\textwidth]{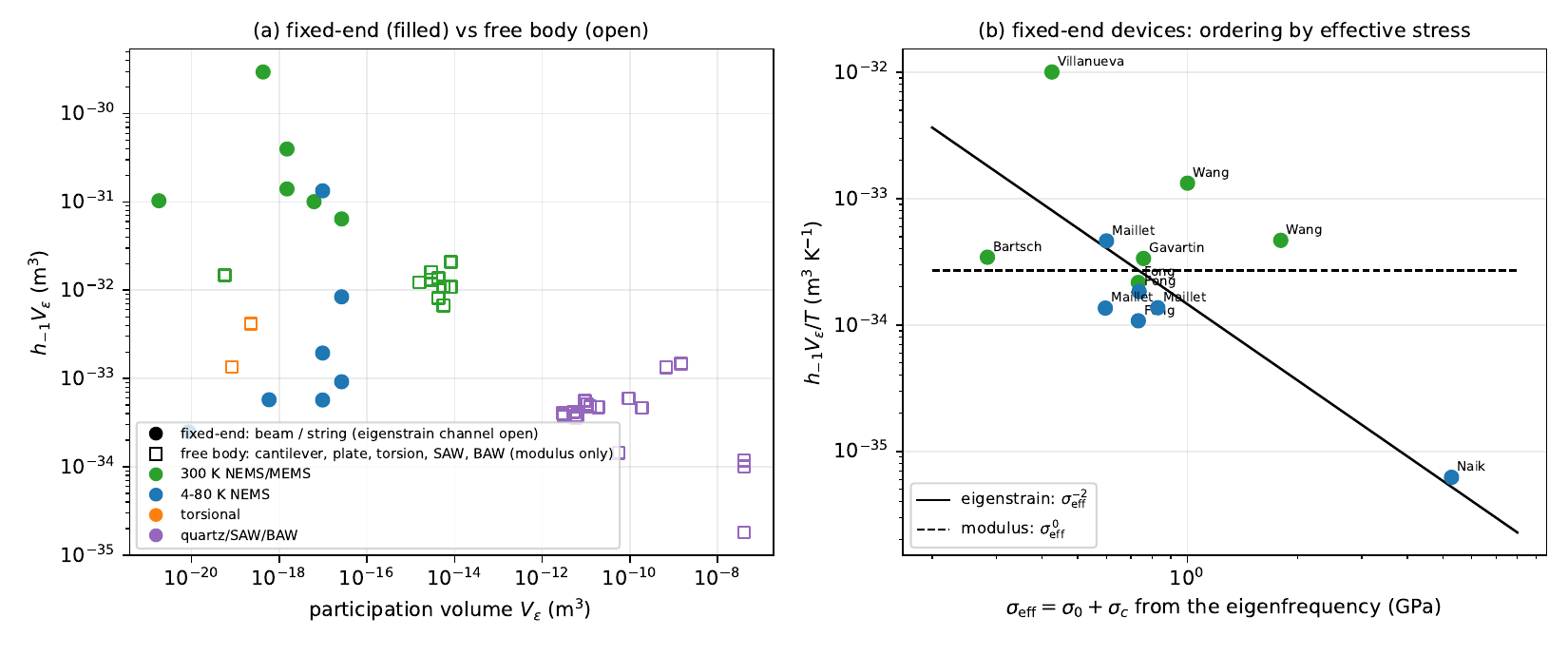}
\caption{\label{fig:split}(a) The invariant $\hV$ against $\Veps$ for the $42$ fitted devices, filled for fixed-end (beam, string) and open for free-body (cantilever, plate, torsion, SAW, BAW) modes. (b) The fixed-end devices alone: $\hV/T$ against the effective stress $\seff$ read from the eigenfrequency, with the $\seff^{-2}$ (eigenstrain) and $\seff^{0}$ (modulus) references through the geometric median. The ordering rests on Naik's $5.3$-GPa beam; the remaining devices span $0.47$~dec of $\seff$.}
\end{figure}

\paragraph{Constraint-class split of the record.} The two-term form of Eq.~(3) of the Letter makes a prediction that the record can test without a new measurement. A device whose span is fixed by its supports --- a string or a doubly clamped beam --- carries both terms, whereas a free body --- a cantilever, a contour-mode plate, a torsion rod, a SAW or BAW substrate --- carries the modulus term alone, because the eigenstrain of a switching defect relaxes when no coordinate is constrained. Two consequences follow. First, within one family at one temperature the fixed-end devices should lie above the free bodies by the eigenstrain term's share of the floor. Second, because the eigenstrain term scales as $T/(\seff^{2}Q_{\rm int})$, dividing the invariant of the fixed-end devices by these variables in turn should reduce their scatter, which the bare invariant $\hV$ knows nothing about. The room-temperature family is the only one that contains both classes, so the first test is made there; the second uses the fixed-end devices of the whole record (Table~\ref{tab:sigmaeff}). Figure~\ref{fig:split} shows both tests and Table~\ref{tab:split} gives the numbers.

The first test returns an offset of $0.88$~dec between the medians of the two classes ($1.06$~dec against bare $V$), in the predicted direction; the torsional and quartz families, which are free bodies, lie lower still, and the cryogenic family contains no free body. The second tightens the scatter of the fixed-end set from $1.33$ to $0.52$~dec, step by step, as the temperature, stress and loss corrections are applied [Fig.~\ref{fig:split}(b)]. Neither test is strong on the present record: the room-temperature free bodies are two clusters, and the $\seff$ lever is $1.27$~dec but rests on one device (Naik, $\sigma_c=5.3$~GPa); without it the lever is $0.47$~dec and no slope is resolved.

\begin{table}[h]
\caption{\label{tab:split}Constraint-class split of the record (\texttt{scripts/constraint\_class\_split.py}). ``Fixed-end'' denotes a device whose span is fixed by its supports (string, doubly clamped beam), which carries both terms of Eq.~(3) of the Letter; ``free body'' denotes one with no constrained coordinate (cantilever, contour-mode plate, torsion rod, SAW, BAW), which carries the modulus term alone. (a)~The room-temperature family --- the only family containing both classes --- by boundary condition, one study-weighted value of $\hV$ per cluster (the convention of Sec.~S17), with the free-body references of the other families for comparison. (b)~Scatter of the fixed-end set of Table~\ref{tab:sigmaeff} (standard deviation of $\log_{10}$, in decades) as the variables of the eigenstrain term are divided out in turn: the temperature, the effective stress $\seff$ read from the eigenfrequency, and the intrinsic quality factor $Q_{\rm int}=Q/D_Q$. Maillet $n{=}1$ is excluded as in Sec.~S1; the last row is restricted to the ten devices with a published $Q$.}
\renewcommand{\arraystretch}{1.3}
\setlength{\tabcolsep}{8pt}
\begin{center}
(a)\\[3pt]
\begin{tabular}{l l l c}
\hline\hline
study cluster & class & device & $\hV$ (m$^{3}$) \\
\hline
Fong 2012        & fixed-end & SiN string                        & $6.4\ee{-32}$ \\
Gavartin 2013    & fixed-end & SiN string                        & $1.0\ee{-31}$ \\
Bartsch 2012     & fixed-end & Si beam                           & $1.0\ee{-31}$ \\
Wang 2020        & fixed-end & SiN tuning forks (two)            & $2.7\ee{-31}$ \\
Villanueva 2011  & fixed-end & AlN/Mo beam                       & $1.9\ee{-30}$ \\
Kim 2017         & free body & AlN contour-mode plates (nine)    & $1.2\ee{-32}$ \\
Mile 2010        & free body & Si cantilever                     & $1.5\ee{-32}$ \\
\hline
median, fixed-end  & & & $1.0\ee{-31}$ \\
median, free body  & & & $1.4\ee{-32}$ \\
fixed-end / free body & & & $0.88$~dec ($1.06$~dec in bare $V$) \\
\hline
torsional family   & free body & torsional paddles (Sec.~\ref{sec:torsion}) & $2.8\ee{-33}$ \\
quartz family      & free body & SAW, BAW                  & $4.2\ee{-34}$ \\
cryogenic family   & \multicolumn{3}{l}{contains no free body} \\
\hline\hline
\end{tabular}

\vspace{8pt}
(b)\\[3pt]
\begin{tabular}{l c c}
\hline\hline
quantity & $n$ & scatter (dec) \\
\hline
$\hV$                              & 12 & $1.33$ \\
$\hV/T$                            & 12 & $0.75$ \\
$\hmo V\seff^{2}/T$                & 12 & $0.62$ \\
$\hmo V\seff^{2}Q_{\rm int}/(ET)$  & 10 & $0.52$ \\
\hline\hline
\end{tabular}
\end{center}
\end{table}

\begin{table}[H]
\caption{\label{tab:sigmaeff}The fixed-end devices of the fitted record: effective stress from the eigenfrequency, dilution factor, intrinsic quality factor, and the fluctuation--dissipation ratio $R$ (\texttt{scripts/constraint\_class\_split.py}). Stated $\sigma_0$ is used for the Wang forks; ``needs'' gives the $\Qel$ the relation would require where $Q$ is unpublished.}
\begin{center}
\begin{tabular}{l c c c c c c}
\hline\hline
device & $T$ (K) & $\seff$ (GPa) & $D_Q$ & $Q_{\rm int}$ & $\hmo V\seff^{2}$ (Pa$^{2}$m$^{3}$) & $R$ \\
\hline
Bartsch 2012, Si beam        & 300 & 0.28 & 1    & --             & $2.8\ee{-14}$ & needs $\Qel=4.1\ee{3}$ \\
Naik 2009, SiC beam          & 40  & 5.3  & 1    & --             & $2.3\ee{-14}$ & needs $\Qel=1.1\ee{3}$ \\
Villanueva 2011, AlN/Mo beam & 295 & 0.43 & 1    & $1.2\ee{3}$    & $1.1\ee{-12}$ & $7.0$ \\
Gavartin 2013, SiN string    & 300 & 0.76 & 83   & $4.5\ee{3}$    & $5.8\ee{-14}$ & $1.6$ \\
Fong 2012, SiN string        & 296 & 0.73 & 106  & $3.8\ee{3}$    & $3.5\ee{-14}$ & $0.80$ \\
Fong 2012, same device       & 78  & 0.73 & 106  & $6.6\ee{3}$    & $4.5\ee{-15}$ & $0.70$ \\
Fong 2012, same device       & 5   & 0.74 & 106  & $2.1\ee{4}$    & $5.0\ee{-16}$ & $3.8$ \\
Maillet 2018, $250~\mu$m, $n{=}3$ & 4.2 & 0.60 & 158 & $2.8\ee{3}$ & $7.0\ee{-16}$ & $1.0$ \\
Maillet 2018, $250~\mu$m, $n{=}5$ & 4.2 & 0.60 & 132 & $3.0\ee{3}$ & $2.0\ee{-16}$ & $0.32$ \\
Maillet 2018, $15~\mu$m      & 4.2 & 0.83 & 10.5 & $1.7\ee{3}$    & $4.0\ee{-16}$ & $0.35$ \\
Wang 2020, compensated fork  & 300 & 1.0  & 12.5 & $2.2\ee{3}$    & $4.0\ee{-13}$ & $5.3$ \\
Wang 2020, uncompensated fork& 300 & 1.8  & 17   & $2.5\ee{3}$    & $4.5\ee{-13}$ & $7.0$ \\
\hline\hline
\end{tabular}
\end{center}
\end{table}

\paragraph{The fluctuation--dissipation relation on the record's strings.} With $R\equiv\hmo V\seff^{2}Q_{\rm int}/(Ek_BT/2\pi)$ and $Q_{\rm int}=Q/D_Q$ from the published $Q$ (Table~\ref{tab:sigmaeff}): Gavartin $1.6$ ($300$~K); Fong $0.80$ ($296$~K), $0.70$ ($78$~K, $Q$ interpolated), $3.8$ ($5$~K); Maillet $1.0$, $0.32$, $0.35$ ($4.2$~K; $n{=}3$, $n{=}5$, $15~\mu$m); Wang forks $5$--$7$ (in-plane squeezing mode; $\seff$ from the stated $1$ and $1.8$~GPa, the eigenfrequency giving no string-mode $\seff$); Villanueva $7.0$ (AlN/Mo stack, $Q=1220$). Because the measured $Q$ cannot exceed $\Qel$, $R$ computed with it bounds the true ratio from below: $R\lesssim1$ is consistent with the relation, $R\gg1$ is a floor the eigenstrain term cannot supply. The bare silicon-nitride strings at $78$--$300$~K satisfy the relation within the $\times2$--$3$ of Ref.~\cite{KanellopulosSchmid2026}; Fong's device at $5$~K exceeds it by $3.8\times$, consistent with their own caveat that structural damping across the two barrier slices fails below $78$~K; the Wang forks and Villanueva's stack exceed it by $5$--$7\times$ and need the modulus term, or a mode-geometry correction, to be explained. For Bartsch and Naik, whose $Q$ is not published, the relation requires $\Qel=4.1\ee{3}$ and $1.1\ee{3}$ respectively, both within the surface-limited range of thin Si and SiC beams.

\paragraph{The modulus channel's share of the loss in a tensioned film.} Under a prestrain $\varepsilon_0=\sigma_0/E$ a modulus fluctuator couples linearly to an added strain, with deformation potential $\gamma_{\rm eff}=v_d\,\delta E\,\varepsilon_0$ [Eq.~(S13) of Ref.~\cite{KanellopulosSchmid2026}], and so contributes a relaxation loss of its own. Eliminating the population between that loss and the modulus term of Eq.~(3) of the Letter, and taking the barrier density of these fluctuators to be the same in the two slices, gives
\[
\phi_{\rm mod}(\sigma_0)=\frac{2\pi\,\sigma_0^{2}\,n\langle v_C^{2}\rangle}{E\,k_BT},
\]
which grows as $\sigma_0^{2}$. Two consequences follow. First, the fluctuation--dissipation form evaluated with the \emph{total} elongation loss already contains the tension-mediated share of the modulus term, $(\sigma_0/\seff)^{2}\,n\langle v_C^{2}\rangle/V$: all of it for a tension-dominated string, none of it for an unstressed beam. In Eq.~(3) of the Letter, $\Qel^{-1}\equiv\phi_{\rm el}$ therefore denotes the loss of the eigenstrain fluctuators alone, $\phi_{\rm el}=Q_{\rm int}^{-1}-\phi_{\rm mod}$; evaluating the second term with the measured $Q_{\rm int}$, as in Table~\ref{tab:sigmaeff}, counts that share twice, which is a further reason that $R$ is a lower bound. Second, since $\phi_{\rm mod}$ cannot exceed the measured loss,
\[
n\langle v_C^{2}\rangle\;\le\;\frac{E\,k_BT}{2\pi\,\sigma_0^{2}\,Q_{\rm int}} .
\]
For stoichiometric silicon nitride at $1$~GPa, with the volume loss $Q_{\rm vol}=2.3\ee{4}$ of Ref.~\cite{KanellopulosSchmid2026} ($E=200$~GPa, $295$~K), the bound is $6\ee{-33}$~m$^{3}$. A cantilever cut from such a film should therefore show $\hV\le6\ee{-33}$~m$^{3}$, more than a decade below the strings of this record ($1.0\ee{-31}$~m$^{3}$), which puts a number on prediction~(4) of the Letter. The free bodies of the record (AlN contour-mode plates, a Si cantilever) sit at $1.4\ee{-32}$~m$^{3}$; they are other materials, and unstressed, so the bound does not apply to them.

\section{Additional comments}
\label{sec:relocated}
\vspace{-0.5cm}
\paragraph{Lorentzian processes are excluded.} --- A Lorentzian process contributes to $h_0$, not to $\hmo$, and thus cannot affect a $\tau$-independent plateau. That single criterion removes equilibrium temperature fluctuations, named in~\cite{Vig1999} as a dominant limit for small resonators (Sec.~S10) --- and is also why the LN-NEMS of~\cite{Yuksel2026,Maksymowych2025}, in which individual defects are resolved, do not appear in Fig.~1 of the Letter. Nothing else is excluded on this ground.

\paragraph{Adsorption--desorption fluctuations are excluded.} --- A signature in $\sA$ arising from adsorption--desorption processes was proposed in \cite{YongVig1989}, and later restated for NEMS in \cite{Vig1999}. Ref.~\cite{Yang2011} examined NEMS devices for this signature and did not find it. Instead, $S_y\propto f^{-1/2}$ was observed and explained theoretically as arising from \emph{adsorbate diffusion} along the surface. Adsorption--desorption processes yield $1/f$ when the residence times are broadly distributed; where diffusion dominates it does not. Accordingly, extrinsic surface-mediated fluctuation mechanisms cannot explain the $1/f$ floor. Ref.~\cite{Sansa2016} reaches the same conclusion by computation, finding both processes to be orders of magnitude below the measured instability.

\paragraph{The audit, quantified.} --- The $\tau$ and mass statistics of the reconstruction are given in Sec.~S2; the one result used here is that a resolved plateau is $\tau$-independent by construction, so the mixed averaging-time convention inflates the apparent scatter (a $\tau^{-1/2}$ branch spans $1.74$ decades) without tilting the fit ($\log\tau$ as a covariate moves the exponent by $+0.005$).

\paragraph{What this compilation can and cannot separate.} --- The record is self-similar: $\log_{10}(S/V)$ and $\log_{10}V$ are $97\%$ collinear ($r=-0.967$, against the $-1/3$ of exact geometric similarity). This separates the two \emph{discrete} surface hypotheses --- $p=-0.661$ for $N\propto S$ and $-0.339$ for site self-averaging, $5.1\,\sigma(p)$ apart and $2.6\,\sigma(p)$ either side of the bulk value --- but leaves an \emph{arbitrary} bulk--surface mixture undetermined (Sec.~S14).

\paragraph{From one defect to the ensemble.} --- Equation~(4) of the Letter models an ensemble of many weak fluctuators, whereas the measurements that fix $\delta_1$ resolve defects one at a time. The bridge between the two regimes, and the conditions under which it holds, are discussed in Sec.~S12.

\paragraph{A barrier-independent prediction.} --- A single species with
barrier $\varepsilon^\ast$ produces a \textit{Debye loss peak} at
$T_Q=\varepsilon^\ast/k_B\ln(1/\omega_0\tau_0)$ and, within a broad distribution of correlation times, contributes its maximum $1/f$ weight at the analysis frequency at
$T_h=\varepsilon^\ast/k_B\ln(\tau_{\rm meas}/\tau_0)$, so
\begin{equation}
\frac{T_Q}{T_h}=\frac{\ln(\tau_{\rm meas}/\tau_0)}{\ln(1/\omega_0\tau_0)}\approx2.8\text{--}3.9 ,
\label{eq:TQTh}
\end{equation}
with $\varepsilon^\ast$ canceling. (Over the plausible attempt-time span, $\tau_0=10^{-14}$--$10^{-12}$~s, the ratio ranges $2.4$--$3.9$; the factor-$\approx\!3$ separation persists; Sec.~S11.) Dissipation for each sub-population should peak at roughly three times the temperature at which the frequency noise peaks. The ratio depends on the device frequency through the denominator alone, so a frequency series converts a single number into a predicted trend with no free parameters.

\paragraph{The dissipation-diluted point.} --- The open star of Fig.~1 of the Letter is drawn but not included in any fit; its provenance and its status as a bound are established in Sec.~S4.

\paragraph{Multimode correlation is not a panacea.} --- Ref.~\cite{Kharbanda2026} shows that the intrinsic flicker of two modes of one membrane resonator can be strongly correlated, and that their difference fluctuates far less than either mode alone. This does not suppress $\hmo$: each mode individually remains on its own $1/f$ floor, and no point of Fig.~1 of the Letter moves. What the correlation permits is a choice of observable --- and the same commonality that cancels the noise in the difference coordinate cancels the signal of any measurand that couples to the two modes in the ratio the fluctuators do. Such a scheme is essentially a gradiometer: it obtains rejection of the
common fluctuations at the price of suppressed responsivity to common signals. It thus wins only for a measurand whose modal coupling ratio differs from that of the noise. The admissible figure of merit is $\sA(\tau)/\mathcal{R}(\tau)$, with $\mathcal{R}$ the responsivity to the intended measurand, measured directly, not assumed (Sec.~S20). By that test, no published two-mode scheme has yet lowered the flicker-limited resolution of a mechanical measurement, though the correlations of~\cite{Kharbanda2026} make the attempt interesting and well-posed.

\paragraph{The cryogenic test of the $Q$ law.} --- The prediction that cooling quartz should lower its flicker floor did not survive cryogenic measurement. Quality factors past $10^{9}$ were attained, but improved stability did \textit{not} follow, and the intrinsic flicker floor above $Q=10^{9}$ remains unresolved. (Sec.~S13.)

\paragraph{Prior art on size scaling.} --- Size scaling of MEMS/NEMS frequency stability was examined theoretically, mechanism by mechanism, in~\cite{Vig1999}, but for other processes --- temperature fluctuations, Johnson noise, adsorption and desorption. What has not been explored is the volume dependence of the intrinsic $1/f$ floor. That is what is tested here, over thirteen decades of volume and five material systems.